\documentclass[preprint,11pt]{elsarticle}
\usepackage{amssymb}
\usepackage{graphicx}
\usepackage{amsthm} 

\usepackage{bm}
\usepackage{rotating}
\usepackage{amssymb}
\usepackage{lscape}
\usepackage{multirow}
\usepackage{natbib}
\usepackage{tfrupee}
\usepackage{subcaption}
\usepackage{amsthm}
\usepackage{comment}
\usepackage[english]{babel}
\usepackage{setspace}
\usepackage{enumerate}
\usepackage{epstopdf}
\usepackage{color}
\usepackage{float}
\usepackage{amsmath}
\allowdisplaybreaks
\usepackage{minted}
\usepackage{scalerel,stackengine}
\usepackage{mathtools}
\usepackage{url}
\usepackage{subcaption}
\usepackage{verbatim}
\usepackage{float}
\usepackage{esvect}
\usepackage{natbib}
\usepackage{amsthm} 

\newtheorem{theorem}{Theorem}

\newtheorem{remark}{Remark}

\usepackage{url}
\usepackage{hyperref}
\hypersetup{colorlinks,linkcolor={blue},citecolor={blue},urlcolor={red}}  
\usepackage{amsmath}
\usepackage{algorithm}
\usepackage{color}
\usepackage{algpseudocode}
\usepackage{amsthm,setspace}
\usepackage{lscape}

\usepackage[numbers]{natbib}
\setcitestyle{numbers,open={[},close={]}}

\usepackage[margin=2.5cm]{geometry}
\usepackage{lineno}

\makeatletter
\def\ps@pprintTitle{%
 \let\@oddhead\@empty
 \let\@evenhead\@empty
 \def\@oddfoot{\reset@font\hfil\thepage\hfil}
 \let\@evenfoot\@oddfoot
}
\makeatletter

\begin{document}
\begin{frontmatter}
\title{Modeling Extreme Events in the Presence of Inliers: A Mixture Approach}
\author[label1]{Shivshankar Nila}
\author[label1]{Ishapathik Das\corref{cor1}}
\ead{ishapathik@iittp.ac.in}
\cortext[cor1]{Corresponding author: Ishapathik Das}
\author[label1]{N. Balakrishna}

\address[label1]{Department of Mathematics and Statistics, Indian Institute of Technology Tirupati, Tirupati, Andhra
Pradesh, India} 
 \begin{abstract}
 In many real-life situations, such as life testing experiments and environmental studies, including rainfall, snowfall, streamflow, and soil moisture, responses contain an excess of zero observations, which are known as inliers at zero. Standard modeling approaches fail to provide expected results if these inliers are not considered in the model. Similar challenges arise in extreme value analysis, where accurate tail estimation depends on the appropriate threshold selection. Inliers can adversely affect both threshold estimation and tail behaviour modeling if not properly considered. Although some extreme value mixture models address threshold and tail estimation, they often ignore inliers or don't account thoroughly, leading to suboptimal results. There is no unified framework for defining extreme value mixture models; the misspecification of the bulk model can affect the threshold, tail estimates, and, particularly, the tail proportion. This paper introduces a framework for modeling extreme events, addressing threshold uncertainty, and capturing inliers at zero. The model parameters are estimated using maximum likelihood estimation, ensuring precision. We compare the proposed model with classical mean excess and parameter stability plot estimates. Theoretical results are established, and the model's utility is shown through real applications. Simulation studies and real data examples demonstrate that the proposed model significantly outperforms traditional methods, which typically neglect inliers.
 \end{abstract}
\end{frontmatter}
\section{Introduction}\label{intro}
\noindent In an age where data drives decision-making, the main purpose of data science is to find patterns in vast datasets. It helps identify patterns in data using various statistical methods to gain insights. However, if data are modeled as realizations of a random variable or vector, then determining the distributions from which they come is important in statistical theory.
 Recently, many new distributions have been introduced, providing more options for data modeling.
 Various statistical tools are used to gain insight into the patterns of the data; however, most of the time, the interest lies in estimating the central characteristics of the population based on a random sample taken from it. Standard analysis tools often disregard or exclude extreme values in conventional data analysis. Extremes mean roughly exceptionally large or small values or high magnitudes with low frequency. However, in some cases, we are interested in estimating extreme effects only, which are the maximum or minimum effects of an event, such as floods, temperature, earthquakes, intensity, and stock prices, as discussed in {\cite{castillo2005extreme}}. 
 
 The main purpose of extreme value theory is to provide a framework to identify and understand the ``statistical pattern'' of exceptionally large or small values and develop tools to handle them. It also helps in characterizing extremal events and in gaining a deeper understanding of the extreme behaviour of variables of interest (i.e., the behaviour of extreme data points), as described in \cite{coles2001introduction}. Even though extreme value theory (EVT) is an asymptotic and valuable approach for modeling and quantifying extreme events, in univariate extreme value modeling, a key challenge lies in selecting an appropriate threshold beyond which EVT models can accurately approximate the tail behaviour of the population as given in  \citep{hu2018evmix, scarrott2012review}. A low threshold leads to a poor approximation of the generalized Pareto distribution (GPD) and biased return level estimates, while a high threshold increases parameter variance due to fewer observations, as given in \cite{roth2016threshold}. Although choosing the right threshold can be challenging, many approaches have been explored in the literature. In the numerical approach method for threshold selection, a common choice for threshold selection in extreme value analysis is the 90th percentile.  Leadbetter \emph{et al.}~\cite{ledbetter1983extremes} demonstrated that the threshold sequence depends on the distribution’s properties for populations in the domain of attraction of the GPD. Several practical rules of thumb have since been proposed based on the convergence of order statistics. The advantage of this method is the ease of threshold selection. Rule of thumb involves selecting the $k$ largest observations, where $k$ can be chosen as  $\sqrt{n}$ or $\frac{\sqrt[2/3]{n}}{\log(\log(n))}$ \cite{benito2023assessing}. It is often unreliable because each data has its own nature. Several graphical diagnostics used for threshold selection in extreme value analysis include the mean residual life (or mean excess) plot, the parameter stability plot, and the Hill plot. The threshold is chosen to be the lowest level at which all higher threshold-based sample mean excesses are consistent with a straight line. The main problem associated with the sample mean excess plot is subjectivity. As Chukwudum \emph{et al.}~\cite{chukwudum2020optimal}
 remark, ``judging from where the graph is approximately linear using only the eyeball inspection approach, is a rather subjective choice so that different viewers of the plot may select different thresholds''.
 In graphical diagnostic plots or rule-of-thumb methods for threshold estimation, the threshold is chosen first, assuming it to be known and constant; the parameters of the GPD are then estimated based on this fixed-threshold assumption, completely ignoring the uncertainty in the threshold. This approach disregards the associated subjectivity and uncertainty in subsequent inferences.  However, Behrens \emph{et al.}~\cite{behrens2004bayesian} proposed an extreme value mixture model (EVMM); in the EVMM, the threshold itself is treated as a parameter to be estimated, considering the uncertainty associated with the threshold estimation; the model and details are given in this paper, in Subsection \ref{evmm}.  Frigessi \emph{et al.}~\cite{frigessi2002dynamic}  proposed a weighted dynamic model that combines a Weibull distribution for modeling the bulk with a GPD for the tail. Various extreme value mixture models have been proposed in the literature, each specifying a bulk distribution complemented by a GPD for the tail. For example, Behrens \emph{et al.}~\cite{behrens2004bayesian} formulated models using the gamma and Weibull distributions for the bulk, paired with a GPD for the upper tail. Similarly, MacDonald \emph{et al.}~\cite{macdonald2011extreme} employed the beta distribution as the bulk component with a GPD for tail modeling. Further, Cabras \emph{et al.}~\cite{cabras2011bayesian} proposed a mixture of the normal distribution with a GPD for extreme values, while Solari \emph{et al.}~\cite{solari2012unified} adopted the log-normal distribution as the bulk with a GPD for the tail.  Lee \emph{et al.}~\cite{lee2012modeling} used an exponential model for data below a threshold and utilized a GPD to model the tail. Slightly earlier, Do Nascimento \emph{et al.}~\cite{do2012semiparametric} proposed a semi-parametric Bayesian method by employing a mixture of gamma distributions below the threshold. Dirichlet process mixture of gamma densities for bulk proposed by F{\'u}quene Pati{\~n}o \emph{et al.}~\cite{fuquene2015semi}. The first proposal to use the full dataset to estimate both the threshold location and the tail of the distribution is due to \cite{behrens2004bayesian}, which used a gamma for the bulk. Since then, a variety of proposals for the bulk have been used, including an infinite mixture of betas \citep{tancredi2006accounting}, a Normal distribution \citep{carreau2009hybrid}, a kernel estimator \citep{macdonald2011extreme}, and a mixture of gamma distributions\citep{do2012semiparametric}.

 There are many random phenomena in real-life situations, such as in life-testing experiments, where an item fails instantaneously, and the observed lifetime is reported as zero. Similarly, in the case of rainfall data, there may be days reported with zero rainfall corresponding to the dry days
during the rainy season, which can lead to severe events or extreme events like drought.  In such contexts, inliers often emerge as a subset of observations that appear statistically inconsistent with the rest of the data under traditional single-component models, particularly those with strictly positive support.  Many real-world data sets exhibit a high concentration of observations near a particular point \( x_{0} \), while the remaining responses follow varying distributions. We refer to such observations as ``early failures'' or ``inliers'' near the point \( x_{0} \). When \( x_{0} = 0 \) and observation at particular \( x_{0} = 0 \), this is specifically termed an ``instantaneous failure'' or an ``inlier at zero''. 
To address such data characteristics, Muralidharan \emph{et al.}~\cite{muralidharan2006analysis} proposed a model combining the Weibull distribution with a degenerate distribution at zero.  Aitchison \emph{et al.}~\cite{aitchison1955distribution}
 investigated and derived efficient estimates for the mean and variance of a distribution that assigns a non-zero probability at zero. In this context, the random variable exhibits a non-zero probability at zero, with the remaining distribution representing a well-known positive-valued distribution, either continuous or discrete. Ramos, P \emph{et al.}~\cite{ramos2019distribution} introduced a new distribution for modeling instantaneous failures in a univariate context. In the bivariate context, Bhattacharya, S \emph{et al.}~\cite{bhattacharya2024modeling} developed a modified bivariate Weibull distribution to model lifetime data with particular consideration for scenarios involving an inlier at origin $[(x_{0},y_{0})=(0,0)]$ and near origin.

 A key consideration in EVMM is that the performance of a given mixture or composite model depends on the characteristics of the underlying data. In other words, the best model depends on the underlying data being used to fit the model and the associated quantile or risk, as shown by Marambakuyana \emph{et al.}~\cite{marambakuyana2024composite}.
It is crucial to study not only extreme events but also non-extreme events that can have extreme effects or severe impacts. Accurate modeling of the entire distribution is important when both extreme and non-extreme data are of interest, as discussed by Andr{\'e} \emph{et al.}~\cite{andre2024joint}. However, even though inliers are a crucial part of the data, EVMMs exclude them from their analytical process, which can lead to incomplete insights. These inlier factors are often overlooked in single-tail models with positive support or are not thoroughly studied when considering the entire support range. Moreover, the literature on inliers, where mixture models are used to incorporate inliers with standard distributions for the remaining data, is not specifically designed to effectively capture extreme values.

Hu \emph{et al.}~\cite{hu2013extreme} discussed in detail that there is no generalized framework for defining these extreme value mixture models. In particular, there is no consistency in terms of how the proportion of the population (say, $\phi_u$) above the threshold is defined in the model. The robustness of the bulk fit and tail fit is a concern. When the bulk model is misspecified, it affects not only the tail fraction but also the threshold estimate and subsequently influences the estimates of the extreme parameters. Behrens \emph{et al.}~\cite{behrens2004bayesian} have discussed, "It is important to analyze if the chosen form (for below the threshold) fits data from different distributions and influences the estimates of the threshold and the extreme parameters, and it is recommended to have a robust model in order to fit several different situations, usually encountered in practice," and the same point is reiterated by \cite{hu2018evmix} also.

\begin{figure}[!htbp]
    \centering    \includegraphics[width=1.1\textwidth,height=16cm]{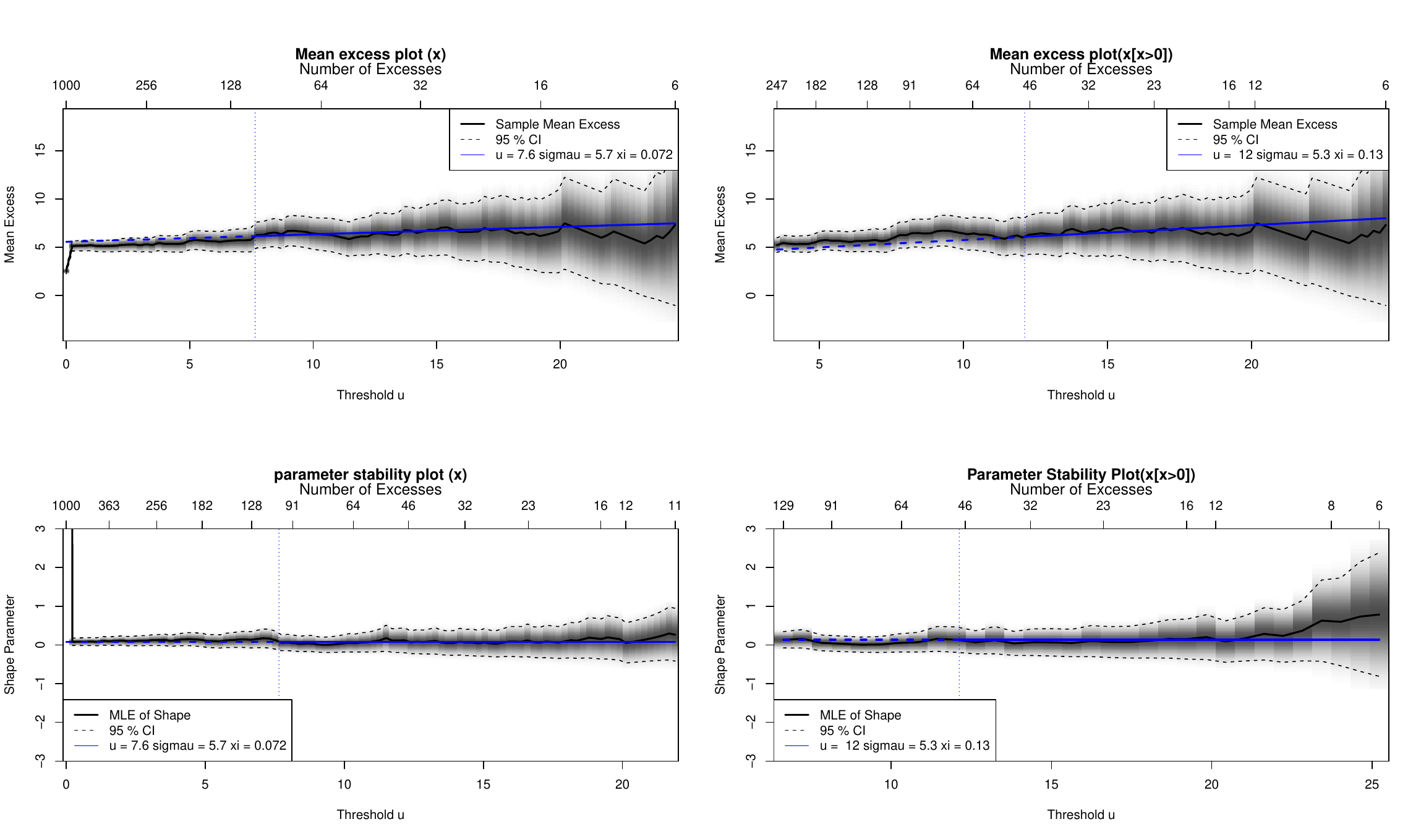}
\caption{ MEP and PSP based on the full dataset including inliers (\(x\)) and the reduced dataset excluding inliers (\(x[x > 0]\)). The plots illustrate how the presence of inliers may influence the estimate of the threshold (\(u\)), subsequently affecting the estimates of the scale (\(\sigma\)) and shape (\(\xi\)) parameters of the GPD.}
    \label{fig:meanexcess}
\end{figure}

One goal of our work is to analyse whether the choice of the distribution for observations below the threshold influences, and how, the threshold estimation. In addition, we are interested in analysing whether the proposed model exhibits good data fitting when compared with other analyses presented in the literature. Another goal of this paper is to study all aspects of robustness, goodness of fit, and parameter estimation. In this study, we compare our proposed  extreme value inlier mixture model (EVIMM) with existing models, specifically the EVMM of \cite{behrens2004bayesian},  the flexible extreme value mixture model (FEVMM) of \cite{macdonald2011flexible}, with respect to the following key aspects:
\begin{enumerate}
    \item Does the proposed model,  EVIMM, provide more accurate and reliable tail estimation compared to existing models?
    \item Does the proposed framework improve the accuracy of threshold estimation?
  \item Does the proposed framework help reduce the bias in the estimation of model parameters?
    \item Does the proposed framework help reduce the mean squared error (MSE) in the estimation of model parameters?
  \item How well does the proposed model perform in goodness-of-fit tests, such as the Anderson–Darling, Cramér–von Mises, and Kolmogorov–Smirnov tests?
\end{enumerate}
To further show the need for a framework like the one we proposed, we first examined the parameter estimates from classical methods, such as the mean excess plot (MEP) and parameter stability plot (PSP), on two versions of the data: one with inliers and one without (i.e., only positive observations). This helped us understand how sensitive these methods are to the presence of inliers, especially for the threshold ($u$), scale ($\sigma$), and shape ($\xi$) parameters of the GPD.  Some detailed results, shown in Tables~\ref{Table-700} to~\ref{Table-1000}, indicate that the MEP and PSP plots give different estimates depending on whether inliers are included. After that, we compared the EVIMM estimates with those of the classical methods using the complete data. EVIMM gave more reliable estimates that were closer to the actual values, especially for the shape parameter of the GPD. 
For example, we compared the estimates obtained from the MEP and PSP using two versions of the same simulated dataset: one including inliers (i.e., the full dataset \(x\)), and the other excluding inliers, denoted by \(x[x > 0]\), which retains only the non-zero values. The variation in estimates shows that inliers might be influencing the results, as illustrated in Figure~\ref{fig:meanexcess}.
\noindent To study the key aspects mentioned above, our proposed EVIMM combines a singular distribution at the origin, a bulk distribution below the threshold (excluding the origin), and a GPD above the threshold, addressing the uncertainty in the threshold. All these aspects of robustness, goodness of fit, and parameter estimation
are considered in this paper.

The rest of this article is organized as follows. In Section \ref{metho}, Subsection \ref{evt} provides an overview of EVT, focusing on two main approaches used in the EVT field. Subsection \ref{evmm} describes the EVMM presented in the literature. Subsection \ref{evimm} introduces the proposed EVIMM, particularly emphasizing its handling of inliers. In Subsection \ref{algorithm}, the algorithm is given to simulate the data of the proposed model. In Subsection \ref{estimation}, we discuss the estimation of the model parameters. To demonstrate the practical application of the proposed model, Section \ref{numericalexample1} presents numerical examples, including simulation studies and real-world data applications. Finally, Section \ref{conclusion} summarizes our findings and suggests directions for future research.

\section{Methodology}\label{metho}

   \subsection{Extreme Value Theory}\label{evt}
\noindent Extreme Value Theory (EVT) is primarily concerned with identifying and analyzing the \emph{statistical behaviour} of exceptionally large or small observations within a dataset, providing tools to model and manage such occurrences. EVT focuses on the asymptotic behaviour of extreme values of random variables and is broadly categorized into two approaches. 

The first approach, known as the \emph{block maxima (BM) method}, utilises the generalized extreme value (GEV) distribution to model the distribution of block-wise maxima.\\
\emph{Fisher–Tippett Theorem:}
Let \( X_1, X_2, \dots, X_n \) be a sequence of independent and identically distributed (i.i.d.) random variables with cumulative distribution function (CDF) \( F(x) = \Pr(X \leq x) \). The maximum of the sample, \( M_n = \max(X_1, X_2, \dots, X_n) \), has the CDF:
\[
\Pr(M_n \leq x) = \Pr(X_1 \leq x, \dots, X_n \leq x) = \prod_{i=1}^n F(x) = [F(x)]^n.
\]
The Fisher–Tippett theorem states that, for appropriate normalizing constants \( a_n > 0 \) and \( b_n \in \mathbb{R} \), the normalized maximum \( Z_n = \frac{M_n - b_n}{a_n} \) converges to a non-degenerate limiting distribution \( H(x) \), known as the generalized extreme value distribution,
$\lim_{n \to \infty} \Pr\left( \frac{M_n - b_n}{a_n} \leq x \right) = H(x)$,  for which an  algebraic expression is given below.\\
\emph{Generalized Extreme Value (GEV) Distribution:}\\
The Gumbel, Frechet, and Weibull families can be combined into a single family of models, represented by a distribution function denoted as $H_{\xi}(x)$, and  can be written in the following form,
\begin{equation}
\label{eq:myotherequation5}
H_{\xi}(x) = \exp\left\{-\left(1 + \xi\left(\frac{x - \mu}{\sigma}\right)\right)^{-\frac{1}{\xi}}\right\}
\end{equation}
where \( 1 + \xi\left(\frac{x - \mu}{\sigma}\right) > 0 \), \( \mu \in \mathbb{R} \), \( \sigma > 0 \), and \( \xi \in \mathbb{R} \).
The shape parameter \( \xi \) governs the tail behaviour: \( \xi = 0 \) corresponds to the Gumbel (light-tailed) distribution, \( \xi > 0 \) to the Fréchet (heavy-tailed) distribution, and \( \xi < 0 \) to the Weibull (bounded support) distribution.
  
The second and popularly used method for modeling extremes is the \textit{Peaks Over Threshold (POT)}, which studies the exceedances over a threshold. This approach leverages data more effectively in the extreme range, providing better estimations in practical applications, as demonstrated by \citep{cunnane1973particular, madsen1997comparison}. Further support for POT's superiority over BM is discussed by  \citep{fischer2018seasonal, wang1991pot}. Additionally, it has been extensively reviewed by \citep{caires2009comparative, scarrott2012review}.  A key result \cite{Pickands1975} states that if a random variable \(X\), with right endpoint \(x_0\), is in the domain of attraction of a generalized extreme value (GEV) distribution, then
\[
\lim_{u \rightarrow x_0} \mathbb{P}(X \leq x + u \mid X > u) = G(x),
\]
where \(G\) is the CDF of the GPD. The \textit{POT method} employs the GPD to characterize values exceeding a specified threshold. The GPD with a scale parameter \( \sigma_u > 0 \) and a shape parameter \( \xi \) is expressed as:
\[
G(x; u, \sigma, \xi) = 
\begin{cases}
    1 - \left(1 + \frac{\xi(x - u)}{\sigma}\right)^{-1/\xi}, & \text{if } \xi \neq 0, \\
    1 - e^{-(x - u)/\sigma}, & \text{if } \xi = 0,
\end{cases}
\]
where \( u, \xi \in \mathbb{R} \) and \( \sigma > 0 \). The support is \( x \geq u \) if \( \xi \geq 0 \), and \( 0 \leq x \leq u - \sigma/\xi \) if \( \xi < 0 \). The GPD is unbounded from above when \( \xi \geq 0 \) and bounded when \( \xi < 0 \).
The GPD is defined
conditionally on being above the threshold $u$.
While this threshold is often explicitly stated in the literature, a third implicit parameter is also involved in most applications, that is, the probability of exceeding the threshold, denoted as \( \phi_u = P(X > u) \). The unconditional survival probability can then be expressed as:  
\begin{equation}
\label{eq:survival_probability}
\Pr(X > x) = \phi_u \left[1 - \Pr(X < x \mid X > u)\right],
\end{equation}
\noexpand where the parameter \( \phi_u \) is typically estimated using the maximum likelihood method, which calculates the proportion of observed data points above the threshold.

\subsection{Extreme Value Mixture Model }\label{evmm}
\noindent To address the challenges of selecting a threshold in univariate extreme value analysis and to account for the related uncertainty, many models known as \textit{extreme value mixture models (EVMM)} have been proposed. The detailed review of various EVMMs available in the literature, see \cite{scarrott2012review}. The \texttt{evmix} R package \cite{hu2018evmix} provides implementations for most of these models. These models do not require manual threshold selection. EVMM combines a flexible
model for the bulk of the data below the threshold, a GPD for the
tail and uncertainty measures for the threshold.
Behrens \emph{et al.}~\cite{behrens2004bayesian} proposed an EVMM of the form \eqref{eq:myotherequation7}; the model splices together a parametric model called a ``bulk model'' for data below the threshold, with a GPD for extreme values beyond the threshold, called as the ``tail model''.

The distribution  function $F$ of an EVMM can be generally defined  as

\begin{equation}
\label{eq:myotherequation7}
F(x|\varphi,\theta)=
\begin{cases}
H(x|\varphi), & \text{if } x \leq u,
\\
H(u|\varphi)+[1-H(u|\varphi)] \cdot G(x|\theta), & \text{if } x > u,
\end{cases}
\end{equation}
\begin{figure}[htbp]
\begin{center}
\includegraphics[scale=0.3]{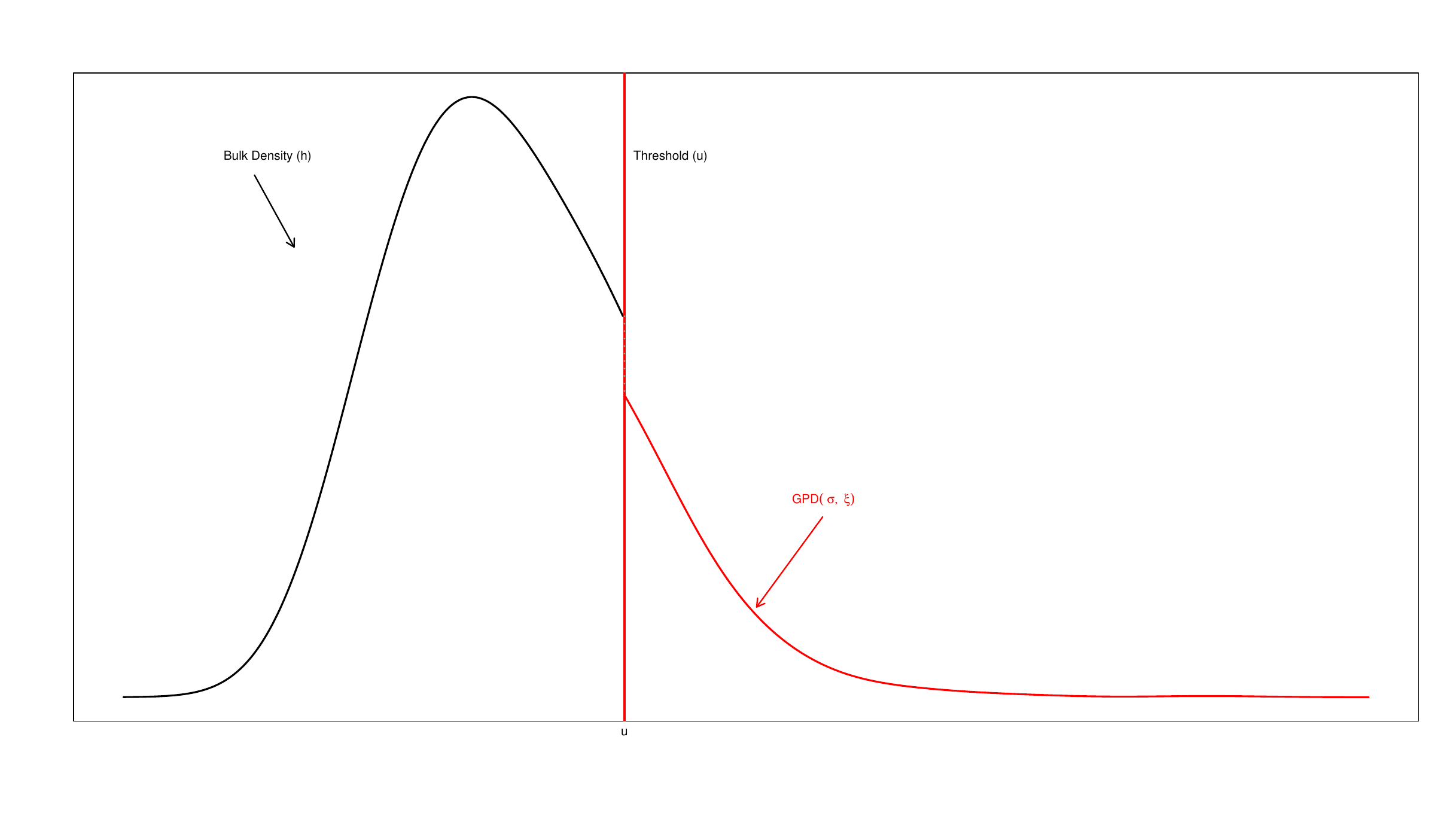}
\end{center}
\vspace{-0.5cm}
\caption{Distribution of an EVMM with bulk density $h$ and a GPD as tail density.}
\label{fig:evmm}
\end{figure}
where \( h \) and \( H \) denote the density and distribution functions, respectively, parameterized by \( \varphi \), representing the bulk distribution of the data (i.e., the portion below the threshold \( u \)). Additionally, \( g \) refers to the GPD density function with parameters \( \theta = \{\xi, \sigma, u\} \), which models the tail of the distribution above the threshold \( u \), and \( G \) denotes its CDF. Figure \ref{fig:evmm} displays the density plot of the EVMM. MacDonald \emph{et al.}~\cite{macdonald2011flexible} extended the EVMM to a more flexible form, known as the FEVMM, in which the tail fraction \( \phi_u \), representing the proportion of the GPD, is treated as a parameter. The tail fraction \( \phi_u \) in equation (\ref{eq:survival_probability}) is defined as the survival probability of the bulk model beyond the threshold, given by \( \phi_u = 1 - H(u|\theta) \). This is referred to as the ``bulk model-based tail fraction'' approach, otherwise called the ``parameterized tail fraction''. For more details, see \citep{hu2018evmix, macdonald2011flexible}.
\subsection{Proposed Model}\label{evimm}
\noindent In this section, we propose a model in which observations below the threshold \( u \) are assumed to come from a specific distribution, either a bulk \(\bm{G}^*(\cdot \mid \varphi)\) or a degenerate distribution, while observations above the threshold \( u \) are assumed to come from a GPD, denoted as \(\bm{G}(. \mid \xi, \sigma, u)\). The CDF \(F(x)\) of the EVIMM model is defined as 
\begin{equation}\label{evimmpdf}
\bm{F}(x \mid \alpha, \varphi, u, \xi, \sigma) = 
\begin{cases} 
\bm{0}, & \text{if } x < 0 , \\
    \bm{\alpha}, & \text{if } x = 0, \\
    \bm{\alpha} + (1 - \bm{\alpha}) \cdot \bm{G}^*(x \mid \varphi ), & \text{if } 0 < x < u, \\
    \bm{\alpha} + (1 - \bm{\alpha}) \cdot \bm{G}^*(u \mid \varphi ) + \Big(1 - \big( \bm{\alpha} + \\
    (1 - \bm{\alpha}) \cdot \bm{G}^*(u \mid \varphi ) \big) \Big) \cdot \bm{G}(x \mid \xi, \sigma, u), & \text{if } x \geq u.
\end{cases}
\end{equation}
\subsection*{Components of the Model:}
\noindent The EVIMM consists of three main components, each designed to capture distinct features of the data distribution:

\paragraph{1. Degenerate Distribution at the Origin:}  
A probability mass \( \alpha \) at \( x = 0 \) explicitly accounts for inliers at the origin, capturing non-extreme observations that are a key feature in mixed distributions.
\paragraph{2. Bulk Distribution:}  
This distribution captures the central tendency of the data, excluding extreme values and inliers at zero. The bulk representing moderate values, excluding extremes and inliers at zero, is modeled using gamma, Weibull, beta, and lognormal distributions. For this paper, we focus on the gamma distribution for simulations and estimations, denoted as \(\bm{G}^*(x \mid \eta, \beta)\), where \(\eta > 0\) is the shape parameter and \(\beta > 0\) is the scale parameter. 
\paragraph{3. Tail Distribution:} 

This distribution models the data above the threshold \( u \) called tail distribution, using the GPD to describe extreme values. The GPD is expressed as \( \bm{G}(x \mid \xi, \sigma, u) \), where \( \xi \) is the shape parameter determining the tail's heaviness, \( \sigma > 0 \) is the scale parameter controlling the distribution's spread, and \( u \) is the threshold separating the bulk from the tail.

\begin{figure}[htbp]
        \centering {\includegraphics[width=18cm]{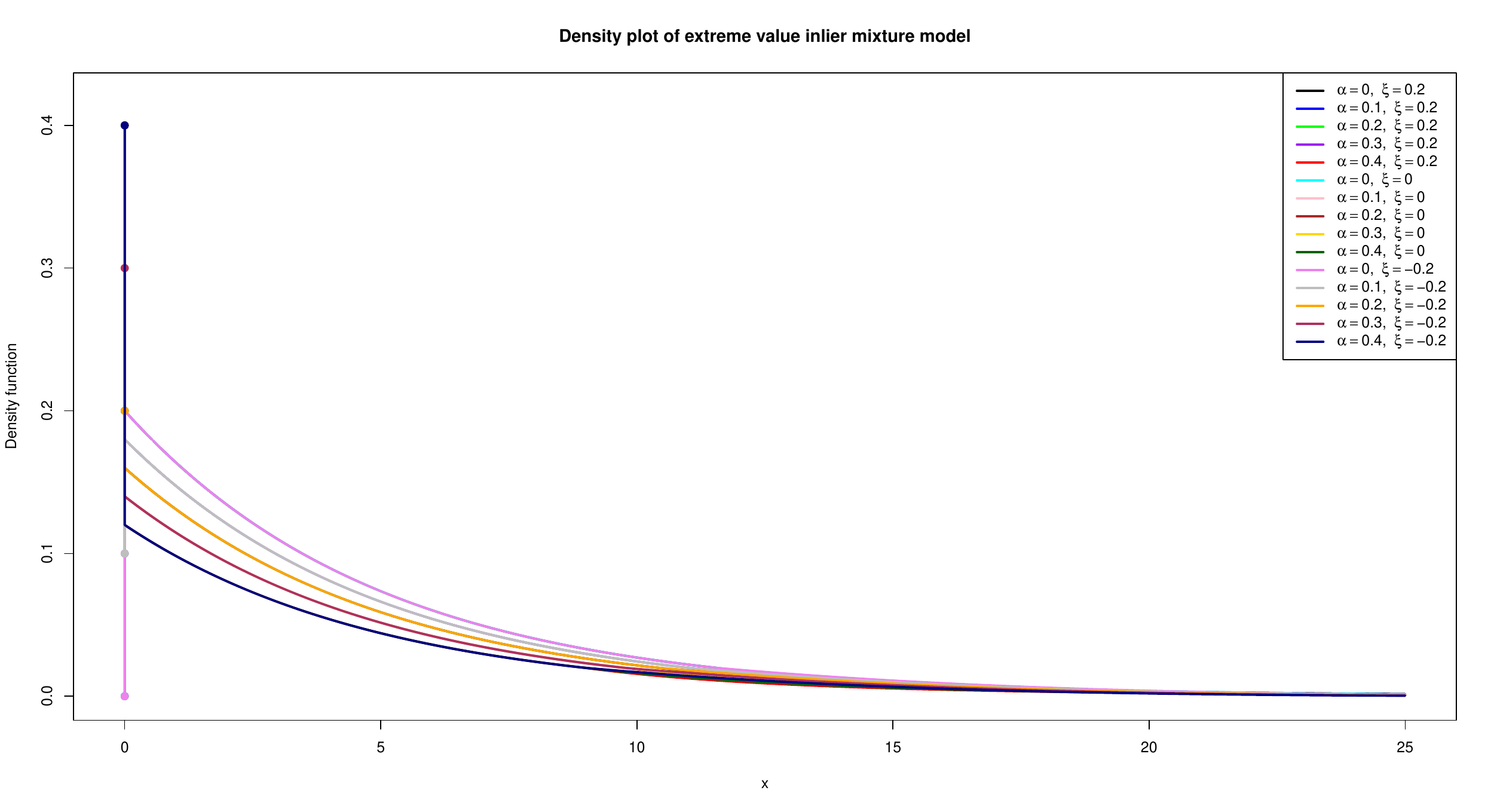}}
		\hspace{6em}
       
\caption{Density plots of the EVIMM for various parameter values: $\alpha \in \{0, 0.1, 0.2, 0.3, 0.4\}$, $\eta = 1$, $\beta = 5$, $\xi \in \{0.2, 0, -0.2\}$, $\sigma = 5$, and thresholds $u \in   \{11.5129,10.9861,10.3972,9.7296,8.9588\}$.}

     \label{fig:desnistyevimm}
\end{figure}

The density function \(f(x)\) of the proposed EVIMM, defined on the domain \(x \ge 0\) and \(0 < \alpha < 1\), is given by:

\begin{equation}
\label{eq:evmmlpdf}
\bm{f}(x \mid \alpha, \eta, \beta, u, \xi, \sigma) = 
\begin{cases}
    \bm{\alpha}, & \text{if } x = 0, \\[10pt]
    \bm{(1-\alpha)} \cdot \bm{g}^*(x \mid  \eta, \beta), & \text{if } 0 < x < u, \\[10pt]
    \Big(1 - \big(\bm{\alpha} + (1 - \bm{\alpha}) \cdot \bm{G}^*(u\mid \eta, \beta)\big)\Big) \cdot \bm{g}(x \mid \xi, \sigma, u), & \text{if } x \geq u.
\end{cases}
\end{equation}
When choosing between parametric and semi/nonparametric families for the bulk distribution in these extreme value mixture models, the usual consideration is that if good prior information is available, better estimates are obtained with an appropriate parametric bulk model. Most of the mixture models use a bulk model-based tail fraction; this approach has the benefit of using the ample data in the bulk of the distribution.
The proposed EVIMM framework offers a key advantage by accounting for inliers concentrated at zero, which improves the modeling of extreme values and enhances the overall accuracy and comprehensiveness of the data analysis. 
The density function in \eqref{eq:evmmlpdf} can be expressed as a three-component mixture model:
\[
f(x) = \phi_1 \cdot \delta_0(x) + \phi_2 \cdot f_1(x) + \phi_3 \cdot f_2(x),
\]
where
\[
\phi_1 = \alpha, \quad 
\phi_2 = (1 - \alpha) \cdot \bm{G}^*(u \mid \Phi), \quad 
\phi_3 = (1 - \alpha) \cdot \left[1 - \bm{G}^*(u \mid \Phi)\right],
\]
\[
f_1(x) = \frac{\bm{g}^*(x \mid \Phi)}{\bm{G}^*(u \mid \Phi)} \cdot \mathbb{I}_{(0, u)}(x), \quad 
f_2(x) = \bm{g}(x \mid \xi, \sigma, u) \cdot \mathbb{I}_{[u, \infty)}(x),
\]
here, \( \delta_0(x) \) represents the Dirac delta function at zero, representing the degenerate distribution at the origin. The density \( \bm{g}^*(\cdot \mid \Phi) \) corresponds to the bulk distribution \( \bm{G}^*(\cdot \mid \Phi) \), while \( g(\cdot \mid u, \sigma, \xi) \) represents the density of GPD. The density plot of the EVIMM for different parameter sets is shown in Figure \ref{fig:desnistyevimm}.
\begin{figure}[htbp]
        \centering {\includegraphics[width=18cm]{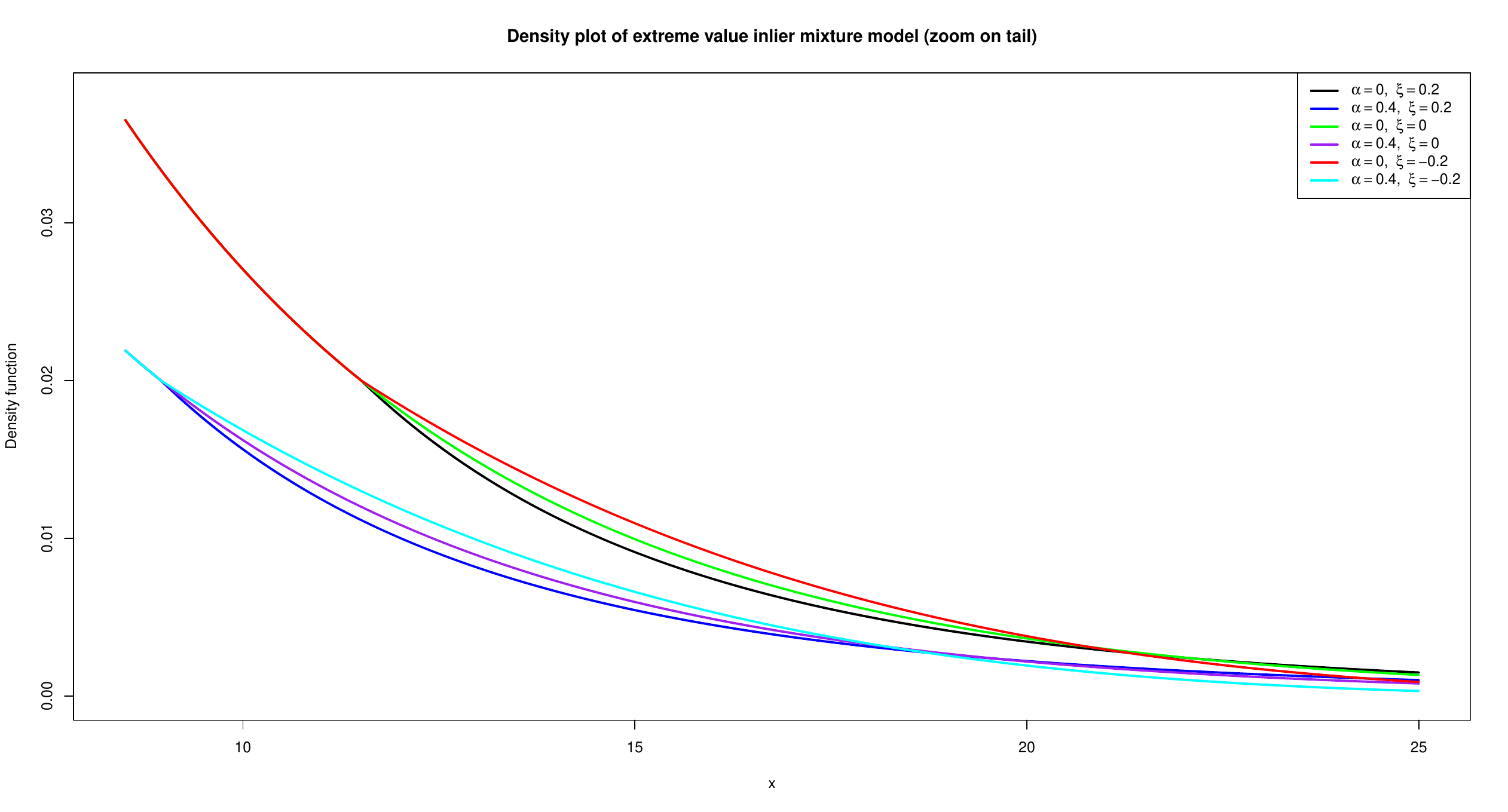}}
		\hspace{6em}
    \caption{Tail behaviour of the EVIMM for various parameter values, with a zoomed-in view of the tail region. Parameters: $\alpha \in \{0, 0.4\}$, $\eta = 1$, $\beta = 5$, $\xi \in \{0.2, 0, -0.2\}$, $\sigma = 5$, and $u \in   \{11.5129,8.9588\}$.}
     \label{fig:desnistyevimm1}
\end{figure}

Figures~\ref{fig:desnistyevimm} and~\ref{fig:desnistyevimm1} show that as the proportion of inliers increases, the shape of the distribution changes across all cases-heavy-tailed, light-tailed, and exponential-tailed. While the total probability remains one, some mass shifts toward the inliers. This redistribution effect has not been thoroughly explored in previous studies within the context of modeling in EVT.

\subsection{Algorithm for Data Simulation}\label{simulation}\label{algorithm}

\noindent This subsection provides an algorithm for simulating data $x_i$, $i = 1, 2, 3, \dots, n$, from the EVIMM with the CDF \(F(x \mid \alpha, \eta, \beta, u, \xi, \sigma)\)(\ref{evimmpdf}). The simulation steps are described below.
\begin{enumerate}  

    \item \text{ A random sample \( U \) is generated from the standard uniform distribution, i.e., \( U \sim \text{uniform}(0,1) \).}  

    \item \text{Sampling based on the value of \( U \).}  

    Depending on the value of \( U \), the value of \( x \) is determined as follows:  

    \begin{itemize}  
        \item \text{Case 1:} If \( U \leq \bm{\alpha} \), set \( x = 0 \).  

        \item \text{Case 2:} If \( \bm{\alpha} < U \leq \bm{\alpha} + (1 - \bm{\alpha})\bm{G}^*(u \mid \eta, \beta) \), compute \( x \) using the inverse of the gamma CDF, given by:  
        \[
        x = \bm{G}^{*-1}\left(\frac{U - \bm{\alpha}}{1 - \bm{\alpha}} \mid \eta, \beta\right),
        \]  
        where \( \bm{G}^{*-1} \) denotes the quantile function of the gamma distribution.  

        \item \text{Case 3:} If \( U > \bm{\alpha} + (1 - \bm{\alpha})\bm{G}^*(u \mid \eta, \beta) \), proceed as follows:  
        \begin{enumerate}  
            \item Calculate the cumulative probability up to \( u \):  
            \[
            C_u = \bm{\alpha} + (1 - \bm{\alpha}) \bm{G}^*(u \mid \eta, \beta).
            \]  

            \item Transform \( U \) to scale it within the region \( x \geq u \):  
            \[
            \frac{U - C_u}{1 - C_u} = \bm{G}(x \mid \xi, \sigma, u).
            \]  

            \item Compute \( x \) by inverting the CDF of the GPD \( \bm{G}(x \mid \xi, \sigma, u) \):  

            \text{For \( \xi \neq 0 \):}  
            \[
            x = u + \frac{\sigma}{\xi} \left[\left(1-\frac{U - C_u}{1 - C_u}\right)^{-\xi} - 1\right].
            \]  

            \text{For \( \xi = 0 \):}  
            \[
            x = u - \sigma \log\left(1-\frac{U - C_u}{1 - C_u}\right),
            \]  

            where \( x \) corresponds to the quantile of the GPD for the scaled value of \( U \).  
        \end{enumerate}  
    \end{itemize}  

    \item \text{Repetition for multiple samples:}  
    Generate a random sample of size \( n \) from the EVIMM and repeat Steps 1 and 2, \( n \) times.

\end{enumerate}  
\subsection{Estimation of Parameters}\label{estimation}
\noindent In this section, we focus on the estimation of the parameters of the EVIMM defined in the previous section. The goal is to derive the (maximum likelihood estimator) MLE for the parameters, let \( \Theta = (\alpha, \eta, \beta, \mathbf{u}, \xi, \sigma) \), which characterize the distribution of the observed data. To achieve this, we specify the likelihood function and then develop the corresponding MLE based on a random sample of size \( n \). 

In MLE, direct likelihood optimization for such mixture models is often challenging due to the multimodal nature of the likelihood surface, especially concerning the threshold parameter. One can use the profile likelihood and grid search approaches to focus on selecting the threshold by maximising the likelihood; however, these methods do not provide uncertainty quantification for the estimated threshold, as discussed in Subsection 2.3 of \cite{hu2018evmix}. In contrast, approaches based on the complete likelihood, where the threshold is estimated jointly with the other model parameters, allow for uncertainty quantification. The complete likelihood approach is adopted in \cite{hu2013extreme}, \cite{hu2018evmix}, and in the \texttt{evmix} R package when accounting for uncertainty is required.

To mitigate the challenges of optimization and uncertainty quantification, we follow the initialization strategies proposed by Hu in his thesis \cite{hu2013extreme} and subsequent work \cite{hu2018evmix}, as implemented in the \texttt{evmix} R package, which provides carefully grounded and empirically derived starting values. Moreover, we increased the maximum number of iterations to 40{,}000 in the \texttt{optim} function in \texttt{R} to ensure that the optimization process is not prematurely terminated. Importantly, we only use the model output for quantile estimation when the optimization algorithm for the likelihood has converged.
\subsection*{Maximum Likelihood Function}\label{mle}
\noindent Let \( \bm{x} = (x_1, x_2, \dots, x_n) \) be a random sample from the population with the CDF \( F \) as defined in equation (\ref{evimmpdf}). The likelihood function \( L(\bm{x}; \Theta) \) is given by: 
\[
L(\bm{x}; \Theta) = \prod_{i=1}^n f(x_i; \alpha, \eta, \beta, u, \xi, \sigma),
\]
where \( f(x_i; \alpha, \eta, \beta, u, \xi, \sigma) \) represents the density function of the EVIMM. We introduce indicator functions that separate the data into distinct subsets to handle different observation ranges. Let 
\[
{A(u) = \{ i \mid 0 < x_i < u \}, \quad B(u) = \{ i \mid x_i \geq u \},}
\]
\[
I_1(x_i; u) = \begin{cases} 
1 & \text{if } x_i = 0, \\
0 & \text{otherwise},
\end{cases} \quad
I_2(x_i; u) = \begin{cases} 
1 & \text{if } 0 < x_i < u, \\
0 & \text{otherwise},
\end{cases} ~~\text{and}~~ 
I_3(x_i; u) = \begin{cases} 
1 & \text{if } x_i \geq u, \\
0 & \text{otherwise}.
\end{cases}
\]
\newcommand{\indone}{\ind{{\color{red}x_i = 0}}}
\newcommand{\indtwo}{\ind{{\color{red}0 < x_i < u}}}
\newcommand{\indthree}{\ind{{\color{red}x_i \geq u}}}

\noindent The likelihood function can be defined as,
\begin{equation}
\begin{aligned}
&\mathcal{L}(\alpha, \eta, \beta, u, \xi, \sigma \mid \bm{x})  \\
&=\alpha^{\sum_{i=1}^{n} I_1(x_i; u)} \cdot (1-\alpha)^{\sum_{i=1}^{n} I_2(x_i; u)} \cdot \prod_{i \in A} \bm{g}^*(x_i \mid \eta, \beta)  \cdot \Big[1 - \big(\alpha + (1 - \alpha)\cdot  \\
&\quad  \bm{G}^*(u \mid \eta, \beta)\big)\Big]^{\sum_{i=1}^{n} I_3(x_i; u)} \cdot \prod_{i \in B} \left\{\frac{1}{\sigma} \left[1 + \xi \frac{(x_i - u)}{\sigma}\right]^{-\frac{1 + \xi}{\xi}}\right\}, ~~  ~~ \quad \text{for } \xi \neq 0,
\end{aligned}
\label{eq:likelihood_xi_nonzero}
\end{equation}
\begin{equation}
\begin{aligned}
&\mathcal{L}(\alpha, \eta, \beta, u, \xi, \sigma \mid \bm{x})  \\
&=\alpha^{\sum_{i=1}^{n} I_1(x_i; u)} \cdot (1-\alpha)^{\sum_{i=1}^{n} I_2(x_i; u)} \cdot \prod_{i \in A} \bm{g}^*(x_i \mid \eta, \beta)  \cdot \Big[1 - \big(\alpha + (1 - \alpha)\cdot  \\
&\quad  \bm{G}^*(u \mid \eta, \beta)\big)\Big]^{\sum_{i=1}^{n} I_3(x_i; u)} \cdot \prod_{i \in B}   \frac{1}{\sigma} 
\exp\left\{\frac{x_i - u}{\sigma}\right\}, \quad \text{for   } \xi = 0.
\end{aligned}
\label{eq:likelihood_xi_zero}
\end{equation}
Here, \( \bm{G}^* \) and \( \bm{g}^* \) represent the CDF and the density functions of the bulk distribution, that is, the gamma distribution.
We use the usual approach of finding the MLEs of parameters by maximizing the log-likelihood function. The likelihood function of the EVIMM is complex and lacks a closed-form solution, necessitating the use of numerical methods to compute the MLEs. The choice of initial parameter values can significantly affect likelihood inference in mixture models due to sometimes multiple local modes.

To ensure fair model comparison, we assumed consistent initial values. Initial values were chosen as follows: 
\(\alpha_{1,\text{init}} = \frac{ \#\{x_i = 0\} }{n}\), 
\( u_{\text{ini}} \) is the 90th percentile of gamma, 
\((\eta_{\text{init}}, \beta_{\text{init}})\) estimated via method of moments for \( x_i \in (0, u) \), and 
\((\xi_{\text{init}}, \sigma_{\text{init}})\) obtained by fitting the GPD to \( x_i \geq u \) using the \texttt{evd} package in \texttt{R}. 
In the numerical examples presented in Section~\ref{numericalexample1}, we use the \texttt{optim} function from \texttt{R's} \texttt{stats} package \cite{team2020ra} to estimate the parameters of the proposed EVIMM, minimizing the negative log-likelihood using the Nelder–Mead method for numerical optimization. 
 \subsection{Asymptotic Distribution of the MLE}
\label{asymptoticdistribution}

\noindent Note that $\bm{f}(x \mid \alpha, \eta, \beta, u, \xi, \sigma)$ given by (see~\eqref{eq:evmmlpdf}) is

\begin{equation}
\bm{f}(x \mid \alpha, \eta, \beta, u, \xi, \sigma) = 
\begin{cases}
    \bm{\alpha}, & \text{if } x = 0, \\[10pt]
    (1 - \bm{\alpha}) \cdot \dfrac{1}{\Gamma(\eta)\beta^\eta} x^{\eta - 1} e^{-x/\beta}, & \text{if } 0 < x < u, \\[12pt]
    \left(1 - \left[\bm{\alpha} + (1 - \bm{\alpha}) \cdot \int_0^u \dfrac{1}{\Gamma(\eta)\beta^\eta} t^{\eta - 1} e^{-t/\beta} \, dt\right]\right) \cdot \bm{g}(x \mid \xi, \sigma, u), & \text{if } x \geq u,
\end{cases}
\end{equation}
Where the function \(g(x)\) denotes the generalized Pareto density and is defined as
\begin{equation*}
\bm{g}(x \mid \xi, \sigma, u) = 
\begin{cases}
    \dfrac{1}{\sigma} \left(1 + \dfrac{\xi(x - u)}{\sigma} \right)^{-\frac{1}{\xi} - 1}, & \text{if } \xi \neq 0, \\[10pt]
    \dfrac{1}{\sigma} \exp\left(-\dfrac{x - u}{\sigma}\right), & \text{if } \xi = 0,
\end{cases}
\end{equation*}
\noindent
where, the parameters $0 < \alpha < 1$, $\eta > 0$, $\beta > 0$, $u > 0$, $\sigma > 0$, and $\xi \in \mathbb{R}$.
\noindent
\begin{remark}[Threshold Assumption]
To apply the standard asymptotic theory of maximum likelihood estimation, we assume that the threshold parameter \( u \) is known and fixed at \( u = u_0 \). If \( u \) were treated as an unknown parameter, it would violate the regularity conditions required for deriving the asymptotic distribution of the MLE. Therefore, the score functions and Fisher information matrix are derived under this simplifying assumption. In practice, including simulation studies, one can estimate the threshold \( u \) using a profile likelihood or grid search approach over a candidate set of values. Once selected, \( u \) is treated as fixed for the estimation of other parameters, thus preserving the validity of the asymptotic inference conditionally on the chosen threshold.
\end{remark}
\subsubsection{Score Functions: Case \(\xi \ne 0\)}
\noindent Let $\boldsymbol{\theta} = (\alpha, \eta, \beta, \xi, \sigma)$ denote the parameter vector. Define $A(u_0) = \{i : 0 < x_i < u_0\}$, $B(u_0) = \{i : x_i \ge u_0\}$, and for $i \in B(u_0)$, let $z_i = (x_i - u_0)/\sigma$. Let $n_1 = |\{i : x_i = 0\}|$, $n_2 = |A(u_0)|$, and $n_3 = |B(u_0)|$, so that $n = n_1 + n_2 + n_3$.
Define
\[
h(\eta, \beta) = \frac{1}{\Gamma(\eta)} \int_0^{u_0} \frac{t^{\eta - 1} e^{-t/\beta}}{\beta^\eta} dt 
= F_{\text{gamma}}(u_0 \mid \eta, \beta), \quad
h_\eta = \frac{\partial h(\eta, \beta)}{\partial \eta}, \quad
h_\beta = \frac{\partial h(\eta, \beta)}{\partial \beta}.
\]
The first derivatives of the log-likelihood \( \ell(\cdot) = \log \mathcal{L}(\cdot) \), where \( \mathcal{L}(\cdot) \) is given in ~\eqref{eq:likelihood_xi_nonzero} are:

\begin{align*}
\frac{\partial \ell}{\partial \alpha} &= \frac{n_1}{\alpha} - \frac{n_2 + n_3}{1 - \alpha}, \\
\frac{\partial \ell}{\partial \eta} &= -n_2 \psi(\eta) - n_2 \log \beta + \sum_{i \in A(u_0)} \log x_i - \frac{n_3 h_\eta}{1 - h}, \\
\frac{\partial \ell}{\partial \beta} &= -\frac{n_2 \eta}{\beta} + \sum_{i \in A(u_0)} \frac{x_i}{\beta^2} - \frac{n_3 h_\beta}{1 - h}, \\
\frac{\partial \ell}{\partial \xi} &= \sum_{i \in B(u_0)} \left[
 \frac{1}{\xi^2} \log(1 + \xi z_i) - \frac{1 + \xi}{\xi} \cdot \frac{z_i}{1 + \xi z_i}
\right], \\
\frac{\partial \ell}{\partial \sigma} &= \sum_{i \in B(u_0)} \left[
- \frac{1}{\sigma} + \frac{(1 + \xi) z_i}{\sigma (1 + \xi z_i)}
\right].
\end{align*}
One can verify that
\begin{multline*}
\mathbb{E} \left[ \frac{\partial}{\partial \alpha} \ell(X) \right] = 0, \quad
\mathbb{E} \left[ \frac{\partial}{\partial \eta} \ell(X) \right] = 0, \quad
\mathbb{E} \left[ \frac{\partial}{\partial \beta} \ell(X) \right] = 0, \\
\mathbb{E} \left[ \frac{\partial}{\partial \xi} \ell(X) \right] = 0, \quad
\mathbb{E} \left[ \frac{\partial}{\partial \sigma} \ell(X) \right] = 0.
\end{multline*}
Here $h = h(\eta, \beta)$, $h_\eta = \frac{\partial h}{\partial \eta}$, $h_\beta = \frac{\partial h}{\partial \beta}$, $h_{\eta\eta} = \frac{\partial^2 h}{\partial \eta^2}$, $h_{\beta\beta} = \frac{\partial^2 h}{\partial \beta^2}$, $h_{\eta\beta} = \frac{\partial^2 h}{\partial \eta \partial \beta}$. Here, \(\psi(\eta) = \frac{d}{d\eta} \log \Gamma(\eta)\) is the digamma function, and \(\psi'(\eta) = \frac{d^2}{d\eta^2} \log \Gamma(\eta)\) is the trigamma function.
The Fisher information matrix is defined as \(\mathcal{I}(\boldsymbol{\theta}) = -\mathbb{E} \left[ \frac{\partial^2 \ell(\boldsymbol{\theta} \mid \mathbf{x})}{\partial \boldsymbol{\theta} \, \partial \boldsymbol{\theta}^T} \right]\), where the \((i, j)\)-th entry is given by \(\mathcal{I}_{ij}(\boldsymbol{\theta}) = -\mathbb{E} \left[ \frac{\partial^2 \ell(\boldsymbol{\theta} \mid \mathbf{x})}{\partial \theta_i \partial \theta_j} \right]\).

\[
\mathcal{I}_{\alpha\alpha} = \frac{n_1}{\alpha^2} + \frac{n_2 + n_3}{(1 - \alpha)^2} ,
\]
\[
\mathcal{I}_{\alpha \eta} = \mathcal{I}_{\eta \alpha} = 0, \quad
\mathcal{I}_{\alpha \beta} = \mathcal{I}_{\beta \alpha} = 0, \quad
\mathcal{I}_{\alpha \xi} = \mathcal{I}_{\xi \alpha} = 0, \quad
\mathcal{I}_{\alpha \sigma} = \mathcal{I}_{\sigma \alpha} = 0,
\]
\[
\mathcal{I}_{\eta\eta} =  n_2 \psi'(\eta) + \frac{n_3}{1 - h} \left( h_{\eta\eta} + \frac{h_\eta^2}{1 - h} \right),
\]
\[
\mathcal{I}_{\eta\beta} =
\frac{n_2}{\beta} 
+ \frac{n_3}{1 - h} \left( h_{\eta\beta} + \frac{h_\eta h_\beta}{1 - h} \right),
\]

\[
\mathcal{I}_{\eta\xi} = 0, \quad
\mathcal{I}_{\eta\sigma} = 0,
\]
\[
\mathcal{I}_{\beta\beta} 
= -\frac{n_2 \eta}{\beta^2} 
+ \frac{2n_2}{\beta^2} \cdot \frac{\gamma(\eta + 1, u_0/\beta)}{\gamma(\eta, u_0/\beta)} 
+ \frac{n_3}{1 - h} \left( h_{\beta\beta} + \frac{h_\beta^2}{1 - h} \right),
\]

\noindent
where \( \gamma(s, x) \) is the \textit{lower incomplete gamma function}, defined as
\( \gamma(s, x) = \textstyle \int_0^x t^{s-1} e^{-t} \, dt. \)
\[
\mathcal{I}_{\beta\xi} = 0, \quad \mathcal{I}_{\beta\sigma} = 0.
\]
When \( X \sim \text{GPD}(\xi, \sigma, u_0) \), then the standardized variable \( Z = \frac{X - u_0}{\sigma} \sim \text{GPD}(\xi, 1) \), i.e., a generalized Pareto distribution with unit scale. 

\[
\mathcal{I}_{\xi\xi} = n_3 \cdot \left[
\frac{2}{\xi^3} \, \mathbb{E}_Z[\log(1 + \xi Z)]
- \frac{2}{\xi^2} \, \mathbb{E}_Z\left(\frac{Z}{1 + \xi Z}\right)
+ (1 + \xi) \, \mathbb{E}_Z\left(\frac{Z^2}{(1 + \xi Z)^2} \right)
\right].
\]
\[
\mathcal{I}_{\xi\sigma} 
= - \frac{n_3}{\sigma} \cdot \mathbb{E}_Z \left[ \frac{Z(1 - Z)}{(1 + \xi Z)^2} \right],
\]
\[
\mathcal{I}_{\sigma\sigma}
= -n_3 \cdot \mathbb{E}_Z \left[ \frac{1}{\sigma^2} \left( 1 - \frac{(1 + \xi) Z (2 + \xi Z)}{(1 + \xi Z)^2} \right) \right],
\]

\[
\mathcal{I}_{\sigma\sigma}
=
\frac{n_3}{\sigma^2}
\left(
-1 + \frac{1 + \xi}{(1 - \xi)(1 - 2\xi)}
\right),
\quad \text{valid for } \xi < \frac{1}{2}.
\]
\begin{theorem}[Asymptotic Normality of the MLE (for known threshold \( u_0 > 0 \))]
Let \( X_1, X_2, \dots, X_n \) be i.i.d. samples from the EVIMM with known threshold \( u_0 > 0 \), and parameter vector.
\[
\bm{\theta} = (\alpha, \eta, \beta, \xi, \sigma)^\top \in \Theta := (0,1) \times (0, \infty)^2 \times \left( (-0.5, \infty) \setminus \{0\} \right) \times (0, \infty).
\]

Assume that the true parameter \( \bm{\theta}_0 \) lies in the interior of the parameter space, and that the regularity conditions for maximum likelihood estimation hold. Let \( \widehat{\bm{\theta}}_n \) be the MLE based on a random sample of size \( n \). Then, as \( n \to \infty \),
\[
\sqrt{n} \left( \widehat{\bm{\theta}}_n - \bm{\theta}_0 \right) 
\xrightarrow{d} \mathcal{N}_5 \left( \bm{0}, \, \mathcal{I}^{-1}(\bm{\theta}_0) \right),
\]
where \( \mathcal{I}(\bm{\theta}_0) \) is theFisher information matrix for the parameter vector \( \bm{\theta} \), evaluated at the true parameter \( \bm{\theta}_0 \), and whose \((i,j)\)-th element is given by
\[
\mathcal{I}_{ij}(\bm{\theta}_0) = - \mathbb{E}_{\bm{\theta}_0} \left[ \frac{\partial^2 \ell(\bm{\theta})}{\partial \theta_i \partial \theta_j} \right],
\]

\[
 \mathcal{I}(\boldsymbol{\theta}_0) =
\begin{bmatrix}
\mathcal{I}_{\alpha\alpha} & 0 & 0 & 0 & 0 \\
0 & \mathcal{I}_{\eta\eta} & \mathcal{I}_{\eta\beta} & 0 & 0 \\
0 & \mathcal{I}_{\beta\eta} & \mathcal{I}_{\beta\beta} & 0 & 0 \\
0 & 0 & 0 & \mathcal{I}_{\xi\xi} & \mathcal{I}_{\xi\sigma} \\
0 & 0 & 0 & \mathcal{I}_{\sigma\xi} & \mathcal{I}_{\sigma\sigma}
\end{bmatrix}.
\]
\end{theorem}

\begin{remark}[Asymptotic Confidence Intervals for the MLE]
Based on the asymptotic normality of the MLE, the approximate \( 100(1 - \alpha)\% \) confidence interval for each parameter 
\( \theta_i \in \{\alpha, \eta, \beta, \xi, \sigma\} \) 
is given by
\[
\widehat{\theta}_i \pm z_{\alpha/2} \sqrt{ \left[ \mathcal{I}^{-1}(\widehat{\boldsymbol{\theta}}) \right]_{ii} },
\]
where \( z_{\alpha/2} \) is the upper \( \alpha/2 \) -quantile of the standard normal distribution, and \( \left[ \mathcal{I}^{-1}(\widehat{\boldsymbol{\theta}}) \right]_{ii} \) denotes the \( i \)-th diagonal element of the inverse Fisher information matrix evaluated at the MLE \( \widehat{\boldsymbol{\theta}} \).
\end{remark}
\subsubsection{Score Functions: Case \(\xi = 0\)}
\noindent When the shape parameter \(\xi = 0\), the tail distribution in the EVIMM reduces to the exponential distribution with rate \(1/\sigma\). In this case, the parameter vector simplifies to 
$\bm{\theta} = (\alpha, \eta, \beta, \sigma)^\top$.
The log-likelihood function is denoted by \( \ell(\cdot) = \log \mathcal{L}(\cdot) \), where \( \mathcal{L}(\cdot) \) is defined in~\eqref{eq:likelihood_xi_zero}. The first derivative of the log-likelihood with respect to \( \sigma \) is given by
\[
\frac{\partial \ell}{\partial \sigma} = -\frac{n_3}{\sigma} + \frac{1}{\sigma^2} \sum_{i \in B(u_0)} (x_i - u_0).
\]
The corresponding Fisher information entry is
\[
\mathcal{I}_{\sigma\sigma}  = \frac{n_3}{\sigma^2}.
\]
The Fisher information cross-terms involving \( \sigma \) and the other parameters are
\[
\mathcal{I}_{\alpha\sigma} = \mathcal{I}_{\eta\sigma} = \mathcal{I}_{\beta\sigma} = 0.
\]
\begin{theorem}[Asymptotic Normality of the MLE (for known threshold \( u_0 > 0 \) and \( \xi = 0 \))]
Let \( X_1, X_2, \dots, X_n \) be i.i.d. from the EVIMM with known threshold \( u_0 > 0 \). Let the parameter vector be\\
$\bm{\theta} = (\alpha, \eta, \beta, \sigma)^\top \in (0,1) \times (0, \infty)^3.$
Assume the true parameter \( \bm{\theta}_0 \) lies in the interior of the parameter space, and that the standard regularity conditions for MLE hold. Let \( \widehat{\bm{\theta}}_n \) be the MLE based on a random sample of size \( n \). Then, as \( n \to \infty \),
\[
\sqrt{n} \left( \widehat{\bm{\theta}}_n - \bm{\theta}_0 \right) 
\xrightarrow{d} \mathcal{N}_4 \left( \bm{0}, \, \mathcal{I}^{-1}(\bm{\theta}_0) \right),
\]
where \( \mathcal{I}(\bm{\theta}_0) \) is theFisher information matrix for the parameter vector \( \bm{\theta} \), evaluated at the true parameter \( \bm{\theta}_0 \), and given by
\[
 \mathcal{I}(\boldsymbol{\theta}_0) =
\begin{bmatrix}
\mathcal{I}_{\alpha\alpha} & 0 & 0 & 0 \\
0 & \mathcal{I}_{\eta\eta} & \mathcal{I}_{\eta\beta} & 0 \\
0 & \mathcal{I}_{\beta\eta} & \mathcal{I}_{\beta\beta} & 0 \\
0 & 0 & 0 & \mathcal{I}_{\sigma\sigma}
\end{bmatrix}.
\]
\end{theorem}
\begin{remark}[Asymptotic Confidence Intervals for \(\xi = 0\)]
Based on the asymptotic normality of the MLE, the approximate \( 100(1 - \alpha)\% \) confidence interval for each parameter 
\( \theta_i \in \{\alpha, \eta, \beta, \sigma\} \) 
in the EVIMM (with \( \xi = 0 \)) is given by
\[
\widehat{\theta}_i \pm z_{\alpha/2} \sqrt{ \left[ \mathcal{I}^{-1}(\widehat{\boldsymbol{\theta}}) \right]_{ii} },
\]
where \( z_{\alpha/2} \) is the upper \( \alpha/2 \)-quantile of the standard normal distribution, and \( \left[ \mathcal{I}^{-1}(\widehat{\boldsymbol{\theta}}) \right]_{ii} \) denotes the \( i \)-th diagonal element of the inverseFisher information matrix evaluated at the MLE \( \widehat{\boldsymbol{\theta}} \).
\end{remark}

\begin{remark}[Regularity Conditions for the GPD Shape Parameter (\(\xi\))]
The asymptotic normality of the MLE for the generalized Pareto distribution (GPD) depends on the existence of finite moments and the regularity of the log-likelihood. When \( \xi \leq -0.5 \), theFisher information matrix becomes infinite, and the standard asymptotic properties of the MLE (such as consistency and asymptotic normality) are violated. Moreover, when \( \xi > 1 \), the likelihood function becomes unbounded and the MLE may not exist (Castillo \emph{et al.}~\cite{castillo1997fitting}). Therefore, to ensure valid asymptotic inference, it is standard practice to restrict \(\xi\) to the interval \( -0.5 < \xi < 1 \).
\end{remark}

\section{Numerical Example}\label{numericalexample1}
 \noindent
This section presents numerical illustrations of the proposed EVIMM framework. We first conduct a simulation study based on data generated from the proposed model to evaluate the performance of the estimation procedure. Then, we apply the methodology to a real-world dataset to demonstrate its practical utility.
    \subsection{Simulation Studies} \label{simulation}
\noindent This section investigates the parameter estimation performance of the EVIMM under various simulation conditions.

We generate data \( x_i \), for \( i = 1, 2, 3, \dots, n \), from the EVIMM~\eqref{evimmpdf}, using the algorithm described in Section~\ref{algorithm}. The parameters used in the simulations are as follows: \( \alpha \in \{0.1, 0.2, 0.3, 0.4\} \), \( \xi \in \{-0.2, 0, 0.2\} \), \( \eta = 1 \), and the scale parameters \( \beta = \sigma = 5 \), with \( u \in \{10.9861, 10.3972, 9.7296, 8.9588\} \). The sample sizes considered are \( n \in \{300, 400, 500, 750, 1000\} \). The simulated data are analyzed using the proposed EVIMM~\eqref{evimmpdf}, assuming that the parameter vector \( \Theta \) is unknown.

The parameter vector \( \Theta = (\alpha, \eta, \beta, \mathbf{u}, \xi, \sigma) \) is treated as unknown and estimated using the methodologies described in Section~\ref{estimation}. The MLEs of \( \Theta \) are obtained by maximizing the log-likelihood function. For each sample size  $n$, 
 a total of 2000 datasets are simulated, and the proposed estimation approach is used to estimate the model parameters for each dataset. Then, we find the sample mean, bias, mean square error (MSE), bootstrap standard error (BSE), and 95\% bootstrap confidence interval (BCI) using 2000 MLEs of each parameter for comparing the estimated values with the true parameter values of the model. A 95\% confidence interval is also computed using the asymptotic results of MLE, and the corresponding coverage probabilities are also found. The coverage probability is a percentage of times when the confidence interval includes the respective true parameter values. The bias and MSE are calculated using the formula 
\begin{equation}
    \begin{aligned}
        Bias(\hat{\theta}) &= \dfrac{1}{N} \sum_{i=1}^{N} \hat{\theta}^{(i)} - \theta, \\
        MSE(\hat{\theta}) &= \dfrac{1}{N} \sum_{i=1}^{N} \left(\hat{\theta}^{(i)} - \theta\right)^2
    \end{aligned}
\end{equation}
\noindent where $\hat{\theta}$ denotes the estimator of the parameter $\theta$ and $\hat{\theta}^{(i)}$ denotes the estimate of $\theta$ for $i$th data set, where $N$ is the number of data generated with sample size $n$.
 The sample mean, coverage probability, bootstrap confidence interval (BCI), MSE and bias using 2000 estimated parameter values with sample sizes $n=300, 400, 500, 750, 1000$ are given in Tables \ref{Table-100} - \ref{Table-600} respectively. We observe that the bias and MSE are small, the estimated parameter values obtained using the proposed methodologies are close to the true parameter values, and the coverage probabilities are approximately equal to the confidence coefficients for each scenario with varying sample sizes. The bootstrap confidence intervals provide expected results for estimating model parameters using the proposed methodologies. In Tables~\ref{Table-100} to \ref{Table-600}, the values in parentheses under the columns \textbf{Sample Mean}, \textbf{MSE}, and \textbf{Bias} represent estimates obtained using the EVMM  (\cite{behrens2004bayesian}). The \texttt{evmix} R package, developed by Hu \emph{et al.}~\cite{hu2018evmix}, was used for this purpose and is available in CRAN. Some tables are provided as supplementary material. The results for different parameter settings have been evaluated and are consistent with expectations.

We compared the estimates by the  MEP and PSP  to both versions of the simulated dataset: one including inliers (i.e., the whole dataset \(x\)) and the other excluding inliers (i.e., \(x[x > 0]\), where only the non-zero values are retained). The corresponding parameter estimates are presented in Tables~\ref{Table-700}--\ref{Table-900}. For reference, the MEP and PSP were generated using the \texttt{evmix} package \cite{hu2018evmix}. We considered a sample size of $1000$ for each table for this simulation study.
\begin{landscape} 
\begin{table*}[htbp]
    \centering
    \scriptsize 
    \setlength{\tabcolsep}{3pt} 
    \renewcommand{\arraystretch}{1.3} 
    \begin{tabular}{|p{2.5cm}|p{1.8cm}|p{1.8cm}|p{2.4cm}|p{1.8cm}|p{2.3cm}|p{2.3cm}|p{2.3cm}|p{1.8cm}|} 
        \hline
        \textbf{Sample Size} & \textbf{Parameters} & \textbf{True Value} & \textbf{Sample Mean} & \textbf{BSE} & \textbf{BCI} & \textbf{MSE} & \textbf{Bias} & \textbf{CP} \\ \hline
        
       \multirow{6}{*}{ 300 } & $\alpha$ &  0.2  &  0.1992  (NA) &  0.024  &  (0.1609, 0.2539)  &  6e-04  (NA) &  -8e-04  (NA) &  96.2  \\ \cline{2-9}
                               & $\eta$   &  1  &  1.0289  ( 1.0175 ) &  0.1058  &  (0.931, 1.3451)  &  0.01  ( 0.0085 ) &  0.0289  ( 0.0175 ) &  95.9  \\ \cline{2-9}
                               & $\beta$  &  5  &  4.8875  ( 4.9435 ) &  0.5624  &  (3.384, 5.5803)  &  0.3996  ( 0.3766 ) &  -0.1125  ( -0.0565 ) &  91.4  \\ \cline{2-9}
                               & $u$      &  10.3972  &  10.3158  ( 11.4119 ) &  1.0412  &  (7.9736, 12.0046)  &  1.2042  ( 5.2352 ) &  -0.0814  ( 1.0147 ) &  94.7  \\ \cline{2-9}
                               & $\xi$    &  0.2  &  0.1354  ( 0.1157 ) &  0.3308  &  (-0.5192, 0.8065)  &  0.1259  ( 0.2992 ) &  -0.0646  ( -0.0843 ) &  92.8  \\ \cline{2-9}
                               & $\sigma$ &  5  &  5.5307  ( 6.3815 ) &  2.1695  &  (1.9613, 10.0033)  &  6.3106  ( 15.7248 ) &  0.5307  ( 1.3815 ) &  94.4  \\ \hline
       \multirow{6}{*}{ 400 } & $\alpha$ &  0.2  &  0.1995  (NA) &  0.0209  &  (0.1665, 0.247)  &  4e-04  (NA) &  -5e-04  (NA) &  96.9  \\ \cline{2-9}
                               & $\eta$   &  1  &  1.0245  ( 1.0149 ) &  0.0864  &  (0.9093, 1.2449)  &  0.0076  ( 0.0065 ) &  0.0245  ( 0.0149 ) &  95.6  \\ \cline{2-9}
                               & $\beta$  &  5  &  4.9072  ( 4.9532 ) &  0.5057  &  (3.6038, 5.6034)  &  0.3052  ( 0.3465 ) &  -0.0928  ( -0.0468 ) &  93.6  \\ \cline{2-9}
                               & $u$      &  10.3972  &  10.2824  ( 11.4483 ) &  0.9204  &  (8.0122, 11.6189)  &  1.1328  ( 5.525 ) &  -0.1148  ( 1.0511 ) &  95.2  \\ \cline{2-9}
                               & $\xi$    &  0.2  &  0.1612  ( 0.1504 ) &  0.2621  &  (-0.3935, 0.6511)  &  0.0819  ( 0.2107 ) &  -0.0388  ( -0.0496 ) &  94.0  \\ \cline{2-9}
                               & $\sigma$ &  5  &  5.3066  ( 6.0092 ) &  1.6734  &  (2.214, 8.3678)  &  4.1129  ( 10.6361 ) &  0.3066  ( 1.0092 ) &  93.6  \\ \hline
         \multirow{6}{*}{ 500 } & $\alpha$ &  0.2  &  0.1993  (NA) &  0.0188  &  (0.1714, 0.2444)  &  3e-04  (NA) &  -7e-04  (NA) &  96.4  \\ \cline{2-9}
                               & $\eta$   &  1  &  1.0199  ( 1.0107 ) &  0.0766  &  (0.9356, 1.2308)  &  0.0058  ( 0.0049 ) &  0.0199  ( 0.0107 ) &  96.5  \\ \cline{2-9}
                               & $\beta$  &  5  &  4.9243  ( 4.9693 ) &  0.4426  &  (3.7429, 5.5234)  &  0.242  ( 0.2313 ) &  -0.0757  ( -0.0307 ) &  93.4  \\ \cline{2-9}
                               & $u$      &  10.3972  &  10.2685  ( 11.5845 ) &  0.8022  &  (8.3052, 11.4774)  &  0.9978  ( 5.1738 ) &  -0.1287  ( 1.1873 ) &  94.6  \\ \cline{2-9}
                               & $\xi$    &  0.2  &  0.1726  ( 0.154 ) &  0.2114  &  (-0.3078, 0.5412)  &  0.0586  ( 0.1735 ) &  -0.0274  ( -0.046 ) &  94.0  \\ \cline{2-9}
                               & $\sigma$ &  5  &  5.2111  ( 5.9488 ) &  1.2364  &  (2.2095, 6.9792)  &  3.0297  ( 9.1987 ) &  0.2111  ( 0.9488 ) &  92.6  \\ \hline
         \multirow{6}{*}{ 750 } & $\alpha$ &  0.2  &  0.1991  (NA) &  0.0159  &  (0.1747, 0.2358)  &  2e-04  (NA) &  -9e-04  (NA) &  97.6  \\ \cline{2-9}
                               & $\eta$   &  1  &  1.017  ( 1.0072 ) &  0.075  &  (1.0228, 1.316)  &  0.0042  ( 0.0032 ) &  0.017  ( 0.0072 ) &  95.5  \\ \cline{2-9}
                               & $\beta$  &  5  &  4.9227  ( 4.9729 ) &  0.361  &  (3.4107, 4.817)  &  0.1767  ( 0.1558 ) &  -0.0773  ( -0.0271 ) &  92.5  \\ \cline{2-9}
                               & $u$      &  10.3972  &  10.1995  ( 11.5133 ) &  0.8744  &  (7.8976, 11.2565)  &  0.9102  ( 4.2843 ) &  -0.1977  ( 1.1161 ) &  95.0  \\ \cline{2-9}
                               & $\xi$    &  0.2  &  0.1923  ( 0.1689 ) &  0.1519  &  (-0.32, 0.2815)  &  0.0387  ( 0.0894 ) &  -0.0077  ( -0.0311 ) &  93.7  \\ \cline{2-9}
                               & $\sigma$ &  5  &  5.0481  ( 5.6428 ) &  1.1584  &  (3.2921, 7.8636)  &  1.8266  ( 4.5047 ) &  0.0481  ( 0.6428 ) &  91.2  \\ \hline
       \multirow{6}{*}{ 1000 } & $\alpha$ &  0.2  &  0.1992  (NA) &  0.0136  &  (0.1809, 0.2336)  &  2e-04  (NA) &  -8e-04  (NA) &  98.1  \\ \cline{2-9}
                               & $\eta$   &  1  &  1.014  ( 1.0054 ) &  0.0618  &  (1.0098, 1.2503)  &  0.0029  ( 0.0024 ) &  0.014  ( 0.0054 ) &  96.5  \\ \cline{2-9}
                               & $\beta$  &  5  &  4.9344  ( 4.9772 ) &  0.3204  &  (3.7062, 4.948)  &  0.131  ( 0.1154 ) &  -0.0656  ( -0.0228 ) &  92.0  \\ \cline{2-9}
                               & $u$      &  10.3972  &  10.1368  ( 11.4291 ) &  0.7977  &  (8.1078, 11.1967)  &  0.8833  ( 3.4362 ) &  -0.2604  ( 1.0319 ) &  94.6  \\ \cline{2-9}
                               & $\xi$    &  0.2  &  0.1977  ( 0.1719 ) &  0.132  &  (-0.2165, 0.2937)  &  0.0284  ( 0.046 ) &  -0.0023  ( -0.0281 ) &  93.1  \\ \cline{2-9}
                               & $\sigma$ &  5  &  4.9992  ( 5.5404 ) &  0.9886  &  (3.2975, 7.1579)  &  1.3209  ( 2.8026 ) &  -8e-04  ( 0.5404 ) &  90.5  \\ \hline
    \end{tabular}
\captionsetup{
  labelfont=bf, 
  textfont=normal 
}
      \caption{Simulation results based on \( N = 2000 \) replications for sample sizes \( n = 300, 400, 500, 750, \) and \( 1000 \), with an inlier proportion of 20\% and heavy-tailed generalized Pareto distributions (GPD). The table presents the sample mean of estimated parameters, bootstrap standard error (BSE), 95\% bootstrap confidence interval (BCI), , mean square error (MSE), bias, and convergence probability (CP).}

    \label{Table-100}
\end{table*}
\end{landscape}

\begin{landscape} 
\begin{table*}[htbp]
    \centering
    \scriptsize 
    \setlength{\tabcolsep}{3pt} 
    \renewcommand{\arraystretch}{1.3} 
    \begin{tabular}{|p{2.5cm}|p{1.8cm}|p{1.8cm}|p{2.3cm}|p{1.8cm}|p{2.3cm}|p{2.3cm}|p{2.3cm}|p{1.8cm}|} 
        \hline
        \textbf{Sample Size} & \textbf{Parameters} & \textbf{True Value} & \textbf{Sample Mean} & \textbf{BSE} & \textbf{BCI} & \textbf{MSE} & \textbf{Bias} & \textbf{CP} \\ \hline

       \multirow{6}{*}{ 300 } & $\alpha$ &  0.4  &  0.4  (NA) &  0.0291  &  (0.3525, 0.4663)  &  8e-04  (NA) &  0  (NA) &  97.6  \\ \cline{2-9}
                               & $\eta$   &  1  &  1.0302  ( 1.0181 ) &  0.1191  &  (0.8815, 1.3461)  &  0.0131  ( 0.0118 ) &  0.0302  ( 0.0181 ) &  96.2  \\ \cline{2-9}
                               & $\beta$  &  5  &  4.8485  ( 4.9609 ) &  0.6659  &  (3.1065, 5.7071)  &  0.5675  ( 0.5484 ) &  -0.1515  ( -0.0391 ) &  92.0  \\ \cline{2-9}
                               & $u$      &  8.9587  &  8.7133  ( 11.3164 ) &  1.073  &  (5.9296, 10.2766)  &  1.3242  ( 10.7526 ) &  -0.2455  ( 2.3576 ) &  95.0  \\ \cline{2-9}
                               & $\xi$    &  0.2  &  0.1096  ( 0.1175 ) &  0.2979  &  (-0.5574, 0.6486)  &  0.1027  ( 0.4329 ) &  -0.0904  ( -0.0825 ) &  92.5  \\ \cline{2-9}
                               & $\sigma$ &  5  &  5.5835  ( 7.0154 ) &  2.2511  &  (2.2667, 10.8567)  &  5.0475  ( 24.1105 ) &  0.5835  ( 2.0154 ) &  96.4  \\ \hline
          \multirow{6}{*}{ 400 } & $\alpha$ &  0.4  &  0.3999  (NA) &  0.0256  &  (0.3802, 0.481)  &  6e-04  (NA) &  -1e-04  (NA) &  96.9  \\ \cline{2-9}
                               & $\eta$   &  1  &  1.0261  ( 1.0139 ) &  0.1093  &  (0.9254, 1.363)  &  0.0096  ( 0.0086 ) &  0.0261  ( 0.0139 ) &  96.4  \\ \cline{2-9}
                               & $\beta$  &  5  &  4.8586  ( 4.9706 ) &  0.6095  &  (3.3088, 5.7001)  &  0.4438  ( 0.4263 ) &  -0.1414  ( -0.0294 ) &  91.2  \\ \cline{2-9}
                               & $u$      &  8.9587  &  8.696  ( 11.2998 ) &  1.1369  &  (6.0103, 10.5532)  &  1.1782  ( 10.1867 ) &  -0.2628  ( 2.341 ) &  93.3  \\ \cline{2-9}
                               & $\xi$    &  0.2  &  0.1421  ( 0.1239 ) &  0.2596  &  (-0.393, 0.6212)  &  0.0678  ( 0.2939 ) &  -0.0579  ( -0.0761 ) &  93.0  \\ \cline{2-9}
                               & $\sigma$ &  5  &  5.3583  ( 6.6913 ) &  1.7307  &  (2.2605, 8.803)  &  3.2556  ( 17.7814 ) &  0.3583  ( 1.6913 ) &  94.3  \\ \hline
         \multirow{6}{*}{ 500 } & $\alpha$ &  0.4  &  0.3999  (NA) &  0.0229  &  (0.375, 0.4654)  &  5e-04  (NA) &  -1e-04  (NA) &  97.1  \\ \cline{2-9}
                               & $\eta$   &  1  &  1.0216  ( 1.0092 ) &  0.091  &  (0.8827, 1.2401)  &  0.0078  ( 0.0066 ) &  0.0216  ( 0.0092 ) &  96.2  \\ \cline{2-9}
                               & $\beta$  &  5  &  4.8767  ( 4.9928 ) &  0.584  &  (3.7188, 6)  &  0.3778  ( 0.3369 ) &  -0.1233  ( -0.0072 ) &  92.9  \\ \cline{2-9}
                               & $u$      &  8.9587  &  8.7006  ( 11.443 ) &  1.1931  &  (6.0534, 10.8587)  &  1.0995  ( 10.6307 ) &  -0.2582  ( 2.4842 ) &  94.0  \\ \cline{2-9}
                               & $\xi$    &  0.2  &  0.1511  ( 0.1364 ) &  0.2406  &  (-0.3053, 0.6644)  &  0.0491  ( 0.2063 ) &  -0.0489  ( -0.0636 ) &  93.2  \\ \cline{2-9}
                               & $\sigma$ &  5  &  5.2972  ( 6.4507 ) &  1.3182  &  (1.9675, 6.9661)  &  2.4827  ( 13.4077 ) &  0.2972  ( 1.4507 ) &  94.5  \\ \hline
           \multirow{6}{*}{ 750 } & $\alpha$ &  0.4  &  0.4001  (NA) &  0.0184  &  (0.3675, 0.4388)  &  3e-04  (NA) &  1e-04  (NA) &  97.4  \\ \cline{2-9}
                               & $\eta$   &  1  &  1.0179  ( 1.006 ) &  0.0683  &  (0.8737, 1.1427)  &  0.0053  ( 0.0044 ) &  0.0179  ( 0.006 ) &  96.6  \\ \cline{2-9}
                               & $\beta$  &  5  &  4.8806  ( 4.9891 ) &  0.4679  &  (3.8961, 5.7176)  &  0.2572  ( 0.2227 ) &  -0.1194  ( -0.0109 ) &  92.0  \\ \cline{2-9}
                               & $u$      &  8.9587  &  8.64  ( 11.4336 ) &  1.1022  &  (6.1418, 10.494)  &  0.9854  ( 9.9639 ) &  -0.3188  ( 2.4748 ) &  94.1  \\ \cline{2-9}
                               & $\xi$    &  0.2  &  0.1686  ( 0.1451 ) &  0.1704  &  (-0.2618, 0.4092)  &  0.0321  ( 0.111 ) &  -0.0314  ( -0.0549 ) &  92.1  \\ \cline{2-9}
                               & $\sigma$ &  5  &  5.1763  ( 6.2327 ) &  1.1285  &  (2.7692, 6.9882)  &  1.5335  ( 10.8403 ) &  0.1763  ( 1.2327 ) &  95.4  \\ \hline
        \multirow{6}{*}{ 1000 } & $\alpha$ &  0.4  &  0.3998  (NA) &  0.0161  &  (0.3693, 0.4322)  &  2e-04  (NA) &  -2e-04  (NA) &  98.1  \\ \cline{2-9}
                               & $\eta$   &  1  &  1.0164  ( 1.0035 ) &  0.0607  &  (0.878, 1.1137)  &  0.0039  ( 0.0031 ) &  0.0164  ( 0.0035 ) &  96.7  \\ \cline{2-9}
                               & $\beta$  &  5  &  4.8772  ( 4.9946 ) &  0.4444  &  (4.0278, 5.7308)  &  0.2035  ( 0.1575 ) &  -0.1228  ( -0.0054 ) &  91.1  \\ \cline{2-9}
                               & $u$      &  8.9587  &  8.6087  ( 11.4059 ) &  1.1426  &  (6.0114, 10.4007)  &  1.034  ( 8.7124 ) &  -0.3501  ( 2.4471 ) &  92.6  \\ \cline{2-9}
                               & $\xi$    &  0.2  &  0.1726  ( 0.153 ) &  0.1447  &  (-0.2024, 0.363)  &  0.0243  ( 0.0709 ) &  -0.0274  ( -0.047 ) &  92.1  \\ \cline{2-9}
                               & $\sigma$ &  5  &  5.1563  ( 6.0406 ) &  1.0031  &  (3.0364, 6.8311)  &  1.1624  ( 4.7637 ) &  0.1563  ( 1.0406 ) &  95.6  \\ \hline
     \end{tabular}
\captionsetup{
  labelfont=bf, 
  textfont=normal 
}
    \caption{Simulation results based on \( N = 2000 \) replications for sample sizes \( n = 300, 400, 500, 750, \) and \( 1000 \), with an inlier proportion of 40\% and heavy-tailed generalized Pareto distributions (GPD). The table presents the sample mean of estimated parameters, bootstrap standard error (BSE), 95\% bootstrap confidence interval (BCI), mean square error (MSE), bias, and convergence probability (CP).}
    \label{tab:simulation_study_parameters_extended}
    \label{Table-200}
\end{table*}
\end{landscape}

\begin{landscape} 
\begin{table*}[htbp]
    \centering
    \scriptsize 
    \setlength{\tabcolsep}{3pt} 
    \renewcommand{\arraystretch}{1.3} 
    \begin{tabular}{|p{2.5cm}|p{1.8cm}|p{1.8cm}|p{2.4cm}|p{1.8cm}|p{2.3cm}|p{2.3cm}|p{2.3cm}|p{1.8cm}|} 
        \hline
        \textbf{Sample Size} & \textbf{Parameters} & \textbf{True Value} & \textbf{Sample Mean} & \textbf{BSE} & \textbf{BCI} & \textbf{MSE} & \textbf{Bias} & \textbf{CP} \\ \hline

   \multirow{6}{*}{ 300 } & $\alpha$ &  0.2  &  0.1992  (NA) &  0.0244  &  (0.1624, 0.2568)  &  6e-04  (NA) &  -8e-04  (NA) &  95.9  \\ \cline{2-9}
                               & $\eta$   &  1  &  1.0288  ( 1.0169 ) &  0.1086  &  (0.9495, 1.3789)  &  0.0098  ( 0.0087 ) &  0.0288  ( 0.0169 ) &  95.8  \\ \cline{2-9}
                               & $\beta$  &  5  &  4.8793  ( 4.9484 ) &  0.5441  &  (3.2482, 5.3923)  &  0.3848  ( 0.3869 ) &  -0.1207  ( -0.0516 ) &  91.6  \\ \cline{2-9}
                               & $u$      &  10.3972  &  10.4997  ( 11.1378 ) &  1.0749  &  (8.074, 12.2973)  &  1.3568  ( 3.5804 ) &  0.1025  ( 0.7406 ) &  95.9  \\ \cline{2-9}
                               & $\xi$    &  -0.2  &  -0.3052  ( -0.2965 ) &  0.2764  &  (-1.0104, 0.1759)  &  0.0982  ( 0.1548 ) &  -0.1052  ( -0.0965 ) &  91.1  \\ \cline{2-9}
                               & $\sigma$ &  5  &  5.5344  ( 5.5835 ) &  1.9929  &  (2.5964, 10.4459)  &  4.2009  ( 7.3345 ) &  0.5344  ( 0.5835 ) &  92.0  \\ \hline
   \multirow{6}{*}{ 400 } & $\alpha$ &  0.2  &  0.1993  (NA) &  0.0209  &  (0.1556, 0.2372)  &  4e-04  (NA) &  -7e-04  (NA) &  97.0  \\ \cline{2-9}
                               & $\eta$   &  1  &  1.0247  ( 1.0144 ) &  0.0895  &  (0.9289, 1.2855)  &  0.0073  ( 0.0064 ) &  0.0247  ( 0.0144 ) &  95.8  \\ \cline{2-9}
                               & $\beta$  &  5  &  4.8966  ( 4.955 ) &  0.4906  &  (3.5098, 5.4602)  &  0.2899  ( 0.2891 ) &  -0.1034  ( -0.045 ) &  93.2  \\ \cline{2-9}
                               & $u$      &  10.3972  &  10.5068  ( 11.1679 ) &  0.9994  &  (8.3798, 12.3087)  &  1.1074  ( 3.5818 ) &  0.1096  ( 0.7706 ) &  96.0  \\ \cline{2-9}
                               & $\xi$    &  -0.2  &  -0.2699  ( -0.2719 ) &  0.2269  &  (-0.9052, 0.043)  &  0.0606  ( 0.1034 ) &  -0.0699  ( -0.0719 ) &  94.3  \\ \cline{2-9}
                               & $\sigma$ &  5  &  5.3348  ( 5.3483 ) &  1.7983  &  (3.1906, 10.4964)  &  2.9058  ( 5.1131 ) &  0.3348  ( 0.3483 ) &  94.2  \\ \hline
          \multirow{6}{*}{ 500 } & $\alpha$ &  0.2  &  0.1988  (NA) &  0.0192  &  (0.1691, 0.2441)  &  4e-04  (NA) &  -0.0012  (NA) &  96.8  \\ \cline{2-9}
                               & $\eta$   &  1  &  1.0209  ( 1.0105 ) &  0.0798  &  (0.9758, 1.29)  &  0.0057  ( 0.0049 ) &  0.0209  ( 0.0105 ) &  95.6  \\ \cline{2-9}
                               & $\beta$  &  5  &  4.9097  ( 4.9734 ) &  0.4154  &  (3.5279, 5.1641)  &  0.2397  ( 0.2293 ) &  -0.0903  ( -0.0266 ) &  93.2  \\ \cline{2-9}
                               & $u$      &  10.3972  &  10.5131  ( 11.2725 ) &  0.8868  &  (8.3472, 11.8133)  &  1.0103  ( 3.2467 ) &  0.1158  ( 0.8753 ) &  96.0  \\ \cline{2-9}
                               & $\xi$    &  -0.2  &  -0.2524  ( -0.2601 ) &  0.1904  &  (-0.6946, 0.0509)  &  0.0433  ( 0.0724 ) &  -0.0524  ( -0.0601 ) &  95.3  \\ \cline{2-9}
                               & $\sigma$ &  5  &  5.2241  ( 5.1985 ) &  1.2939  &  (2.8502, 7.7989)  &  2.1072  ( 3.7663 ) &  0.2241  ( 0.1985 ) &  94.4  \\ \hline
     \multirow{6}{*}{ 750 } & $\alpha$ &  0.2  &  0.1984  (NA) &  0.0159  &  (0.1758, 0.2366)  &  2e-04  (NA) &  -0.0016  (NA) &  97.5  \\ \cline{2-9}
                               & $\eta$   &  1  &  1.0153  ( 1.0068 ) &  0.066  &  (0.9974, 1.2547)  &  0.0038  ( 0.0031 ) &  0.0153  ( 0.0068 ) &  96.8  \\ \cline{2-9}
                               & $\beta$  &  5  &  4.9317  ( 4.9768 ) &  0.3411  &  (3.6328, 4.9661)  &  0.1604  ( 0.1546 ) &  -0.0683  ( -0.0232 ) &  93.2  \\ \cline{2-9}
                               & $u$      &  10.3972  &  10.5074  ( 11.3359 ) &  0.7856  &  (8.4779, 11.5657)  &  0.9157  ( 2.6543 ) &  0.1102  ( 0.9386 ) &  96.4  \\ \cline{2-9}
                               & $\xi$    &  -0.2  &  -0.2223  ( -0.2355 ) &  0.1438  &  (-0.5799, -0.0109)  &  0.0258  ( 0.0406 ) &  -0.0223  ( -0.0355 ) &  94.5  \\ \cline{2-9}
                               & $\sigma$ &  5  &  5.0522  ( 5.0148 ) &  1.0076  &  (3.205, 7.0334)  &  1.3408  ( 2.1811 ) &  0.0522  ( 0.0148 ) &  94.1  \\ \hline
    \multirow{6}{*}{ 1000 } & $\alpha$ &  0.2  &  0.1984  (NA) &  0.014  &  (0.1789, 0.2337)  &  2e-04  (NA) &  -0.0016  (NA) &  98.3  \\ \cline{2-9}
                               & $\eta$   &  1  &  1.0133  ( 1.0058 ) &  0.0585  &  (1.0112, 1.2406)  &  0.0029  ( 0.0024 ) &  0.0133  ( 0.0058 ) &  96.7  \\ \cline{2-9}
                               & $\beta$  &  5  &  4.9354  ( 4.9754 ) &  0.3014  &  (3.7318, 4.9243)  &  0.1301  ( 0.1133 ) &  -0.0646  ( -0.0246 ) &  93.4  \\ \cline{2-9}
                               & $u$      &  10.3972  &  10.4608  ( 11.3997 ) &  0.7667  &  (8.6327, 11.6795)  &  0.8055  ( 2.2838 ) &  0.0636  ( 1.0025 ) &  96.9  \\ \cline{2-9}
                               & $\xi$    &  -0.2  &  -0.2144  ( -0.2292 ) &  0.1196  &  (-0.5347, -0.0668)  &  0.0177  ( 0.026 ) &  -0.0144  ( -0.0292 ) &  94.1  \\ \cline{2-9}
                               & $\sigma$ &  5  &  5.0199  ( 4.9571 ) &  0.9057  &  (3.5088, 7.0605)  &  0.943  ( 1.4207 ) &  0.0199  ( -0.0429 ) &  94.8  \\ \hline
    \end{tabular}
\captionsetup{
  labelfont=bf, 
  textfont=normal 
}
    \caption{Simulation results based on \( N = 2000 \) replications for sample sizes \( n = 300, 400, 500, 750, \) and \( 1000 \), with an inlier proportion of 20\% and light-tailed generalized Pareto distributions (GPD). The table presents the sample mean of estimated parameters, bootstrap standard error (BSE), 95\% bootstrap confidence interval (BCI), mean square error (MSE), bias, and convergence probability (CP).}
   \label{Table-300}
\end{table*}
\end{landscape}

\begin{landscape} 
\begin{table*}[htbp]
    \centering
    \scriptsize 
    \setlength{\tabcolsep}{3pt} 
    \renewcommand{\arraystretch}{1.3} 
    \begin{tabular}{|p{2.5cm}|p{1.8cm}|p{1.8cm}|p{2.3cm}|p{1.8cm}|p{2.3cm}|p{2.3cm}|p{2.3cm}|p{1.8cm}|} 
        \hline
        \textbf{Sample Size} & \textbf{Parameters} & \textbf{True Value} & \textbf{Sample Mean} & \textbf{BSE} & \textbf{BCI} & \textbf{MSE} & \textbf{Bias} & \textbf{CP} \\ \hline

  \multirow{6}{*}{ 300 } & $\alpha$ &  0.4  &  0.3996  (NA) &  0.0292  &  (0.3551, 0.47)  &  8e-04  (NA) &  -4e-04  (NA) &  97.1  \\ \cline{2-9}
                               & $\eta$   &  1  &  1.0338  ( 1.0204 ) &  0.1313  &  (0.9505, 1.4669)  &  0.0133  ( 0.0116 ) &  0.0338  ( 0.0204 ) &  95.9  \\ \cline{2-9}
                               & $\beta$  &  5  &  4.819  ( 4.9449 ) &  0.5923  &  (2.9003, 5.2078)  &  0.5608  ( 0.5333 ) &  -0.181  ( -0.0551 ) &  90.1  \\ \cline{2-9}
                               & $u$      &  8.9587  &  8.9747  ( 11.0208 ) &  1.0885  &  (5.9752, 10.1572)  &  1.5974  ( 7.6277 ) &  0.0159  ( 2.062 ) &  94.3  \\ \cline{2-9}
                               & $\xi$    &  -0.2  &  -0.2894  ( -0.2689 ) &  0.307  &  (-0.6598, 0.5701)  &  0.0923  ( 0.2045 ) &  -0.0894  ( -0.0689 ) &  91.0  \\ \cline{2-9}
                               & $\sigma$ &  5  &  5.4877  ( 5.1836 ) &  1.593  &  (1.6459, 7.612)  &  4.3261  ( 8.1771 ) &  0.4877  ( 0.1836 ) &  92.9  \\ \hline
    \multirow{6}{*}{ 400 } & $\alpha$ &  0.4  &  0.3999  (NA) &  0.0257  &  (0.3832, 0.4845)  &  6e-04  (NA) &  -1e-04  (NA) &  96.8  \\ \cline{2-9}
                               & $\eta$   &  1  &  1.0257  ( 1.0147 ) &  0.1087  &  (0.9316, 1.3575)  &  0.0096  ( 0.0082 ) &  0.0257  ( 0.0147 ) &  96.0  \\ \cline{2-9}
                               & $\beta$  &  5  &  4.8655  ( 4.9593 ) &  0.6017  &  (3.2613, 5.6381)  &  0.4265  ( 0.3881 ) &  -0.1345  ( -0.0407 ) &  90.7  \\ \cline{2-9}
                               & $u$      &  8.9587  &  9.0523  ( 11.1 ) &  1.3567  &  (6.1694, 11.458)  &  1.4944  ( 7.6043 ) &  0.0935  ( 2.1412 ) &  93.9  \\ \cline{2-9}
                               & $\xi$    &  -0.2  &  -0.2597  ( -0.2569 ) &  0.2369  &  (-0.7854, 0.1842)  &  0.0566  ( 0.1451 ) &  -0.0597  ( -0.0569 ) &  94.0  \\ \cline{2-9}
                               & $\sigma$ &  5  &  5.2751  ( 4.9828 ) &  1.5753  &  (2.4229, 8.423)  &  2.7296  ( 6.042 ) &  0.2751  ( -0.0172 ) &  94.6  \\ \hline
       \multirow{6}{*}{ 500 } & $\alpha$ &  0.4  &  0.3998  (NA) &  0.0228  &  (0.3756, 0.4648)  &  5e-04  (NA) &  -2e-04  (NA) &  97.5  \\ \cline{2-9}
                               & $\eta$   &  1  &  1.0197  ( 1.0112 ) &  0.0903  &  (0.8896, 1.2385)  &  0.0073  ( 0.0066 ) &  0.0197  ( 0.0112 ) &  95.9  \\ \cline{2-9}
                               & $\beta$  &  5  &  4.8914  ( 4.9714 ) &  0.5751  &  (3.7067, 5.9437)  &  0.3382  ( 0.3164 ) &  -0.1086  ( -0.0286 ) &  89.2  \\ \cline{2-9}
                               & $u$      &  8.9587  &  9.0819  ( 11.157 ) &  1.3393  &  (6.5309, 11.6468)  &  1.4409  ( 7.5632 ) &  0.1231  ( 2.1982 ) &  92.3  \\ \cline{2-9}
                               & $\xi$    &  -0.2  &  -0.2472  ( -0.2515 ) &  0.2073  &  (-0.632, 0.1879)  &  0.0405  ( 0.1101 ) &  -0.0472  ( -0.0515 ) &  94.8  \\ \cline{2-9}
                               & $\sigma$ &  5  &  5.1982  ( 4.8894 ) &  1.3041  &  (2.1003, 6.9638)  &  1.9874  ( 4.6703 ) &  0.1982  ( -0.1106 ) &  94.6  \\ \hline
      \multirow{6}{*}{ 750 } & $\alpha$ &  0.4  &  0.3999  (NA) &  0.0183  &  (0.3675, 0.4386)  &  3e-04  (NA) &  -1e-04  (NA) &  97.2  \\ \cline{2-9}
                               & $\eta$   &  1  &  1.0142  ( 1.0071 ) &  0.0687  &  (0.8799, 1.1486)  &  0.0049  ( 0.0045 ) &  0.0142  ( 0.0071 ) &  95.9  \\ \cline{2-9}
                               & $\beta$  &  5  &  4.914  ( 4.9791 ) &  0.4569  &  (3.8765, 5.6521)  &  0.2278  ( 0.2099 ) &  -0.086  ( -0.0209 ) &  92.1  \\ \cline{2-9}
                               & $u$      &  8.9587  &  9.1508  ( 11.2966 ) &  1.0836  &  (6.6961, 10.8782)  &  1.1856  ( 7.5062 ) &  0.1921  ( 2.3378 ) &  90.9  \\ \cline{2-9}
                               & $\xi$    &  -0.2  &  -0.2244  ( -0.2403 ) &  0.1524  &  (-0.5358, 0.0651)  &  0.024  ( 0.0635 ) &  -0.0244  ( -0.0403 ) &  92.7  \\ \cline{2-9}
                               & $\sigma$ &  5  &  5.0493  ( 4.7481 ) &  0.9993  &  (2.6942, 6.5048)  &  1.2266  ( 2.985 ) &  0.0493  ( -0.2519 ) &  94.0  \\ \hline
    \multirow{6}{*}{ 1000 } & $\alpha$ &  0.4  &  0.3998 (NA)  &  0.016  &  (0.3718, 0.4342)  &  3e-04 (NA)  &  -2e-04 (NA) &  96.9  \\ \cline{2-9}
                               & $\eta$   &  1  &  1.0119 (1.0054)  &  0.0611  &  (0.8961, 1.1346)  &  0.0036 (0.0030)  &  0.0119 (0.0054)  &  96.6  \\ \cline{2-9}
                               & $\beta$  &  5  &  4.9201 (4.9773)  &  0.4169  &  (3.8178, 5.4377)  &  0.1747 (0.1483)  &  -0.0799 (-0.0226) &  92.2  \\ \cline{2-9}
                               & $u$      &  8.9587  &  9.2069 ( 11.3484) &  1.0663  &  (6.66, 10.7907)  &  1.134 (7.3956) &  0.2481 (2.3896)  &  91.6  \\ \cline{2-9}
                               & $\xi$    &  -0.2  &  -0.2158 (-0.2212 ) &  0.1209  &  (-0.4943, -0.0201)  &  0.0161 (0.0392)  &  -0.0158 (-0.0212)  &  93.5  \\ \cline{2-9}
                               & $\sigma$ &  5  &  4.9852 (4.6002 )  &  0.9334  &  (3.2454, 6.8039)  &  0.8917 (2.0285)   &  -0.0148 (-0.3997) &  94.1  \\ \hline
    \end{tabular}
\captionsetup{
  labelfont=bf, 
  textfont=normal 
}
     \caption{Simulation results based on \( N = 2000 \) replications for sample sizes \( n = 300, 400, 500, 750, \) and \( 1000 \), with an inlier proportion of 40\% and light-tailed generalized Pareto distributions (GPD). The table presents the sample mean of estimated parameters, bootstrap standard error (BSE), 95\% bootstrap confidence interval (BCI), mean square error (MSE), bias, and convergence probability (CP).}
   \label{Table-400}
\end{table*}
\end{landscape}


\begin{landscape} 
\begin{table*}[htbp]
    \centering
    \scriptsize 
    \setlength{\tabcolsep}{3pt} 
    \renewcommand{\arraystretch}{1.3} 
    \begin{tabular}{|p{2.5cm}|p{1.8cm}|p{1.8cm}|p{2.5cm}|p{1.8cm}|p{2.3cm}|p{2.3cm}|p{2.3cm}|p{1.8cm}|} 
        \hline
        \textbf{Sample Size} & \textbf{Parameters} & \textbf{True Value} & \textbf{Sample Mean} & \textbf{BSE} & \textbf{BCI} & \textbf{MSE} & \textbf{Bias} & \textbf{CP} \\ \hline

   \multirow{6}{*}{ 300 } & $\alpha$ &  0.2  &  0.1993  (NA) &  0.0244  &  (0.1625, 0.2559)  &  6e-04  (NA) &  -7e-04  (NA) &  95.3  \\ \cline{2-9}
                               & $\eta$   &  1  &  1.0275  ( 1.0169 ) &  0.1087  &  (0.9538, 1.3833)  &  0.0098  ( 0.0086 ) &  0.0275  ( 0.0169 ) &  95.9  \\ \cline{2-9}
                               & $\beta$  &  5  &  4.8919  ( 4.9501 ) &  0.5401  &  (3.2357, 5.3219)  &  0.388  ( 0.389 ) &  -0.1081  ( -0.0499 ) &  91.3  \\ \cline{2-9}
                               & $u$      &  10.3972  &  10.3537  ( 11.2879 ) &  1.0186  &  (7.9238, 12.0314)  &  1.15  ( 4.8305 ) &  -0.0435  ( 0.8907 ) &  96.3  \\ \cline{2-9}
                               & $\xi$    &  0  &  -0.084  ( -0.1079 ) &  0.2906  &  (-0.8132, 0.3704)  &  0.1044  ( 0.2356 ) &  -0.084  ( -0.1079 ) &  93.5  \\ \cline{2-9}
                               & $\sigma$ &  5  &  5.5427  ( 6.0494 ) &  2.2795  &  (2.5864, 11.5353)  &  5.2195  ( 11.512 ) &  0.5427  ( 1.0494 ) &  93.4  \\ \hline
      \multirow{6}{*}{ 400 } & $\alpha$ &  0.2  &  0.1997  (NA) &  0.0211  &  (0.1648, 0.246)  &  4e-04  (NA) &  -3e-04  (NA) &  96.9  \\ \cline{2-9}
                               & $\eta$   &  1  &  1.0216  ( 1.0136 ) &  0.0885  &  (0.9311, 1.2737)  &  0.007  ( 0.0064 ) &  0.0216  ( 0.0136 ) &  96.5  \\ \cline{2-9}
                               & $\beta$  &  5  &  4.9197  ( 4.9669 ) &  0.4857  &  (3.4503, 5.3734)  &  0.2899  ( 0.3162 ) &  -0.0803  ( -0.0331 ) &  93.4  \\ \cline{2-9}
                               & $u$      &  10.3972  &  10.3365  ( 11.3217 ) &  0.9332  &  (7.9685, 11.6104)  &  1.061  ( 4.8945 ) &  -0.0607  ( 0.9245 ) &  95.8  \\ \cline{2-9}
                               & $\xi$    &  0  &  -0.0554  ( -0.0749 ) &  0.2425  &  (-0.6057, 0.3337)  &  0.0655  ( 0.1611 ) &  -0.0554  ( -0.0749 ) &  94.3  \\ \cline{2-9}
                               & $\sigma$ &  5  &  5.326  ( 5.7328 ) &  1.619  &  (2.485, 8.5055)  &  3.3196  ( 8.0523 ) &  0.326  ( 0.7328 ) &  93.4  \\ \hline
    \multirow{6}{*}{ 500 } & $\alpha$ &  0.2  &  0.1991  (NA) &  0.0189  &  (0.1695, 0.2434)  &  4e-04  (NA) &  -9e-04  (NA) &  96.3  \\ \cline{2-9}
                               & $\eta$   &  1  &  1.0187  ( 1.0103 ) &  0.0754  &  (0.9366, 1.2276)  &  0.0058  ( 0.0049 ) &  0.0187  ( 0.0103 ) &  95.7  \\ \cline{2-9}
                               & $\beta$  &  5  &  4.9268  ( 4.9756 ) &  0.4279  &  (3.781, 5.4684)  &  0.2432  ( 0.2397 ) &  -0.0732  ( -0.0244 ) &  91.6  \\ \cline{2-9}
                               & $u$      &  10.3972  &  10.3045  ( 11.4473 ) &  0.792  &  (8.4235, 11.5052)  &  0.9333  ( 4.7137 ) &  -0.0927  ( 1.0501 ) &  95.1  \\ \cline{2-9}
                               & $\xi$    &  0  &  -0.0473  ( -0.0649 ) &  0.203  &  (-0.4438, 0.3646)  &  0.0491  ( 0.1107 ) &  -0.0473  ( -0.0649 ) &  94.1  \\ \cline{2-9}
                               & $\sigma$ &  5  &  5.2789  ( 5.592 ) &  1.1389  &  (2.2416, 6.5979)  &  2.5781  ( 6.1772 ) &  0.2789  ( 0.592 ) &  93.1  \\ \hline
      \multirow{6}{*}{ 750 } & $\alpha$ &  0.2  &  0.1989  (NA) &  0.0157  &  (0.1708, 0.2312)  &  2e-04  (NA) &  -0.0011  (NA) &  97.7  \\ \cline{2-9}
                               & $\eta$   &  1  &  1.0133  ( 1.0065 ) &  0.0671  &  (0.9916, 1.251)  &  0.0037  ( 0.0032 ) &  0.0133  ( 0.0065 ) &  96.9  \\ \cline{2-9}
                               & $\beta$  &  5  &  4.9477  ( 4.9798 ) &  0.3583  &  (3.6604, 5.05)  &  0.1668  ( 0.156 ) &  -0.0523  ( -0.0202 ) &  93.8  \\ \cline{2-9}
                               & $u$      &  10.3972  &  10.2265  ( 11.4912 ) &  0.8525  &  (8.2064, 11.5566)  &  0.8173  ( 4.2088 ) &  -0.1707  ( 1.094 ) &  96.5  \\ \cline{2-9}
                               & $\xi$    &  0  &  -0.0196  ( -0.0472 ) &  0.1593  &  (-0.4155, 0.2124)  &  0.0294  ( 0.0672 ) &  -0.0196  ( -0.0472 ) &  93.8  \\ \cline{2-9}
                               & $\sigma$ &  5  &  5.098  ( 5.3924 ) &  1.0813  &  (2.9287, 7.0016)  &  1.5584  ( 3.6978 ) &  0.098  ( 0.3924 ) &  92.3  \\ \hline
    \multirow{6}{*}{ 1000 } & $\alpha$ &  0.2  &  0.1987  (NA) &  0.0138  &  (0.1774, 0.2315)  &  2e-04  (NA) &  -0.0013  (NA) &  98.9  \\ \cline{2-9}
                               & $\eta$   &  1  &  1.0116  ( 1.0046 ) &  0.059  &  (1.0045, 1.2376)  &  0.0029  ( 0.0024 ) &  0.0116  ( 0.0046 ) &  96.7  \\ \cline{2-9}
                               & $\beta$  &  5  &  4.9493  ( 4.9863 ) &  0.3144  &  (3.7647, 4.9867)  &  0.1266  ( 0.1159 ) &  -0.0507  ( -0.0137 ) &  93.6  \\ \cline{2-9}
                               & $u$      &  10.3972  &  10.1802  ( 11.4895 ) &  0.8231  &  (8.2599, 11.4968)  &  0.7409  ( 3.7661 ) &  -0.217  ( 1.0923 ) &  94.2  \\ \cline{2-9}
                               & $\xi$    &  0  &  -0.0144  ( -0.0326 ) &  0.1308  &  (-0.3624, 0.1479)  &  0.0208  ( 0.0419 ) &  -0.0144  ( -0.0326 ) &  93.0  \\ \cline{2-9}
                               & $\sigma$ &  5  &  5.0666  ( 5.2705 ) &  0.9283  &  (3.213, 6.7283)  &  1.1155  ( 2.3853 ) &  0.0666  ( 0.2705 ) &  90.5  \\ \hline
    \end{tabular}
    \captionsetup{
  labelfont=bf, 
  textfont=normal 
}
    \caption{Simulation results based on \( N = 2000 \) replications for sample sizes \( n = 300, 400, 500, 750, \) and \( 1000 \), with an inlier proportion of 20\% and exponential-tailed generalized Pareto distributions (GPD). The table presents the sample mean of estimated parameters, bootstrap standard error (BSE), 95\% bootstrap confidence interval (BCI), mean square error (MSE), bias, and convergence probability (CP).}
   \label{Table-5000}
\end{table*}
\end{landscape}

\begin{landscape} 
\begin{table*}[htbp]
    \centering
    \scriptsize 
    \setlength{\tabcolsep}{3pt} 
    \renewcommand{\arraystretch}{1.3} 
    \begin{tabular}{|p{2.5cm}|p{1.8cm}|p{1.8cm}|p{2.3cm}|p{1.8cm}|p{2.3cm}|p{2.3cm}|p{2.3cm}|p{1.8cm}|} 
        \hline
        \textbf{Sample Size} & \textbf{Parameters} & \textbf{True Value} & \textbf{Sample Mean} & \textbf{BSE} & \textbf{BCI} & \textbf{MSE} & \textbf{Bias} & \textbf{CP} \\ \hline

   \multirow{6}{*}{ 300 } & $\alpha$ &  0.4  &  0.3997  (NA) &  0.0293  &  (0.3584, 0.4739)  &  8e-04  (NA) &  -3e-04  (NA) &  97.3  \\ \cline{2-9}
                               & $\eta$   &  1  &  1.0313  ( 1.0193 ) &  0.1161  &  (0.8539, 1.309)  &  0.0129  ( 0.0116 ) &  0.0313  ( 0.0193 ) &  96.8  \\ \cline{2-9}
                               & $\beta$  &  5  &  4.836  ( 4.9564 ) &  0.7138  &  (3.4024, 6.2097)  &  0.5518  ( 0.5738 ) &  -0.164  ( -0.0436 ) &  91.0  \\ \cline{2-9}
                               & $u$      &  8.9587  &  8.7807  ( 11.2088 ) &  1.2365  &  (6.1081, 11.0685)  &  1.3721  ( 9.6523 ) &  -0.1781  ( 2.25 ) &  95.2  \\ \cline{2-9}
                               & $\xi$    &  0  &  -0.0917  ( -0.0963 ) &  0.3163  &  (-0.5957, 0.6753)  &  0.094  ( 0.3313 ) &  -0.0917  ( -0.0963 ) &  92.7  \\ \cline{2-9}
                               & $\sigma$ &  5  &  5.5858  ( 6.2152 ) &  1.8127  &  (1.6771, 8.5549)  &  4.5346  ( 16.7675 ) &  0.5858  ( 1.2152 ) &  94.7  \\ \hline
     \multirow{6}{*}{ 400 } & $\alpha$ &  0.4  &  0.3997  (NA) &  0.0254  &  (0.3823, 0.482)  &  6e-04  (NA) &  -3e-04  (NA) &  97.3  \\ \cline{2-9}
                               & $\eta$   &  1  &  1.0272  ( 1.0141 ) &  0.1107  &  (0.9459, 1.377)  &  0.0097  ( 0.0087 ) &  0.0272  ( 0.0141 ) &  96.1  \\ \cline{2-9}
                               & $\beta$  &  5  &  4.8503  ( 4.9709 ) &  0.5848  &  (3.2274, 5.4909)  &  0.4431  ( 0.4205 ) &  -0.1497  ( -0.0291 ) &  90.4  \\ \cline{2-9}
                               & $u$      &  8.9587  &  8.7999  ( 11.3024 ) &  1.1977  &  (5.9503, 10.7534)  &  1.3562  ( 10.0359 ) &  -0.1589  ( 2.3436 ) &  93.5  \\ \cline{2-9}
                               & $\xi$    &  0  &  -0.0608  ( -0.0851 ) &  0.2405  &  (-0.5525, 0.429)  &  0.0601  ( 0.2408 ) &  -0.0608  ( -0.0851 ) &  94.2  \\ \cline{2-9}
                               & $\sigma$ &  5  &  5.3766  ( 5.9402 ) &  1.57  &  (2.3482, 8.2123)  &  2.9116  ( 11.9615 ) &  0.3766  ( 0.9402 ) &  94.7  \\ \hline
    \multirow{6}{*}{ 500 } & $\alpha$ &  0.4  &  0.3999  (NA) &  0.0228  &  (0.3771, 0.467)  &  5e-04  (NA) &  -1e-04  (NA) &  96.9  \\ \cline{2-9}
                               & $\eta$   &  1  &  1.0228  ( 1.0099 ) &  0.0905  &  (0.8873, 1.238)  &  0.0075  ( 0.0066 ) &  0.0228  ( 0.0099 ) &  96.9  \\ \cline{2-9}
                               & $\beta$  &  5  &  4.8651  ( 4.9881 ) &  0.562  &  (3.6951, 5.8743)  &  0.355  ( 0.3206 ) &  -0.1349  ( -0.0119 ) &  91.6  \\ \cline{2-9}
                               & $u$      &  8.9587  &  8.7941  ( 11.3255 ) &  1.2202  &  (6.1838, 11.1064)  &  1.206  ( 9.4272 ) &  -0.1647  ( 2.3667 ) &  94.1  \\ \cline{2-9}
                               & $\xi$    &  0  &  -0.0501  ( -0.0753 ) &  0.2124  &  (-0.4429, 0.4062)  &  0.043  ( 0.1891 ) &  -0.0501  ( -0.0753 ) &  93.8  \\ \cline{2-9}
                               & $\sigma$ &  5  &  5.317  ( 5.7655 ) &  1.2042  &  (2.1605, 6.7659)  &  2.1614  ( 8.8878 ) &  0.317  ( 0.7655 ) &  94.9  \\ \hline
           \multirow{6}{*}{ 750 } & $\alpha$ &  0.4  &  0.3997  (NA) &  0.0182  &  (0.368, 0.4384)  &  3e-04  (NA) &  -3e-04  (NA) &  97.0  \\ \cline{2-9}
                               & $\eta$   &  1  &  1.019  ( 1.0061 ) &  0.0697  &  (0.8889, 1.1628)  &  0.0052  ( 0.0044 ) &  0.019  ( 0.0061 ) &  96.3  \\ \cline{2-9}
                               & $\beta$  &  5  &  4.8664  ( 4.9899 ) &  0.4545  &  (3.7429, 5.5289)  &  0.25  ( 0.2229 ) &  -0.1336  ( -0.0101 ) &  91.1  \\ \cline{2-9}
                               & $u$      &  8.9587  &  8.7472  ( 11.441 ) &  1.0016  &  (6.5426, 10.3564)  &  1.1993  ( 9.8333 ) &  -0.2116  ( 2.4822 ) &  91.1  \\ \cline{2-9}
                               & $\xi$    &  0  &  -0.0329  ( -0.0543 ) &  0.1459  &  (-0.4196, 0.1604)  &  0.026  ( 0.1073 ) &  -0.0329  ( -0.0543 ) &  91.1  \\ \cline{2-9}
                               & $\sigma$ &  5  &  5.1979  ( 5.5299 ) &  1.0684  &  (3.1408, 7.2901)  &  1.3427  ( 5.3156 ) &  0.1979  ( 0.5299 ) &  95.4  \\ \hline
       \multirow{6}{*}{ 1000 } & $\alpha$ &  0.4  &  0.3997  (NA) &  0.0162  &  (0.3681, 0.4322)  &  3e-04  (NA) &  -3e-04  (NA) &  98.0  \\ \cline{2-9}
                               & $\eta$   &  1  &  1.0171  ( 1.0039 ) &  0.0576  &  (0.8708, 1.0969)  &  0.0038  ( 0.003 ) &  0.0171  ( 0.0039 ) &  96.5  \\ \cline{2-9}
                               & $\beta$  &  5  &  4.869  ( 4.9927 ) &  0.4231  &  (4.1305, 5.7466)  &  0.1939  ( 0.1546 ) &  -0.131  ( -0.0073 ) &  89.9  \\ \cline{2-9}
                               & $u$      &  8.9587  &  8.7508  ( 11.5243 ) &  1.0704  &  (6.7931, 10.7824)  &  1.0668  ( 9.7531 ) &  -0.208  ( 2.5655 ) &  89.1  \\ \cline{2-9}
                               & $\xi$    &  0  &  -0.0256  ( -0.045 ) &  0.1299  &  (-0.3346, 0.1874)  &  0.0191  ( 0.0684 ) &  -0.0256  ( -0.045 ) &  91.6  \\ \cline{2-9}
                               & $\sigma$ &  5  &  5.1712  ( 5.4092 ) &  0.9218  &  (3.0288, 6.4961)  &  1.0254  ( 3.7208 ) &  0.1712  ( 0.4092 ) &  94.5  \\ \hline
    \end{tabular}
    \captionsetup{
  labelfont=bf, 
  textfont=normal 
}
     \caption{Simulation results based on \( N = 2000 \) replications for sample sizes \( n = 300, 400, 500, 750, \) and \( 1000 \), with an inlier proportion of 40\% and exponential-tailed generalized Pareto distributions (GPD). The table presents the sample mean of estimated parameters, bootstrap standard error (BSE), 95\% bootstrap confidence interval (BCI), mean square error (MSE), bias, and convergence probability (CP).}
    \label{Table-600}
\end{table*}
\end{landscape}
\begin{table}[!htbp]
\centering
\small
\setlength{\tabcolsep}{4pt} 
\renewcommand{\arraystretch}{1.2}

\begin{tabular}{lcccccc}
\hline
\textbf{Parameter} 
& \textbf{True}  & \textbf{EVIMM} & \textbf{MEP ($x$)} 
& \textbf{MEP (\(x[x > 0]\))} & \textbf{PSP ($x$)} & \textbf{PSP (\(x[x > 0]\))} \\ \hline
\(\alpha\)  & 0.200   & 0.208 & N/A    & N/A    & N/A    & N/A    \\
\(\eta\)    & 1.000   & 1.056 & N/A    & N/A    & N/A    & N/A    \\
\(\beta\)   & 5.000 & 4.612 & N/A    & N/A    & N/A    & N/A    \\
\(u\)       & 10.397   & 9.394 & 9.700    & 11.000    & 9.700   & 11.00    \\
\(\sigma\)  & 5.000    & 3.929  & 4.700    & 4.600    & 4.700    & 4.600    \\
\(\xi\)     &  0.200   & 0.156 & 0.089   & 0.110  & 0.089   & 0.11  \\ \hline

\end{tabular}
\caption{Parameter estimates using different methods: MEP, PSP, and the proposed EVIMM. Results are shown for both with and without inliers.}
\label{Table-700}
\end{table}
\begin{table}[!htbp]

\centering
\small
\setlength{\tabcolsep}{4pt} 
\renewcommand{\arraystretch}{1.2}
\begin{tabular}{lcccccc}
\hline
\textbf{Parameter} 
& \textbf{True}  & \textbf{EVIMM} & \textbf{MEP ($x$)} 
& \textbf{MEP (\(x[x > 0]\))} & \textbf{PSP ($x$)}  & \textbf{PSP (\(x[x > 0]\))} \\ \hline
\(\alpha\)  & 0.400   & 0.3806 & N/A    & N/A    & N/A    & N/A    \\
\(\eta\)    & 1.000   & 1.0914 & N/A    & N/A    & N/A    & N/A    \\
\(\beta\)   & 5.000 & 4.1632 & N/A    & N/A    & N/A    & N/A    \\
\(u\)       & 8.958   & 8.6229 & 8.600    & 12.00    & 8.600   & 12.00    \\
\(\sigma\)  & 5.000    & 5.0497  & 5.700    & 5.700    & 5.700    & 5.700    \\
\(\xi\)     &  0.200   & 0.1916 & 0.072   & 0.077  & 0.072   & 0.072 \\ \hline

\end{tabular}
\caption{Parameter estimates using different methods: MEP, PSP, and the proposed EVIMM. Results are shown for both with and without inliers.}
\label{Table-800}
\end{table}
\begin{table}[!htbp]

\centering
\begin{tabular}{lcccccc}
\hline
\textbf{Parameter} 
& \textbf{True}  & \textbf{EVIMM} & \textbf{MEP ($x$)} 
& \textbf{MEP (\(x[x > 0]\))} & \textbf{PSP ($x$)}  & \textbf{PSP (\(x[x > 0]\))} \\ \hline
\(\alpha\)  & 0.200   & 0.2075 & N/A    & N/A    & N/A    & N/A    \\
\(\eta\)    & 1.000   & 1.1029 & N/A    & N/A    & N/A    & N/A    \\
\(\beta\)   & 5.000 & 4.3600 & N/A    & N/A    & N/A    & N/A    \\
\(u\)       & 10.397   & 10.1265   & 10.000    & 11.000    & 10.000       & 11.000    \\
\(\sigma\)  & 5.000    & 5.0787  & 4.900    & 4.500    & 4.900    & 4.500     \\
\(\xi\)     &  -0.200   & -0.2790 & -0.2600 & -0.240  & -0.2600  & -0.240  \\ \hline

\end{tabular}
\caption{Parameter estimates using different methods: MEP, PSP, and the proposed EVIMM. Results are shown for both with and without inliers.}
\label{Table-900}
\end{table}
\begin{table}[!htbp]

\centering
\begin{tabular}{lcccccc}
\hline
\textbf{Parameter} 
& \textbf{True}  & \textbf{EVIMM} & \textbf{MEP ($x$)} 
& \textbf{MEP (\(x[x > 0]\))} & \textbf{PSP ($x$)}  & \textbf{PSP (\(x[x > 0]\))} \\ \hline
\(\alpha\)  & 0.400   & 0.4020 & N/A    & N/A    & N/A    & N/A    \\
\(\eta\)    & 1.000   & 1.0001 & N/A    & N/A    & N/A    & N/A    \\
\(\beta\)   & 5.000 & 4.6254 & N/A    & N/A    & N/A    & N/A    \\
\(u\)       & 8.9587   & 6.8500   & 8.500    & 11.000    & 8.500       & 11.000    \\
\(\sigma\)  & 5.000    & 5.3168  & 4.900    & 3.900    & 4.900    & 3.900     \\
\(\xi\)     &  -0.200   & -0.2357 & -0.2600 & -0.200  & -0.2600  & -0.2000  \\ \hline
\end{tabular}
\caption{Parameter estimates using different methods: MEP, PSP, and the proposed EVIMM. Results are shown for both with and without inliers.}
\label{Table-1000}
\end{table}
\subsection{Assessing Bias and MSE in Parameter Estimations: EVIMM vs. EVMM} \label{realdata}

\begin{figure}[!htbp]
    \centering
    \includegraphics[width=18cm]{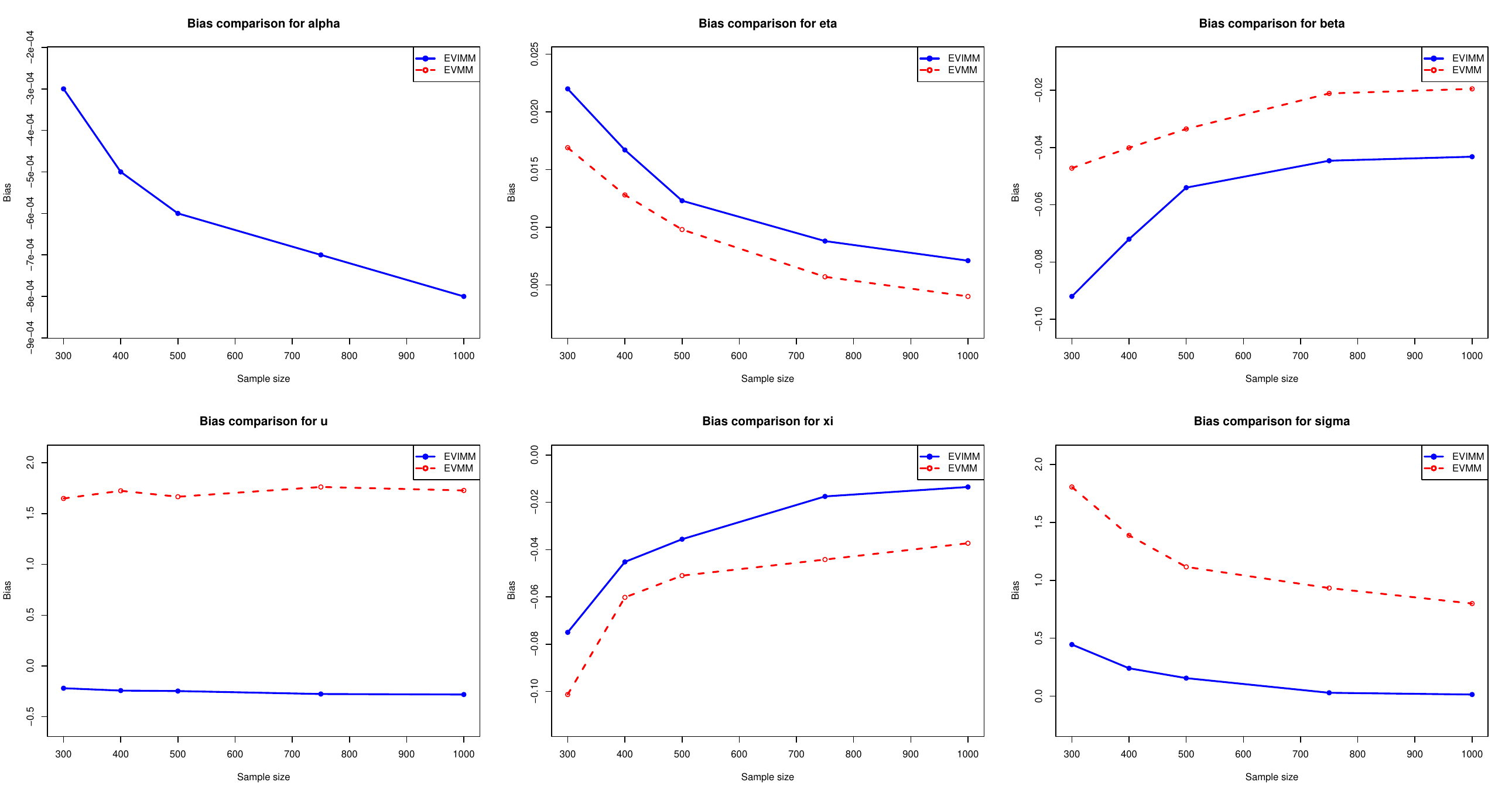}
    \caption{Bias comparison of parameter estimates under the EVIMM and EVMM.}
    \label{fig:bias_comparison}
\end{figure}

\begin{figure}[!htbp]
    \centering
    \includegraphics[width=18cm]{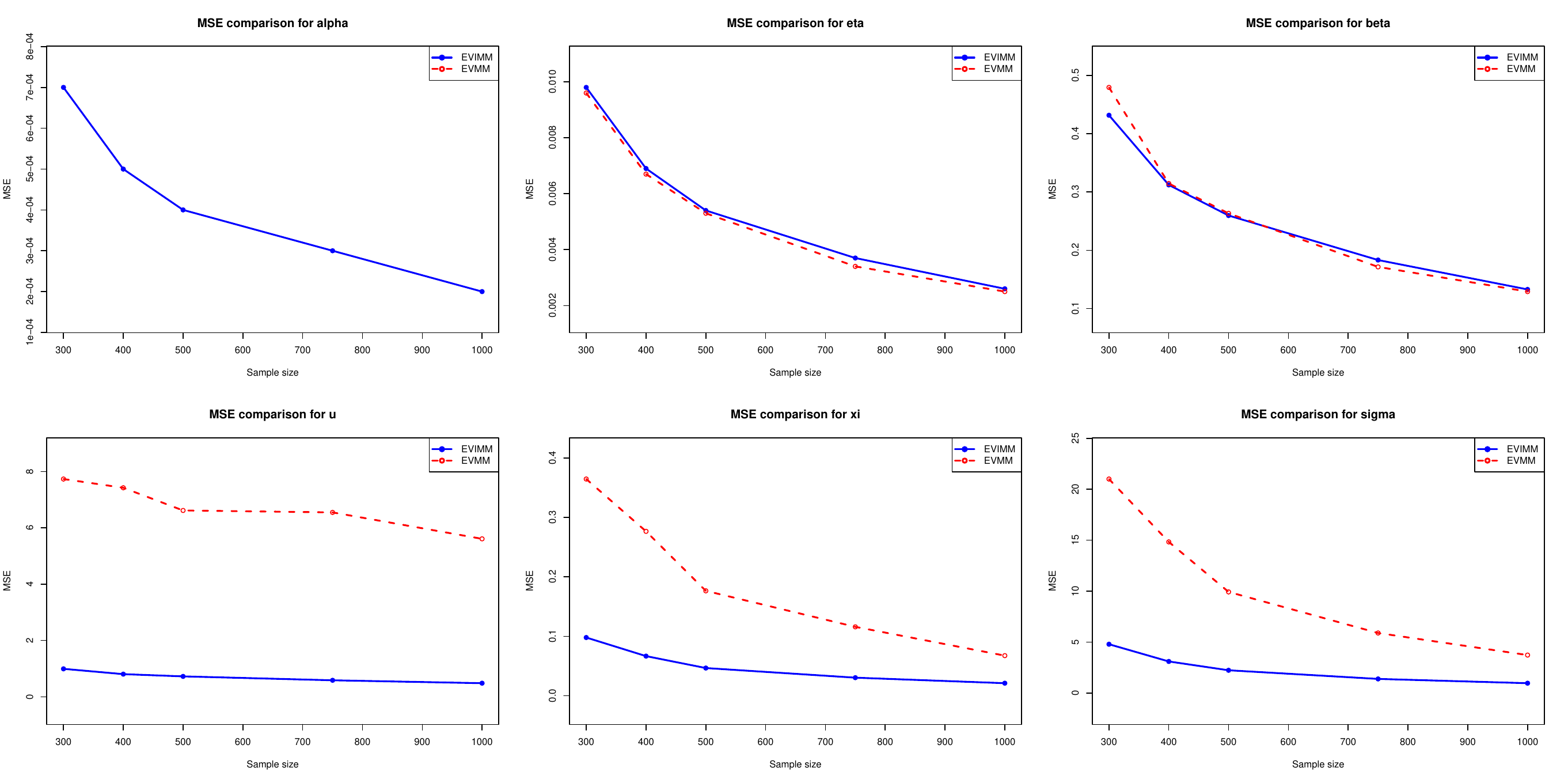}
    \caption{MSE comparison of parameter estimates between the EVIMM and EVMM.}
    \label{fig:mse_comparison}
\end{figure}

\noindent To evaluate the bias and MSE in the parameter estimation process of the EVIMM and the EVMM, we simulated data from the EVIMM with parameters \(\alpha = 0.3\), \(\eta = 1\), \(\beta = 5\), \(u = 9.7295\), \(\xi = 0.2\), and \(\sigma = 5\). Simulations were conducted for sample sizes \(n = 300, 400, 500, 750, 1000\). The bias and MSE were computed using the methodologies described earlier for both EVIMM and EVMM, and the results are presented in Figures~\ref{fig:bias_comparison} and~\ref{fig:mse_comparison}, respectively.

\noindent The bias and MSE in the estimated parameters (for different sets of parameters) from EVIMM are observed to be lower compared to those of the EVMM, particularly as the proportion of inliers increases.
\subsection{Comparison of Return Level Estimates: EVIMM, EVMM, and FEVMM}
\label{realdata11}
\begin{figure}[!htbp]
    \centering
    \begin{subfigure}{0.45\textwidth}
        \centering
        \includegraphics[width=8.5cm, height=7cm]{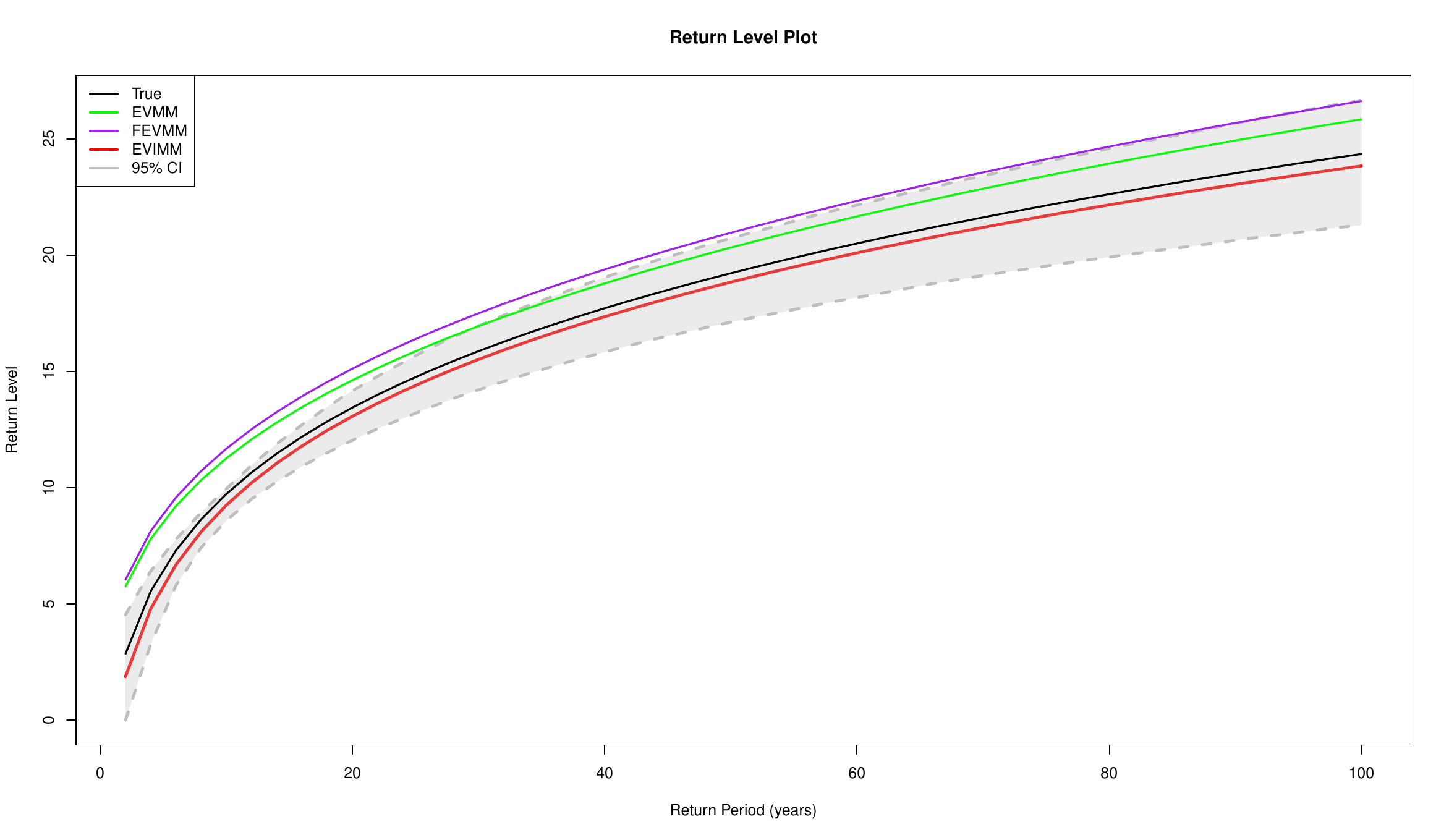}
        \caption{Return level estimates for a heavy-tailed GPD $(\xi=0.2)$.}
        \label{fig:bias_comparison1}
    \end{subfigure}
    \hfill
    \begin{subfigure}{0.45\textwidth}
        \centering
        \includegraphics[width=8.5cm, height=7cm]{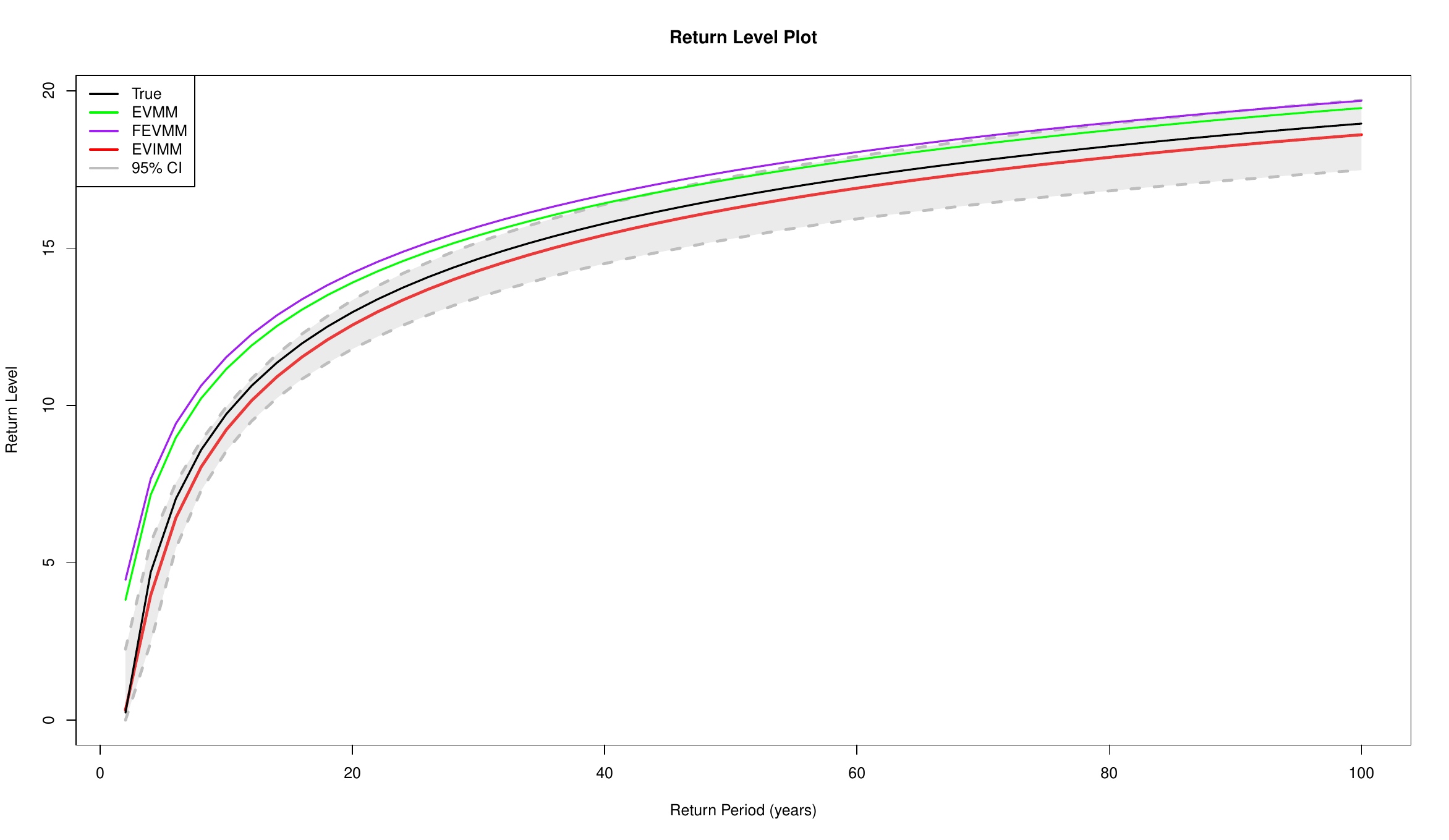}
        \caption{Return level estimates for a lighter-tailed GPD $(\xi=-0.2)$.}
        \label{fig:bias_comparison2}
    \end{subfigure}
   \caption{Return level estimates for  EVMM, FEVMM and the true values, EVIMM, along with 95\% confidence intervals for EVIMM.}
    \label{fig:return_comparisonsimu}
\end{figure}

\noindent \textbf{Return Level:} The value expected to be equal to  or exceeded on average once every interval of the time period \( T \) (with probability \( 1/T \)). Return levels and return periods are used to describe and quantify risk. We simulated data from the EVIMM using parameters \(\alpha = 0.3\), \(\eta = 1\), \(\beta = 5\), \(u = 9.7295\), \(\sigma = 5\) and \(\xi = 0.2,-0.2\) with sample size \(n = 2000\).\
The 95\% confidence interval for the return level is estimated using the bootstrap method. Figure \ref{fig:return_comparisonsimu} shows the return level versus the return period, along with the 95\% confidence interval and the actual return level. 
 As the sample size and inlier proportion increase, EVIMM provides better return level estimates compared to EVMM, FEVMM.

\subsection{Sensitivity Analysis under Varying Inlier Proportions}

\noindent 
To evaluate the robustness of the proposed EVIMM under different proportions of inliers, 
a sensitivity analysis was conducted by simulating data from the EVIMM~\eqref{evimmpdf} 
with varying levels of inlier proportions: \( \alpha \in \{0.2, 0.3, 0.4, 0.5, 0.6\} \). 
The remaining model parameters were fixed at \( \eta = 1 \), \( \beta = \sigma = 5 \), \( \xi = 0.2 \), and \( u \in \{10.3972, 9.7296, 8.9588, 8.0472, 6.9315\} \).  
Simulations were performed for two sample sizes, \( n \in \{500, 1000\} \). 
For each inlier proportion, 2000 datasets (replications) were simulated, and the parameters were estimated using the MLE procedure described in Section~\ref{estimation}. 
The bias and MSE for each parameter were then estimated.

\begin{figure}[!htbp]
    \centering
    \includegraphics[width=18cm]{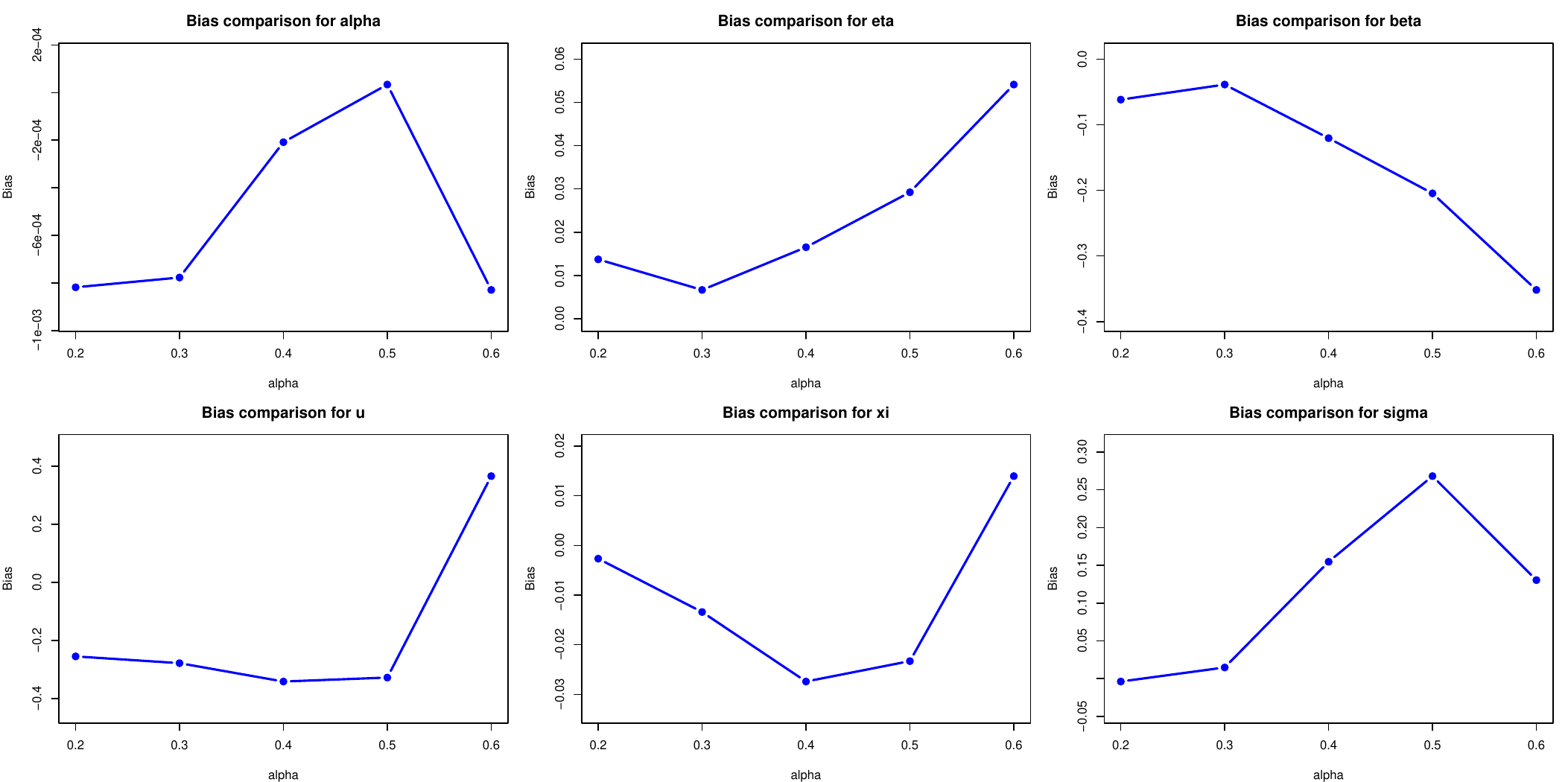}
  \caption{Bias of the EVIMM parameter estimates under varying inlier proportions (alpha).}
   \label{fig:inlier_sensitivity1}
\end{figure}

\begin{figure}[!htbp]
    \centering
    \includegraphics[width=18cm]{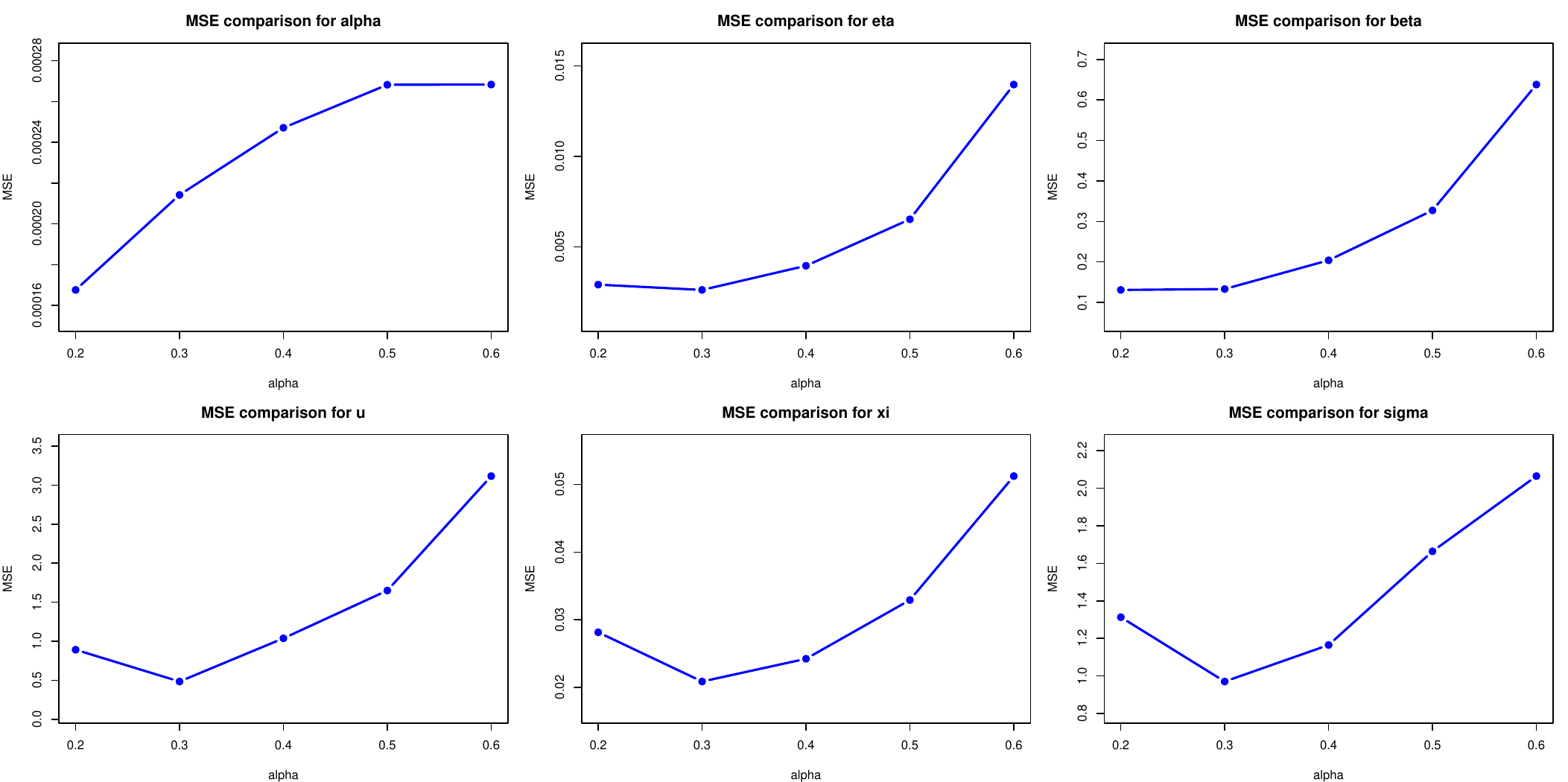}
   \caption{MSE of the EVIMM parameter estimates under varying inlier proportions (alpha).}
   \label{fig:inlier_sensitivity2}
\end{figure}

Figures~\ref{fig:inlier_sensitivity1} and~\ref{fig:inlier_sensitivity2} present the variation in bias and MSE for the estimated parameters across different levels of inlier proportions for sample size $n=1000$. The results show that the bias and MSE of the EVIMM parameter estimates remain relatively stable for moderate inlier proportions but tend to increase as the proportion of inliers becomes larger. This is because the data with higher proportions of inliers have fewer non-zero observations, thereby reducing the available information for estimating bulk distribution and tail distribution parameters. 
Although the bias and MSE of the parameter estimates of the EVIMM increase with higher inlier proportions, they remain consistently lower than those of the EVMM and FEVMM, which do not account for inliers. This highlights the enhanced robustness of the proposed EVIMM for data in the presence of inliers.

\subsection{Application to Real Data} \label{realdata}
\noindent We have estimated the parameters, particularly the threshold and the parameters of the GPD, using various well-known methods: the 90th percentile (rule of thumb), the MEP, the Hill plot, PSP, the EVMM (bulk-based model), the EVMM (proportion of tail fraction as the parameter), and our proposed model, EVIMM. In the Tables, \ref{Table-21} - \ref{Table-22}, the values in parentheses represent the standard errors corresponding to each estimated parameter. 

\subsubsection*{ Comparative Analysis 1}
\noindent This study considers the real-life dataset \texttt{cps91} from the \texttt{wooldridge} R package. Compiled from the May 1991 U.S. Current Population Survey, the dataset contains 5,634 observations. For our analysis, we considered the last 2000 data points. The variable  \texttt{earns} represents the weekly earnings of wives, providing insight into the income distribution of married women in 1991. The estimated parameters are given in Table \ref{Table-21}.

\begin{table}[h!]
\centering
\label{tab:parameters_methods}
\begin{tabular}{lcccccc}
\hline
\textbf{Method} & \(\alpha\) & \(\eta\) & \(\beta\) & \(u\) & \(\sigma\) & \(\xi\) \\ \hline
90th Percentile & N/A & N/A & N/A & 558  & 186.0460 & 0.1090 \\
                 &     &     &     &     & (N/A)   & (N/A)    \\
mean excess plot & N/A & N/A & N/A & 560 & 190 & 0.11 \\
                 &     &     &     &     & (N/A)   & (N/A)    \\
Hill Plot        & N/A & N/A & N/A & 640 & (N/A) & 0.27 \\
                 &     &     &     &     & (N/A)   & (N/A)    \\
Parameter Stability Plot & N/A & N/A & N/A & 560 & 190 & 0.11 \\
                         &     &     &     &     & (N/A)   & (N/A)    \\
EVM Model (Bulk)  & N/A  & 2.5765 & 137.4781 & 640.1537 & 208.9931 & 0.0835 \\
&       & (0.0019) & (0.1173) & (0.3735) & (0.4890) & (0.0017)   \\
EVM Model (Para.) & N/A  & 2.4757 & 149.5479 & 615.0008 & 208.9209 & 0.0771 \\
                  &      & (0.1329) & (12.7141) & (0.0025) & (24.9188) & (0.0745) \\

EVIMM        & 0.3782 &  2.5383 & 140.1094 & 556.1742 &  187.6737 & 0.1019 \\
& (0.0002) & (0.0034) & (0.1986) & (0.6471) & (0.6220)  & (0.0022) \\ \hline

\end{tabular}
\captionsetup{
  labelfont=bf, 
  textfont=normal 
}
\caption{Parameter estimates using different methods and models, with standard errors provided in parentheses (or brackets) below the corresponding values.}
\label{Table-21}
\end{table}

\subsubsection*{ Comparative Analysis 2}
\noindent This study analyzed monthly rainfall data from 1971 to 2007, having a length of $n=441$, recorded at the Modikuppam rainfall station near Gudiyattam in Vellore district, Tamil Nadu. This station, representing the Malattar sub-watershed area spanning 163 km\textsuperscript{2}, recorded heavy rainfall during October and November, leading to substantial runoff. The data for this study are taken from \cite{amutha2009estimation}. The estimated parameters are given in Table \ref{Table-22}.
\begin{table}[h!]
\centering

\label{tab:parameters_methods}
\begin{tabular}{lcccccc}
\hline
\textbf{Method} & \(\alpha\) & \(\eta\) & \(\beta\) & \(u\) & \(\sigma\) & \(\xi\) \\ \hline
90th Percentile & N/A & N/A & N/A & 220 & 75.960 & -0.0737 \\
                 &     &     &     &     & (N/A)   & (N/A)    \\
Mean Excess Plot & N/A & N/A & N/A & 220 & 76.000 & -0.0720 \\
                 &     &     &     &     & (N/A)   & (N/A)    \\
Hill Plot        & N/A & N/A & N/A & 240 & (N/A) & 0.2300 \\
                 &     &     &     &     & (N/A)   & (N/A)    \\
Parameter Stability Plot & N/A & N/A & N/A & 220 & 76.000 & -0.0720 \\
                         &     &     &     &     & (N/A)   & (N/A)    \\
EVM Model (Bulk) & N/A  & 1.0106 & 118.4097 & 173.9996 & 67.4980 & 0.0040 \\

&       & (0.0017) & (0.2490) & (0.6272) & (0.4114) & (0.0045)\\

EVM Model (Para.) & N/A  & 0.9816 & 145.6462 & 258.7006 & 72.1900 & -0.0599 \\
                  &      & (0.1387) & (1.0025) & (0.3989) & (0.4510) & (0.0184) \\
EVIMM        & 0.2130 & 1.0426 & 110.9591 & 218.8999 & 58.6405 & 0.1588 \\
             & (0.0004) & (0.0019) & (0.2357) & (0.2413) & (0.3861) &  (0.0050)\\
             \hline
                
\end{tabular}

\captionsetup{
  labelfont=bf, 
  textfont=normal 
}
\caption{Parameter estimates using different methods and models for the Modikuppam rainfall station, with standard errors provided in parentheses (or brackets) below the corresponding values.}
\label{Table-22}
\end{table}

\begin{figure}[!htbp]
    \centering
    \begin{subfigure}{0.48\textwidth}
        \centering
        \includegraphics[width=7.5cm, height=7cm]{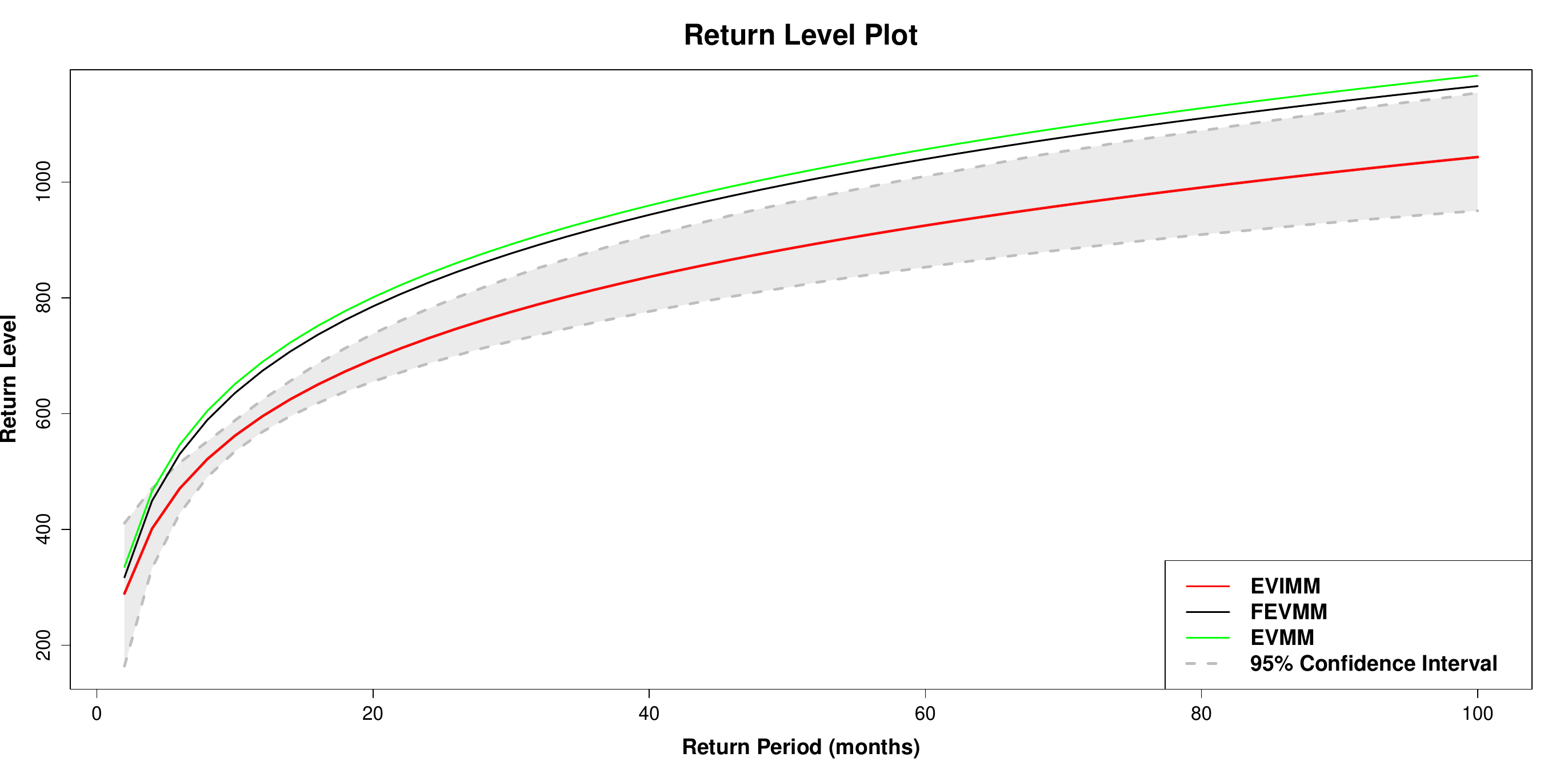}
        \caption{Return level estimates for the real-life dataset \textit{cps91} from the \texttt{wooldridge} R package.}
        \label{fig:return_comparison}
    \end{subfigure}
    \hfill
    \begin{subfigure}{0.48\textwidth}
        \centering
        \includegraphics[width=7.5cm, height=7cm]{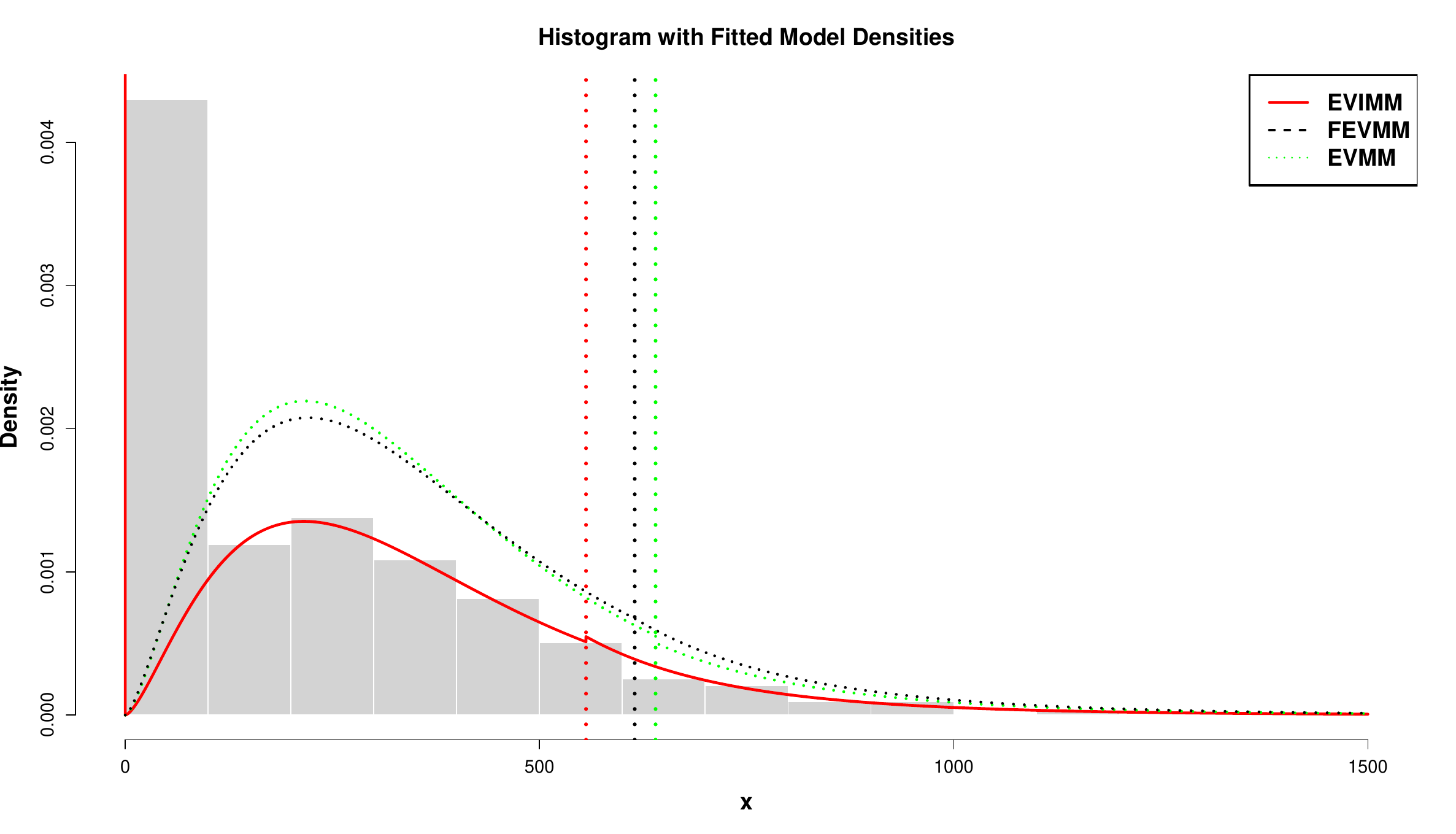}
        \caption{Histogram with fitted model densities. Vertical dashed lines pass through the model-estimated thresholds (\textit{cps91} dataset).}
        \label{fig:hist_syphilis}
    \end{subfigure}
    \\
    \begin{subfigure}{0.48\textwidth}
        \centering
        \includegraphics[width=7.5cm, height=7cm]{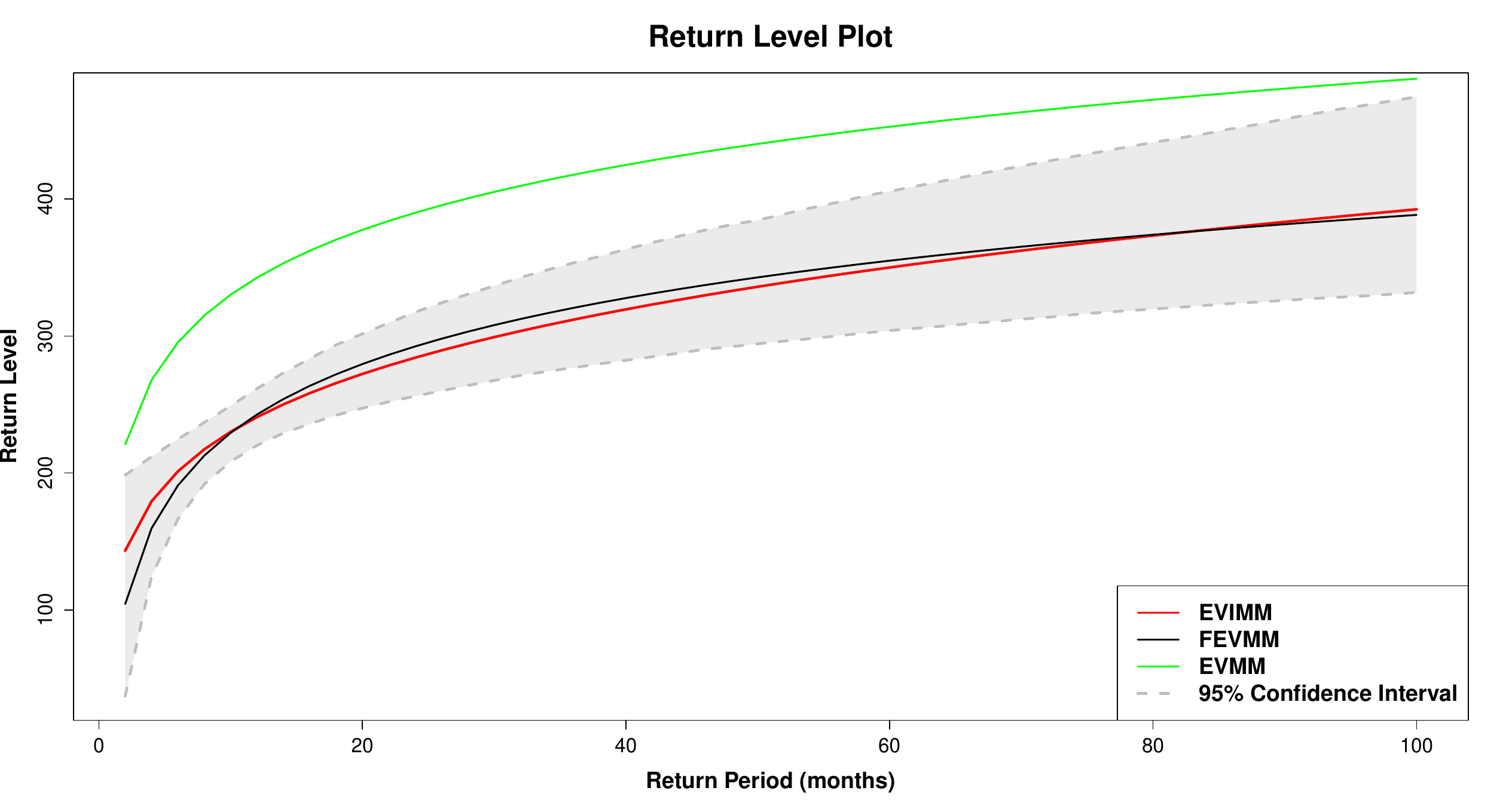}
        \caption{ Return level estimates for the monthly rainfall data (1971–2007) from Modikuppam station, in Vellore District, Tamil Nadu.}
        \label{fig:bias_comparison2}
    \end{subfigure}
    \hfill
    \begin{subfigure}{0.48\textwidth}
        \centering
        \includegraphics[width=7.5cm, height=7cm]{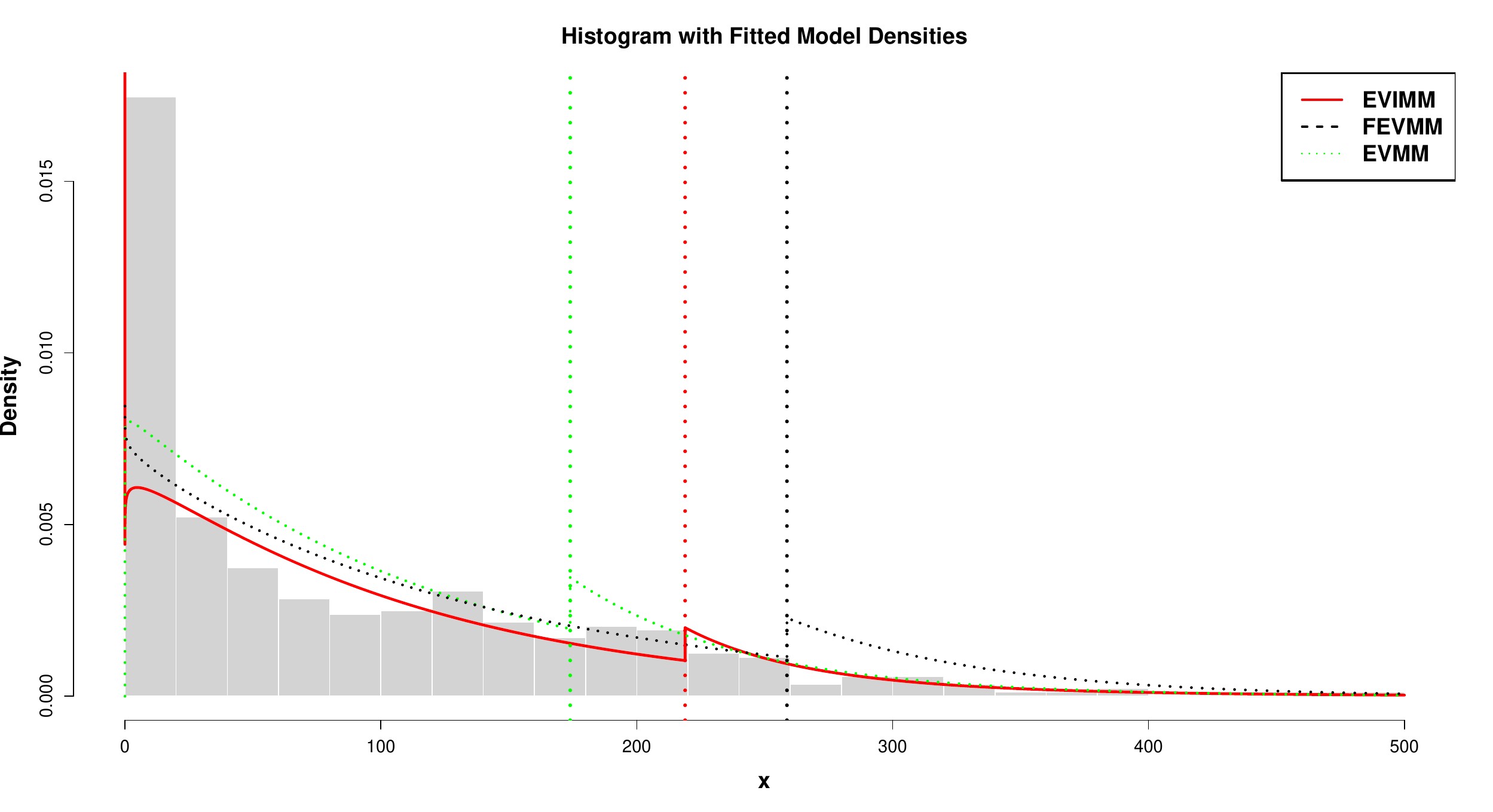}
        \caption{Histogram with fitted model densities. Vertical dashed lines pass through the model-estimated thresholds (rainfall dataset).}
        \label{fig:hist_rainfall}
    \end{subfigure}
    \caption{Return level plots and fitted densities for the \textit{cps91} and rainfall datasets, comparing the proposed EVIMM  with EVMM and FEVMM.}
    \label{fig:return_comparison1}
\end{figure}

Figure \ref{fig:return_comparison} presents the return level plotted against the return period, with the 95\% confidence interval, for both real-life data.
Hence, researchers may achieve better performance using the proposed model, EVIMM, in the presence of inliers, compared to the traditional EVMM, which completely ignores them.
\paragraph{Model Evaluation and Comparison:}
To assess the performance of the proposed EVIMM  against existing models (FEVMM and EVMM), we consider both the normalized Akaike Information Criterion (nAIC) \cite{cohen2021normalized} and standard goodness-of-fit tests. The estimated parameters, along with nAIC values, are presented in Tables~\ref{Table-21} and \ref{Table-22} for the \textit{cps91} and rainfall datasets, respectively. A lower nAIC indicates a better model fit, and in both datasets, the EVIMM achieves the lowest nAIC value among all considered models.

 Model validation is carried out using standard goodness-of-fit tests, namely the Anderson–Darling ($A^2_n$), Cramér–von Mises ($W^2_n$), and Kolmogorov–Smirnov ($D$) statistics. The null hypothesis ($H_0$) in each of these tests states that the observed data are drawn from the specified fitted distribution, implying that the model adequately fits the data. The corresponding test statistics and $p$-values are reported in Table~\ref{tab:gof_results}. For each test, the model with the lowest test statistic and the highest $p$-value is considered to provide a better fit to the observed data. As reported in Table~\ref{tab:gof_results}, the EVIMM  provides the best fit according to the three goodness-of-fit tests. 
 
Furthermore, graphical assessments are provided in Figure \ref{fig:return_comparison}. Figures \ref{fig:hist_syphilis} and \ref{fig:hist_rainfall}
show the histogram of the  \textit{cps91} and rainfall data sets overlaid with the fitted densities, highlighting the visual agreement between the data and the model fits. Figures~\ref{fig:return_comparison} and \ref{fig:bias_comparison2} present the return level plots against the return period, along with the 95\% confidence intervals, demonstrating the reliability of EVIMM in estimating extreme quantiles relevant for real-world risk assessment. Based on these criteria, the EVIMM  outperforms both FEVMM and EVMM, demonstrating its superior ability to capture key distributional features such as extreme events, threshold estimation, tail fraction, inliers, and the bulk of the distribution.
\begin{table}[H]
\centering
\scriptsize
\setlength{\tabcolsep}{8pt}
\renewcommand{\arraystretch}{1.2}
\begin{tabular}{|c|c|c|c|c|}
\hline
\textbf{Model} & \boldmath$A^2_n$ \textbf{($p$-value)} & \boldmath$W^2_n$ \textbf{($p$-value)} & \boldmath$D$ \textbf{($p$-value)} & \textbf{nAIC} \\
\hline
\multicolumn{5}{|c|}{\textbf{\texttt{Cps91}} Dataset} \\
\hline
EVMM  & 16249.76 ($<$0.0001) & 247.2503 ($<$0.0001) & 0.3805 ($<$0.0001) & 13.4020\\
FEVMM & 16664.74 ($<$0.0001) & 264.2363 ($<$0.0001) & 0.3797 ($<$0.0001) & 13.4003 \\
EVIMM & 194.1135 (0.8340) & 36.1086 (0.7918) & 0.0133 (0.7497)  & 9.6570 \\
\hline
\multicolumn{4}{|c|}{\textbf{Rainfall Dataset}} \\
\hline
EVMM  & 3403.101 ($<$0.0001) & 44.5000 ($<$0.0001) & 0.2200 ($<$0.0001) & 11.3973 \\
FEVMM & 3835.447 ($<$0.0001) & 56.6216 ($<$0.0001) & 0.2200 ($<$0.0001) & 11.4083 \\
EVIMM & 12.3224 (0.4716) & 1.5460 (0.4190) & 0.0391 (0.4443) & 9.9677 \\
\hline
\end{tabular}
\caption{Summary of goodness-of-fit tests and normalized AIC (nAIC) values for the two datasets.}
\label{tab:gof_results}
\end{table}

\section{Conclusion and Future Work}\label{conclusion}
\noindent In this article, we proposed a model that effectively captures extreme data, as well as inliers at the origin. This is important when we aim to obtain an accurate representation of the data. The model has the advantage of capturing the extreme and inliers, addressing the common occurrence of instantaneous events or inliers in random phenomena, often discarding the assumptions of unimodal distributions. The simulation study shows that the proposed EVIMM  estimates the true parameters more accurately specially threshold ($u$), shape parameter ($\xi$), scale parameter ($\sigma$) of GPD, with lower bias and MSE compared to the existing models, EVMM and FEVMM, in the presence of inliers, as shown in Figures~\ref{fig:bias_comparison} and~\ref{fig:mse_comparison}, and given in Tables~\ref{Table-100}-\ref{Table-600}. As seen in Table~\ref{tab:gof_results} and Figure~\ref{fig:return_comparison1}, the EVIMM provides the best fit to the data based on test results, the density-histogram plot, and nAIC, highlighting its superior performance in handling such scenarios. To sum up, the results in Tables~\ref{Table-700}-\ref{Table-1000} show that classical methods like MEP and PSP may be affected by the presence of inliers, as they give different parameter estimates depending on whether inliers are included. This highlights the need for a more robust approach. Our proposed EVIMM  handled the complete data better and gave more accurate estimates, especially for the shape parameter of the GPD. These findings support the usefulness of the EVIMM framework in situations where inliers may affect estimates of the extreme parameters or may themselves lead to severe effects. For example, heavy rainfall can cause floods, but many dry days \text{(i.e., inliers)} during seasonal periods can also lead to drought, which is an extreme effect. From Figure~\ref{fig:return_comparisonsimu}, it is clear that the EVIMM  provides better return level estimates, which are closer to the true values compared to those from EVMM and FEVMM.

Overall, this study shows the importance of having a robust and flexible framework for modeling extreme values, especially when inliers are present. As discussed by  Hu \emph{et al.}\cite{hu2013extreme}, there is no general framework for defining EVMMs, and the way the proportion above the threshold is handled can vary across models. Also, the choice of the distribution below the threshold can affect the threshold estimate and the extreme value parameters, as pointed out by Behrens \emph{et al.}~\cite{behrens2004bayesian} and \citet{hu2018evmix}. Our comparison with existing models like EVMM and FEVMM, MEP, PSP shows that EVIMM gives better results in terms of robustness, parameter estimation, and overall fit. This highlights the need to study such models carefully, especially in real-world situations where inliers can affect the estimation of extreme events.

Existing literature on the modeling of extremes that incorporates inliers in the distribution below the threshold is almost nonexistent. To the best of our knowledge, this specific area remains unexplored, indicating a significant opportunity for further research.
A natural extension of the EVIMM framework is to address scenarios where the inliers are concentrated around a specific point rather than at a specific point, such as zero, while others follow a different distribution. This can be achieved by incorporating a finite-range distribution, like a uniform distribution near the origin.
One potential extension is to study how inliers, which occur at a point or multiple points, affect the threshold location and subsequently GPD parameter estimation, potentially having a significant impact on the tail parameter estimation. 
 In this context, different bulk distributions can be explored to identify the optimal formulation.
Another potential direction of the EVIMM is to incorporate multimodality in the bulk distribution. One can use a mixture of distributions (such as a mixture of gamma distributions) as a bulk distribution. The advantage of this model formulation is flexibility; the information criterion can decide the number of components.
For greater flexibility and to accommodate a wide variety of shapes and spreads in the bulk part, one can extend and utilise the Dirichlet process of a mixture of gamma densities for the bulk distribution.
Finally, extending the EVIMM to a change-point setup represents another promising research avenue. The change point models, which explicitly study distributional changes in the structure of the extreme values, are very limited in the literature.
 Developing models to handle such cases could provide further insights. Additionally, further analysis could explore advanced estimation methods, such as Bayesian techniques, to more effectively estimate the model's unknown parameters. These directions represent promising opportunities for extending the scope and applicability of the proposed model. It could involve expanding the applicability of the proposed model to real-life data in fields such as reliability, environmental data (rainfall or drought), and finance, as well as comparing different estimation techniques to achieve optimal model performance.

\section*{Acknowledgements}\label{conclusion1}
\noindent The authors thank the referees for their constructive comments on an earlier version of the article. A preprint of this article is available on arXiv; see \cite{nila2025modeling}.

 \bibliographystyle{apalike}
 \bibliography{01ref}

\newpage
\section*{Supplementary Material: Additional Simulation Tables for Sample Mean, BSE, BCI, MSE, Bias, CP}
\begin{landscape} 
\begin{table*}[htbp]
    \centering
    \scriptsize 
    \setlength{\tabcolsep}{3pt} 
    \renewcommand{\arraystretch}{1.3} 
    \begin{tabular}{|p{2.5cm}|p{1.8cm}|p{1.8cm}|p{2.40cm}|p{1.8cm}|p{2.3cm}|p{2.3cm}|p{2.3cm}|p{1.8cm}|} 
        \hline
        \textbf{Sample Size} & \textbf{Parameters} & \textbf{True Value} & \textbf{Sample Mean} & \textbf{BSE} & \textbf{BCI} & \textbf{MSE} & \textbf{Bias} & \textbf{CP} \\ \hline
   \multirow{6}{*}{ 300 } & $\alpha$ &  0.1  &  0.1006  (NA) &  0.0164  &  (0.0537, 0.1175)  &  3e-04  (NA) &  6e-04  (NA) &  97.6  \\ \cline{2-9}
                               & $\eta$   &  1  &  1.0228  ( 1.0098 ) &  0.0831  &  (0.8384, 1.167)  &  0.0081  ( 0.007 ) &  0.0228  ( 0.0098 ) &  96.9  \\ \cline{2-9}
                               & $\beta$  &  5  &  4.897  ( 4.9771 ) &  0.5587  &  (3.7207, 5.9153)  &  0.3264  ( 0.3338 ) &  -0.103  ( -0.0229 ) &  92.2  \\ \cline{2-9}
                               & $u$      &  10.9861  &  11.3489  ( 11.4183 ) &  1.1752  &  (8.6897, 13.4214)  &  1.8504  ( 4.3832 ) &  0.3628  ( 0.4322 ) &  94.3  \\ \cline{2-9}
                               & $\xi$    &  0.2  &  0.1496  ( 0.1442 ) &  0.3339  &  (-0.6739, 0.6399)  &  0.1585  ( 0.3001 ) &  -0.0504  ( -0.0558 ) &  93.7  \\ \cline{2-9}
                               & $\sigma$ &  5  &  5.6882  ( 6.0615 ) &  2.8003  &  (2.2598, 13.1484)  &  7.7169  ( 14.235 ) &  0.6882  ( 1.0615 ) &  95.1  \\ \hline
     \multirow{6}{*}{ 400 } & $\alpha$ &  0.1  &  0.1007  (NA) &  0.015  &  (0.0692, 0.1269)  &  2e-04  (NA) &  7e-04  (NA) &  97.4  \\ \cline{2-9}
                               & $\eta$   &  1  &  1.0215  ( 1.0075 ) &  0.0784  &  (0.9183, 1.2214)  &  0.0062  ( 0.0053 ) &  0.0215  ( 0.0075 ) &  96.1  \\ \cline{2-9}
                               & $\beta$  &  5  &  4.9006  ( 4.9872 ) &  0.4447  &  (3.5472, 5.2964)  &  0.2599  ( 0.2642 ) &  -0.0994  ( -0.0128 ) &  93.0  \\ \cline{2-9}
                               & $u$      &  10.9861  &  11.4155  ( 11.52 ) &  1.0114  &  (8.4317, 12.5374)  &  1.8287  ( 4.3341 ) &  0.4294  ( 0.5339 ) &  93.1  \\ \cline{2-9}
                               & $\xi$    &  0.2  &  0.1793  ( 0.162 ) &  0.2676  &  (-0.5236, 0.5463)  &  0.1133  ( 0.1998 ) &  -0.0207  ( -0.038 ) &  94.1  \\ \cline{2-9}
                               & $\sigma$ &  5  &  5.474  ( 5.7329 ) &  2.2325  &  (2.6011, 11.4788)  &  5.1572  ( 8.5738 ) &  0.474  ( 0.7329 ) &  93.4  \\ \hline
       \multirow{6}{*}{ 500 } & $\alpha$ &  0.1  &  0.1006  (NA) &  0.013  &  (0.066, 0.1171)  &  2e-04  (NA) &  6e-04  (NA) &  97.0  \\ \cline{2-9}
                               & $\eta$   &  1  &  1.0198  ( 1.0072 ) &  0.0643  &  (0.8742, 1.1277)  &  0.0049  ( 0.0042 ) &  0.0198  ( 0.0072 ) &  96.7  \\ \cline{2-9}
                               & $\beta$  &  5  &  4.901  ( 4.9815 ) &  0.4284  &  (4.0824, 5.7653)  &  0.2062  ( 0.2047 ) &  -0.099  ( -0.0185 ) &  92.3  \\ \cline{2-9}
                               & $u$      &  10.9861  &  11.5179  ( 11.542 ) &  1.2317  &  (8.9944, 13.8289)  &  2.1158  ( 3.9504 ) &  0.5317  ( 0.5559 ) &  91.1  \\ \cline{2-9}
                               & $\xi$    &  0.2  &  0.1887  ( 0.1555 ) &  0.2605  &  (-0.401, 0.5732)  &  0.0893  ( 0.1207 ) &  -0.0113  ( -0.0445 ) &  94.6  \\ \cline{2-9}
                               & $\sigma$ &  5  &  5.4306  ( 5.6809 ) &  1.6648  &  (2.1885, 8.041)  &  4.1245  ( 6.8022 ) &  0.4306  ( 0.6809 ) &  92.0  \\ \hline
            \multirow{6}{*}{ 750 } & $\alpha$ &  0.1  &  0.1004  (NA) &  0.011  &  (0.0744, 0.1168)  &  1e-04  (NA) &  4e-04  (NA) &  98.0  \\ \cline{2-9}
                               & $\eta$   &  1  &  1.0175  ( 1.006 ) &  0.054  &  (0.9114, 1.1257)  &  0.0035  ( 0.0029 ) &  0.0175  ( 0.006 ) &  95.2  \\ \cline{2-9}
                               & $\beta$  &  5  &  4.9047  ( 4.9795 ) &  0.3431  &  (4.1173, 5.4733)  &  0.152  ( 0.138 ) &  -0.0953  ( -0.0205 ) &  91.9  \\ \cline{2-9}
                               & $u$      &  10.9861  &  11.5728  ( 11.4956 ) &  1.3182  &  (8.9372, 14.0195)  &  2.0889  ( 2.9793 ) &  0.5867  ( 0.5094 ) &  88.3  \\ \cline{2-9}
                               & $\xi$    &  0.2  &  0.2023  ( 0.1782 ) &  0.209  &  (-0.3034, 0.4807)  &  0.0564  ( 0.0818 ) &  0.0023  ( -0.0218 ) &  92.6  \\ \cline{2-9}
                               & $\sigma$ &  5  &  5.2926  ( 5.4044 ) &  1.2955  &  (2.4515, 7.0479)  &  2.4831  ( 3.5538 ) &  0.2926  ( 0.4044 ) &  92.9  \\ \hline
          \multirow{6}{*}{ 1000 } & $\alpha$ &  0.1  &  0.1003  (NA) &  0.0098  &  (0.0814, 0.1199)  &  1e-04  (NA) &  3e-04  (NA) &  98.4  \\ \cline{2-9}
                               & $\eta$   &  1  &  1.0145  ( 1.0048 ) &  0.0488  &  (0.9439, 1.1344)  &  0.0025  ( 0.0021 ) &  0.0145  ( 0.0048 ) &  95.6  \\ \cline{2-9}
                               & $\beta$  &  5  &  4.9197  ( 4.9828 ) &  0.3017  &  (4.0964, 5.2875)  &  0.1114  ( 0.1033 ) &  -0.0803  ( -0.0172 ) &  92.1  \\ \cline{2-9}
                               & $u$      &  10.9861  &  11.626  ( 11.5208 ) &  1.4013  &  (8.8489, 14.2067)  &  2.1303  ( 2.28 ) &  0.6398  ( 0.5347 ) &  90.1  \\ \cline{2-9}
                               & $\xi$    &  0.2  &  0.2057  ( 0.1819 ) &  0.1681  &  (-0.2703, 0.3726)  &  0.0402  ( 0.0405 ) &  0.0057  ( -0.0181 ) &  72.8  \\ \cline{2-9}
                               & $\sigma$ &  5  &  5.2655  ( 5.3122 ) &  1.2151  &  (3.0731, 7.5284)  &  1.8942  ( 2.0743 ) &  0.2655  ( 0.3122 ) &  91.1  \\ \hline
     \end{tabular}
    \captionsetup{
  labelfont=bf, 
  textfont=normal 
}
     \caption{Simulation results based on \( N = 2000 \) replications for sample sizes \( n = 300, 400, 500, 750, \) and \( 1000 \), with an inlier proportion of 10\% and heavy-tailed generalized Pareto distributions (GPD). The table presents the sample mean of estimated parameters, bootstrap standard error (BSE), 95\% bootstrap confidence interval (BCI), mean square error (MSE), bias, and convergence probability (CP).}

    \label{Table-1100}
\end{table*}
\end{landscape}

    \begin{landscape} 
\begin{table*}[htbp]
    \centering
    \scriptsize 
    \setlength{\tabcolsep}{3pt} 
    \renewcommand{\arraystretch}{1.3} 
    \begin{tabular}{|p{2.5cm}|p{1.8cm}|p{1.8cm}|p{2.3cm}|p{1.8cm}|p{2.3cm}|p{2.3cm}|p{2.3cm}|p{1.8cm}|} 
        \hline
        \textbf{Sample Size} & \textbf{Parameters} & \textbf{True Value} & \textbf{Sample Mean} & \textbf{BSE} & \textbf{BCI} & \textbf{MSE} & \textbf{Bias} & \textbf{CP} \\ \hline

        \multirow{6}{*}{ 300 } & $\alpha$ &  0.3  &  0.2997  (NA) &  0.0273  &  (0.2681, 0.3749)  &  7e-04  (NA) &  -3e-04  (NA) &  96.4  \\ \cline{2-9}
                               & $\eta$   &  1  &  1.022  ( 1.0169 ) &  0.1214  &  (0.9976, 1.4702)  &  0.0098  ( 0.0096 ) &  0.022  ( 0.0169 ) &  97.1  \\ \cline{2-9}
                               & $\beta$  &  5  &  4.908  ( 4.9528 ) &  0.5401  &  (2.9843, 5.1016)  &  0.4318  ( 0.4796 ) &  -0.092  ( -0.0472 ) &  92.8  \\ \cline{2-9}
                               & $u$      &  9.7295  &  9.5101  ( 11.3791 ) &  0.911  &  (7.0607, 10.6418)  &  0.9939  ( 7.7344 ) &  -0.2194  ( 1.6495 ) &  95.7  \\ \cline{2-9}
                               & $\xi$    &  0.2  &  0.125  ( 0.0987 ) &  0.2878  &  (-0.5414, 0.6101)  &  0.0979  ( 0.3648 ) &  -0.075  ( -0.1013 ) &  93.7  \\ \cline{2-9}
                               & $\sigma$ &  5  &  5.4454  ( 6.8068 ) &  1.937  &  (2.1855, 9.6555)  &  4.7926  ( 21.0096 ) &  0.4454  ( 1.8068 ) &  94.4  \\ \hline
       \multirow{6}{*}{ 400 } & $\alpha$ &  0.3  &  0.2995  (NA) &  0.0239  &  (0.2787, 0.3716)  &  5e-04  (NA) &  -5e-04  (NA) &  97.3  \\ \cline{2-9}
                               & $\eta$   &  1  &  1.0167  ( 1.0128 ) &  0.0933  &  (0.9136, 1.2747)  &  0.0069  ( 0.0067 ) &  0.0167  ( 0.0128 ) &  97.8  \\ \cline{2-9}
                               & $\beta$  &  5  &  4.928  ( 4.9599 ) &  0.5279  &  (3.5476, 5.605)  &  0.3122  ( 0.3145 ) &  -0.072  ( -0.0401 ) &  94.0  \\ \cline{2-9}
                               & $u$      &  9.7295  &  9.4869  ( 11.4543 ) &  0.8234  &  (7.3466, 10.621)  &  0.8054  ( 7.4234 ) &  -0.2427  ( 1.7248 ) &  95.4  \\ \cline{2-9}
                               & $\xi$    &  0.2  &  0.1548  ( 0.1398 ) &  0.2612  &  (-0.3202, 0.699)  &  0.0666  ( 0.2767 ) &  -0.0452  ( -0.0602 ) &  93.5  \\ \cline{2-9}
                               & $\sigma$ &  5  &  5.2403  ( 6.3886 ) &  1.4873  &  (1.8952, 7.4652)  &  3.1008  ( 14.8354 ) &  0.2403  ( 1.3886 ) &  93.8  \\ \hline
  \multirow{6}{*}{ 500 } & $\alpha$ &  0.3  &  0.2994  (NA) &  0.0214  &  (0.275, 0.3589)  &  4e-04  (NA) &  -6e-04  (NA) &  97.2  \\ \cline{2-9}
                               & $\eta$   &  1  &  1.0123  ( 1.0098 ) &  0.0774  &  (0.8978, 1.2002)  &  0.0054  ( 0.0053 ) &  0.0123  ( 0.0098 ) &  96.7  \\ \cline{2-9}
                               & $\beta$  &  5  &  4.946  ( 4.9665 ) &  0.4955  &  (3.8134, 5.7652)  &  0.2596  ( 0.2633 ) &  -0.054  ( -0.0335 ) &  93.1  \\ \cline{2-9}
                               & $u$      &  9.7295  &  9.4823  ( 11.3961 ) &  0.8031  &  (7.6783, 10.8667)  &  0.7257  ( 6.6181 ) &  -0.2472  ( 1.6666 ) &  93.3  \\ \cline{2-9}
                               & $\xi$    &  0.2  &  0.1644  ( 0.149 ) &  0.2095  &  (-0.32, 0.5237)  &  0.0465  ( 0.1765 ) &  -0.0356  ( -0.051 ) &  93.3  \\ \cline{2-9}
                               & $\sigma$ &  5  &  5.1552  ( 6.1164 ) &  1.2669  &  (2.3836, 7.1814)  &  2.2346  ( 9.9278 ) &  0.1552  ( 1.1164 ) &  93.7  \\ \hline
         \multirow{6}{*}{ 750 } & $\alpha$ &  0.3  &  0.2993  (NA) &  0.0175  &  (0.2759, 0.343)  &  3e-04  (NA) &  -7e-04  (NA) &  97.9  \\ \cline{2-9}
                               & $\eta$   &  1  &  1.0088  ( 1.0057 ) &  0.0668  &  (0.9707, 1.2309)  &  0.0037  ( 0.0034 ) &  0.0088  ( 0.0057 ) &  96.7  \\ \cline{2-9}
                               & $\beta$  &  5  &  4.9554  ( 4.9789 ) &  0.371  &  (3.6694, 5.118)  &  0.1832  ( 0.1714 ) &  -0.0446  ( -0.0211 ) &  93.5  \\ \cline{2-9}
                               & $u$      &  9.7295  &  9.4531  ( 11.4923 ) &  0.6141  &  (7.7371, 10.1361)  &  0.5873  ( 6.5489 ) &  -0.2765  ( 1.7627 ) &  94.4  \\ \cline{2-9}
                               & $\xi$    &  0.2  &  0.1825  ( 0.1558 ) &  0.1474  &  (-0.2752, 0.3114)  &  0.0303  ( 0.1159 ) &  -0.0175  ( -0.0442 ) &  92.6  \\ \cline{2-9}
                               & $\sigma$ &  5  &  5.0286  ( 5.9335 ) &  1.0331  &  (3.2022, 7.1882)  &  1.3865  ( 5.8942 ) &  0.0286  ( 0.9335 ) &  93.0  \\ \hline
        \multirow{6}{*}{ 1000 } & $\alpha$ &  0.3  &  0.2992  (NA) &  0.0149  &  (0.2745, 0.3328)  &  2e-04  (NA) &  -8e-04  (NA) &  98.3  \\ \cline{2-9}
                               & $\eta$   &  1  &  1.0071  ( 1.004 ) &  0.0555  &  (0.9299, 1.1455)  &  0.0026  ( 0.0025 ) &  0.0071  ( 0.004 ) &  96.8  \\ \cline{2-9}
                               & $\beta$  &  5  &  4.9568  ( 4.9805 ) &  0.3483  &  (4.0258, 5.371)  &  0.1327  ( 0.129 ) &  -0.0432  ( -0.0195 ) &  93.4  \\ \cline{2-9}
                               & $u$      &  9.7295  &  9.448  ( 11.4581 ) &  0.6045  &  (7.9167, 10.3136)  &  0.484  ( 5.6107 ) &  -0.2815  ( 1.7285 ) &  93.3  \\ \cline{2-9}
                               & $\xi$    &  0.2  &  0.1865  ( 0.1627 ) &  0.1237  &  (-0.2224, 0.2664)  &  0.021  ( 0.0674 ) &  -0.0135  ( -0.0373 ) &  91.6  \\ \cline{2-9}
                               & $\sigma$ &  5  &  5.0135  ( 5.7996 ) &  0.9093  &  (3.5399, 7.0413)  &  0.9672  ( 3.7267 ) &  0.0135  ( 0.7996 ) &  94.1  \\ \hline
     \end{tabular}
\captionsetup{
  labelfont=bf, 
  textfont=normal 
}
     \caption{Simulation results based on \( N = 2000 \) replications for sample sizes \( n = 300, 400, 500, 750, \) and \( 1000 \), with an inlier proportion of 30\% and heavy-tailed generalized Pareto distributions (GPD). The table presents the sample mean of estimated parameters, bootstrap standard error (BSE), 95\% bootstrap confidence interval (BCI), mean square error (MSE), bias, and convergence probability (CP).}
    \label{tab:simulation_study_parameters_extended}
    \label{Table-1200}
\end{table*}
\end{landscape}

\begin{landscape} 
\begin{table*}[htbp]
    \centering
    \scriptsize 
    \setlength{\tabcolsep}{3pt} 
    \renewcommand{\arraystretch}{1.3} 
    \begin{tabular}{|p{2.5cm}|p{1.8cm}|p{1.8cm}|p{2.4cm}|p{1.8cm}|p{2.3cm}|p{2.3cm}|p{2.3cm}|p{1.8cm}|} 
        \hline
        \textbf{Sample Size} & \textbf{Parameters} & \textbf{True Value} & \textbf{Sample Mean} & \textbf{BSE} & \textbf{BCI} & \textbf{MSE} & \textbf{Bias} & \textbf{CP} \\ \hline
   \multirow{6}{*}{ 300 } & $\alpha$ &  0.1  &  0.1003  (NA) &  0.0167  &  (0.0535, 0.1191)  &  3e-04  (NA) &  3e-04  (NA) &  96.3  \\ \cline{2-9}
                               & $\eta$   &  1  &  1.0245  ( 1.0114 ) &  0.0891  &  (0.8658, 1.2154)  &  0.0081  ( 0.0071 ) &  0.0245  ( 0.0114 ) &  95.7  \\ \cline{2-9}
                               & $\beta$  &  5  &  4.8861  ( 4.9671 ) &  0.5397  &  (3.5417, 5.6621)  &  0.3151  ( 0.3443 ) &  -0.1139  ( -0.0329 ) &  90.4  \\ \cline{2-9}
                               & $u$      &  10.9861  &  11.431  ( 11.1255 ) &  1.1892  &  (8.6954, 13.5307)  &  1.678  ( 3.1 ) &  0.4449  ( 0.1394 ) &  94.4  \\ \cline{2-9}
                               & $\xi$    &  -0.2  &  -0.3065  ( -0.2903 ) &  0.3078  &  (-1.0734, 0.2349)  &  0.1245  ( 0.1313 ) &  -0.1065  ( -0.0903 ) &  93.5  \\ \cline{2-9}
                               & $\sigma$ &  5  &  5.5694  ( 5.6387 ) &  2.4102  &  (2.6069, 12.4937)  &  5.4306  ( 6.7653 ) &  0.5694  ( 0.6387 ) &  91.1  \\ \hline
       \multirow{6}{*}{ 400 } & $\alpha$ &  0.1  &  0.1004  (NA) &  0.0145  &  (0.0615, 0.1185)  &  2e-04  (NA) &  4e-04  (NA) &  97.7  \\ \cline{2-9}
                               & $\eta$   &  1  &  1.0217  ( 1.0077 ) &  0.0732  &  (0.8603, 1.1453)  &  0.0063  ( 0.0053 ) &  0.0217  ( 0.0077 ) &  96.1  \\ \cline{2-9}
                               & $\beta$  &  5  &  4.9026  ( 4.9848 ) &  0.4816  &  (3.8562, 5.7628)  &  0.2506  ( 0.2615 ) &  -0.0974  ( -0.0152 ) &  91.7  \\ \cline{2-9}
                               & $u$      &  10.9861  &  11.4603  ( 11.2114 ) &  1.1725  &  (8.8729, 13.4797)  &  1.5935  ( 2.7607 ) &  0.4742  ( 0.2253 ) &  91.8  \\ \cline{2-9}
                               & $\xi$    &  -0.2  &  -0.2667  ( -0.2691 ) &  0.2643  &  (-0.8894, 0.2116)  &  0.0854  ( 0.087 ) &  -0.0667  ( -0.0691 ) &  93.8  \\ \cline{2-9}
                               & $\sigma$ &  5  &  5.3155  ( 5.4342 ) &  1.84  &  (2.4929, 9.5964)  &  3.5544  ( 4.6116 ) &  0.3155  ( 0.4342 ) &  91.1  \\ \hline
     \multirow{6}{*}{ 500 } & $\alpha$ &  0.1  &  0.1003  (NA) &  0.0133  &  (0.0666, 0.1193)  &  2e-04  (NA) &  3e-04  (NA) &  97.1  \\ \cline{2-9}
                               & $\eta$   &  1  &  1.018  ( 1.0074 ) &  0.0652  &  (0.8736, 1.1319)  &  0.005  ( 0.0043 ) &  0.018  ( 0.0074 ) &  94.5  \\ \cline{2-9}
                               & $\beta$  &  5  &  4.918  ( 4.9825 ) &  0.4224  &  (4.0714, 5.7513)  &  0.1978  ( 0.2147 ) &  -0.082  ( -0.0175 ) &  90.0  \\ \cline{2-9}
                               & $u$      &  10.9861  &  11.5102  ( 11.294 ) &  1.0635  &  (9.2155, 13.3659)  &  1.6045  ( 2.4295 ) &  0.5241  ( 0.3078 ) &  92.2  \\ \cline{2-9}
                               & $\xi$    &  -0.2  &  -0.2535  ( -0.26 ) &  0.2395  &  (-0.6792, 0.2556)  &  0.0606  ( 0.0633 ) &  -0.0535  ( -0.06 ) &  94.7  \\ \cline{2-9}
                               & $\sigma$ &  5  &  5.2074  ( 5.3341 ) &  1.4181  &  (2.1816, 7.2949)  &  2.6715  ( 3.4877 ) &  0.2074  ( 0.3341 ) &  91.1  \\ \hline
          \multirow{6}{*}{ 750 } & $\alpha$ &  0.1  &  0.1  (NA) &  0.0111  &  (0.0738, 0.1171)  &  1e-04  (NA) &  0  (NA) &  97.7  \\ \cline{2-9}
                               & $\eta$   &  1  &  1.0172  ( 1.0061 ) &  0.0559  &  (0.9182, 1.1358)  &  0.0036  ( 0.0029 ) &  0.0172  ( 0.0061 ) &  95.5  \\ \cline{2-9}
                               & $\beta$  &  5  &  4.9107  ( 4.979 ) &  0.3416  &  (4.0475, 5.4021)  &  0.1453  ( 0.1399 ) &  -0.0893  ( -0.021 ) &  89.0  \\ \cline{2-9}
                               & $u$      &  10.9861  &  11.521  ( 11.3775 ) &  1.0957  &  (9.0228, 13.2515)  &  1.6194  ( 2.0043 ) &  0.5349  ( 0.3913 ) &  93.0  \\ \cline{2-9}
                               & $\xi$    &  -0.2  &  -0.2224  ( -0.2365 ) &  0.1787  &  (-0.5993, 0.1125)  &  0.0366  ( 0.0361 ) &  -0.0224  ( -0.0365 ) &  93.0  \\ \cline{2-9}
                               & $\sigma$ &  5  &  5.0192  ( 5.1405 ) &  1.1691  &  (2.6499, 6.8333)  &  1.6592  ( 2.0507 ) &  0.0192  ( 0.1405 ) &  90.9  \\ \hline
       \multirow{6}{*}{ 1000 } & $\alpha$ &  0.1  &  0.0994  (NA) &  0.0101  &  (0.08, 0.1195)  &  1e-04  (NA) &  -6e-04  (NA) &  98.2  \\ \cline{2-9}
                               & $\eta$   &  1  &  1.0142  ( 1.0048 ) &  0.0494  &  (0.9377, 1.1335)  &  0.0025  ( 0.0021 ) &  0.0142  ( 0.0048 ) &  95.6  \\ \cline{2-9}
                               & $\beta$  &  5  &  4.9227  ( 4.9833 ) &  0.2973  &  (4.1721, 5.3589)  &  0.1048  ( 0.1054 ) &  -0.0773  ( -0.0167 ) &  90.3  \\ \cline{2-9}
                               & $u$      &  10.9861  &  11.5441  ( 11.4225 ) &  1.0475  &  (9.212, 13.2607)  &  1.5883  ( 1.5418 ) &  0.558  ( 0.4363 ) &  92.6  \\ \cline{2-9}
                               & $\xi$    &  -0.2  &  -0.2184  ( -0.2264 ) &  0.1528  &  (-0.5135, 0.0846)  &  0.0249  ( 0.0223 ) &  -0.0184  ( -0.0264 ) &  91.8  \\ \cline{2-9}
                               & $\sigma$ &  5  &  4.9929  ( 5.0559 ) &  1.0216  &  (2.6277, 6.3973)  &  1.1997  ( 1.2731 ) &  -0.0071  ( 0.0559 ) &  90.8  \\ \hline
    \end{tabular}
\captionsetup{
  labelfont=bf, 
  textfont=normal 
}
      \caption{Simulation results based on \( N = 2000 \) replications for sample sizes \( n = 300, 400, 500, 750, \) and \( 1000 \), with an inlier proportion of 10\% and light-tailed generalized Pareto distributions (GPD). The table presents the sample mean of estimated parameters, bootstrap standard error (BSE), 95\% bootstrap confidence interval (BCI), mean square error (MSE), bias, and convergence probability (CP).}
    \label{Table-1300}
\end{table*}
\end{landscape}

\begin{landscape} 
\begin{table*}[htbp]
    \centering
    \scriptsize 
    \setlength{\tabcolsep}{3pt} 
    \renewcommand{\arraystretch}{1.3} 
    \begin{tabular}{|p{2.5cm}|p{1.8cm}|p{1.8cm}|p{2.3cm}|p{1.8cm}|p{2.3cm}|p{2.3cm}|p{2.3cm}|p{1.8cm}|} 
        \hline
        \textbf{Sample Size} & \textbf{Parameters} & \textbf{True Value} & \textbf{Sample Mean} & \textbf{BSE} & \textbf{BCI} & \textbf{MSE} & \textbf{Bias} & \textbf{CP} \\ \hline

   \multirow{6}{*}{ 300 } & $\alpha$ &  0.3  &  0.2997  (NA) &  0.0273  &  (0.2668, 0.374)  &  7e-04  (NA) &  -3e-04  (NA) &  97.2  \\ \cline{2-9}
                               & $\eta$   &  1  &  1.0231  ( 1.0168 ) &  0.1151  &  (0.9672, 1.4157)  &  0.0096  ( 0.0097 ) &  0.0231  ( 0.0168 ) &  96.7  \\ \cline{2-9}
                               & $\beta$  &  5  &  4.893  ( 4.9506 ) &  0.5501  &  (3.2253, 5.3756)  &  0.4112  ( 0.4376 ) &  -0.107  ( -0.0494 ) &  91.5  \\ \cline{2-9}
                               & $u$      &  9.7295  &  9.6492  ( 11.0965 ) &  0.8691  &  (7.5382, 10.9653)  &  0.9768  ( 5.0175 ) &  -0.0804  ( 1.3669 ) &  94.4  \\ \cline{2-9}
                               & $\xi$    &  -0.2  &  -0.2995  ( -0.2819 ) &  0.2793  &  (-0.8193, 0.3094)  &  0.0891  ( 0.1795 ) &  -0.0995  ( -0.0819 ) &  91.8  \\ \cline{2-9}
                               & $\sigma$ &  5  &  5.5385  ( 5.3934 ) &  1.64  &  (2.0657, 8.3462)  &  4.1475  ( 7.8519 ) &  0.5385  ( 0.3934 ) &  92.2  \\ \hline
  \multirow{6}{*}{ 400 } & $\alpha$ &  0.3  &  0.2998  (NA) &  0.0241  &  (0.2808, 0.3755)  &  5e-04  (NA) &  -2e-04  (NA) &  97.2  \\ \cline{2-9}
                               & $\eta$   &  1  &  1.016  ( 1.0121 ) &  0.0844  &  (0.8622, 1.1947)  &  0.0067  ( 0.0071 ) &  0.016  ( 0.0121 ) &  97.0  \\ \cline{2-9}
                               & $\beta$  &  5  &  4.9268  ( 4.9676 ) &  0.5635  &  (3.9329, 6.1321)  &  0.3048  ( 0.3366 ) &  -0.0732  ( -0.0324 ) &  93.1  \\ \cline{2-9}
                               & $u$      &  9.7295  &  9.6721  ( 11.1071 ) &  0.9046  &  (7.8315, 11.3974)  &  0.8096  ( 5.0201 ) &  -0.0575  ( 1.3776 ) &  93.8  \\ \cline{2-9}
                               & $\xi$    &  -0.2  &  -0.259  ( -0.2599 ) &  0.2244  &  (-0.6565, 0.229)  &  0.0511  ( 0.124 ) &  -0.059  ( -0.0599 ) &  95.3  \\ \cline{2-9}
                               & $\sigma$ &  5  &  5.2721  ( 5.1545 ) &  1.3352  &  (2.2968, 7.3168)  &  2.4281  ( 5.3747 ) &  0.2721  ( 0.1545 ) &  94.6  \\ \hline
   \multirow{6}{*}{ 500 } & $\alpha$ &  0.3  &  0.2994  (NA) &  0.0213  &  (0.2747, 0.358)  &  4e-04  (NA) &  -6e-04  (NA) &  97.4  \\ \cline{2-9}
                               & $\eta$   &  1  &  1.0127  ( 1.008 ) &  0.0759  &  (0.8994, 1.1969)  &  0.0052  ( 0.0051 ) &  0.0127  ( 0.008 ) &  96.3  \\ \cline{2-9}
                               & $\beta$  &  5  &  4.9401  ( 4.982 ) &  0.4763  &  (3.8592, 5.7171)  &  0.2455  ( 0.2538 ) &  -0.0599  ( -0.018 ) &  91.8  \\ \cline{2-9}
                               & $u$      &  9.7295  &  9.6694  ( 11.2109 ) &  0.8173  &  (7.8551, 11.0899)  &  0.6748  ( 4.6265 ) &  -0.0601  ( 1.4814 ) &  93.0  \\ \cline{2-9}
                               & $\xi$    &  -0.2  &  -0.2452  ( -0.2572 ) &  0.1793  &  (-0.6018, 0.0904)  &  0.0357  ( 0.0896 ) &  -0.0452  ( -0.0572 ) &  96.2  \\ \cline{2-9}
                               & $\sigma$ &  5  &  5.194  ( 5.0608 ) &  1.1605  &  (2.633, 6.9814)  &  1.7614  ( 4.1574 ) &  0.194  ( 0.0608 ) &  94.3  \\ \hline
     \multirow{6}{*}{ 750 } & $\alpha$ &  0.3  &  0.2995  (NA) &  0.0175  &  (0.2761, 0.3436)  &  3e-04  (NA) &  -5e-04  (NA) &  97.6  \\ \cline{2-9}
                               & $\eta$   &  1  &  1.0089  ( 1.0055 ) &  0.066  &  (0.9638, 1.2199)  &  0.0035  ( 0.0035 ) &  0.0089  ( 0.0055 ) &  96.2  \\ \cline{2-9}
                               & $\beta$  &  5  &  4.9526  ( 4.9802 ) &  0.3732  &  (3.6979, 5.1677)  &  0.1684  ( 0.1725 ) &  -0.0474  ( -0.0198 ) &  92.7  \\ \cline{2-9}
                               & $u$      &  9.7295  &  9.6758  ( 11.2946 ) &  0.6848  &  (7.9644, 10.6759)  &  0.5769  ( 4.4615 ) &  -0.0538  ( 1.565 ) &  92.3  \\ \cline{2-9}
                               & $\xi$    &  -0.2  &  -0.2251  ( -0.2335 ) &  0.1326  &  (-0.5607, -0.0453)  &  0.0201  ( 0.0483 ) &  -0.0251  ( -0.0335 ) &  93.0  \\ \cline{2-9}
                               & $\sigma$ &  5  &  5.083  ( 4.8666 ) &  0.9574  &  (3.325, 7.0527)  &  1.0449  ( 2.4929 ) &  0.083  ( -0.1334 ) &  93.4  \\ \hline
    \multirow{6}{*}{ 1000 } & $\alpha$ &  0.3  &  0.2993  (NA) &  0.0153  &  (0.2772, 0.337)  &  2e-04  (NA) &  -7e-04  (NA) &  96.9  \\ \cline{2-9}
                               & $\eta$   &  1  &  1.0067  ( 1.0036 ) &  0.0543  &  (0.9258, 1.1389)  &  0.0025  ( 0.0025 ) &  0.0067  ( 0.0036 ) &  96.6  \\ \cline{2-9}
                               & $\beta$  &  5  &  4.9582  ( 4.9841 ) &  0.3486  &  (4.0399, 5.4046)  &  0.1241  ( 0.1208 ) &  -0.0418  ( -0.0159 ) &  92.2  \\ \cline{2-9}
                               & $u$      &  9.7295  &  9.6623  ( 11.3597 ) &  0.666  &  (8.1226, 10.7544)  &  0.5072  ( 4.178 ) &  -0.0673  ( 1.6302 ) &  91.6  \\ \cline{2-9}
                               & $\xi$    &  -0.2  &  -0.2176  ( -0.2235 ) &  0.108  &  (-0.5147, -0.0892)  &  0.0135  ( 0.0316 ) &  -0.0176  ( -0.0235 ) &  93.5  \\ \cline{2-9}
                               & $\sigma$ &  5  &  5.0583  ( 4.7835 ) &  0.8214  &  (3.5848, 6.75)  &  0.7161  ( 1.6974 ) &  0.0583  ( -0.2165 ) &  94.1  \\ \hline
    \end{tabular}
\captionsetup{
  labelfont=bf, 
  textfont=normal 
}
     \caption{Simulation results based on \( N = 2000 \) replications for sample sizes \( n = 300, 400, 500, 750, \) and \( 1000 \), with an inlier proportion of 30\% and light-tailed generalized Pareto distributions (GPD). The table presents the sample mean of estimated parameters, bootstrap standard error (BSE), 95\% bootstrap confidence interval (BCI), mean square error (MSE), bias, and convergence probability (CP).}
   \label{Table-1400}
\end{table*}
\end{landscape}

\begin{landscape} 
\begin{table*}[htbp]
    \centering
    \scriptsize 
    \setlength{\tabcolsep}{3pt} 
    \renewcommand{\arraystretch}{1.3} 
    \begin{tabular}{|p{2.5cm}|p{1.8cm}|p{1.8cm}|p{2.4cm}|p{1.8cm}|p{2.3cm}|p{2.3cm}|p{2.3cm}|p{1.8cm}|} 
        \hline
        \textbf{Sample Size} & \textbf{Parameters} & \textbf{True Value} & \textbf{Sample Mean} & \textbf{BSE} & \textbf{BCI} & \textbf{MSE} & \textbf{Bias} & \textbf{CP} \\ \hline

   \multirow{6}{*}{ 300 } & $\alpha$ &  0.1  &  0.1004  (NA) &  0.0162  &  (0.0496, 0.112)  &  3e-04  (NA) &  4e-04  (NA) &  97.1  \\ \cline{2-9}
                               & $\eta$   &  1  &  1.0244  ( 1.0095 ) &  0.0852  &  (0.8592, 1.1888)  &  0.0081  ( 0.0069 ) &  0.0244  ( 0.0095 ) &  96.6  \\ \cline{2-9}
                               & $\beta$  &  5  &  4.8909  ( 4.9837 ) &  0.5308  &  (3.6111, 5.6984)  &  0.3227  ( 0.3419 ) &  -0.1091  ( -0.0163 ) &  92.4  \\ \cline{2-9}
                               & $u$      &  10.9861  &  11.3779  ( 11.3334 ) &  1.175  &  (8.6262, 13.3931)  &  1.8911  ( 4.0259 ) &  0.3918  ( 0.3473 ) &  93.5  \\ \cline{2-9}
                               & $\xi$    &  0  &  -0.0687  ( -0.0936 ) &  0.3298  &  (-0.9366, 0.4759)  &  0.141  ( 0.1965 ) &  -0.0687  ( -0.0936 ) &  93.8  \\ \cline{2-9}
                               & $\sigma$ &  5  &  5.5914  ( 5.9055 ) &  2.6967  &  (2.3367, 13.2405)  &  6.3694  ( 10.0406 ) &  0.5914  ( 0.9055 ) &  92.9  \\ \hline
   \multirow{6}{*}{ 400 } & $\alpha$ &  0.1  &  0.1003  (NA) &  0.0147  &  (0.0643, 0.1202)  &  2e-04  (NA) &  3e-04  (NA) &  97.5  \\ \cline{2-9}
                               & $\eta$   &  1  &  1.0219  ( 1.0067 ) &  0.0755  &  (0.875, 1.173)  &  0.0063  ( 0.0052 ) &  0.0219  ( 0.0067 ) &  95.7  \\ \cline{2-9}
                               & $\beta$  &  5  &  4.9032  ( 4.9996 ) &  0.4914  &  (3.7926, 5.755)  &  0.256  ( 0.2786 ) &  -0.0968  ( -4e-04 ) &  91.3  \\ \cline{2-9}
                               & $u$      &  10.9861  &  11.425  ( 11.4268 ) &  1.2205  &  (8.816, 13.7358)  &  1.8532  ( 4.3821 ) &  0.4389  ( 0.4406 ) &  92.7  \\ \cline{2-9}
                               & $\xi$    &  0  &  -0.0427  ( -0.07 ) &  0.2583  &  (-0.7305, 0.341)  &  0.0953  ( 0.1439 ) &  -0.0427  ( -0.07 ) &  94.3  \\ \cline{2-9}
                               & $\sigma$ &  5  &  5.3886  ( 5.6639 ) &  2.0395  &  (2.7665, 10.9676)  &  4.341  ( 7.2994 ) &  0.3886  ( 0.6639 ) &  92.7  \\ \hline
   \multirow{6}{*}{ 500 } & $\alpha$ &  0.1  &  0.1  (NA) &  0.0129  &  (0.0632, 0.1137)  &  2e-04  (NA) &  0  (NA) &  97.1  \\ \cline{2-9}
                               & $\eta$   &  1  &  1.0187  ( 1.0063 ) &  0.0632  &  (0.8625, 1.1126)  &  0.0048  ( 0.0042 ) &  0.0187  ( 0.0063 ) &  96.3  \\ \cline{2-9}
                               & $\beta$  &  5  &  4.9132  ( 4.9942 ) &  0.4099  &  (4.1425, 5.7675)  &  0.2  ( 0.2198 ) &  -0.0868  ( -0.0058 ) &  92.0  \\ \cline{2-9}
                               & $u$      &  10.9861  &  11.5504  ( 11.4795 ) &  1.1283  &  (9.1287, 13.5132)  &  2.1184  ( 4.143 ) &  0.5643  ( 0.4934 ) &  91.2  \\ \cline{2-9}
                               & $\xi$    &  0  &  -0.0282  ( -0.0666 ) &  0.2645  &  (-0.4856, 0.5089)  &  0.0725  ( 0.1091 ) &  -0.0282  ( -0.0666 ) &  94.7  \\ \cline{2-9}
                               & $\sigma$ &  5  &  5.2854  ( 5.5703 ) &  1.4156  &  (1.9517, 7.1467)  &  3.4103  ( 6.0285 ) &  0.2854  ( 0.5703 ) &  91.1  \\ \hline
        \multirow{6}{*}{ 750 } & $\alpha$ &  0.1  &  0.0997  (NA) &  0.0111  &  (0.0734, 0.1166)  &  1e-04  (NA) &  -3e-04  (NA) &  97.8  \\ \cline{2-9}
                               & $\eta$   &  1  &  1.0172  ( 1.0052 ) &  0.0563  &  (0.9318, 1.1507)  &  0.0033  ( 0.0028 ) &  0.0172  ( 0.0052 ) &  96.0  \\ \cline{2-9}
                               & $\beta$  &  5  &  4.9135  ( 4.9875 ) &  0.338  &  (3.938, 5.2803)  &  0.1387  ( 0.1401 ) &  -0.0865  ( -0.0125 ) &  91.9  \\ \cline{2-9}
                               & $u$      &  10.9861  &  11.5867  ( 11.5061 ) &  1.2677  &  (8.7386, 13.6877)  &  2.1461  ( 3.0262 ) &  0.6005  ( 0.52 ) &  90.8  \\ \cline{2-9}
                               & $\xi$    &  0  &  -0.0053  ( -0.0393 ) &  0.1997  &  (-0.4799, 0.3132)  &  0.0467  ( 0.0536 ) &  -0.0053  ( -0.0393 ) &  90.9  \\ \cline{2-9}
                               & $\sigma$ &  5  &  5.127  ( 5.3248 ) &  1.3277  &  (2.6725, 7.543)  &  2.1452  ( 3.0423 ) &  0.127  ( 0.3248 ) &  90.1  \\ \hline
  \multirow{6}{*}{ 1000 } & $\alpha$ &  0.1  &  0.0993  (NA) &  0.0102  &  (0.0798, 0.12)  &  1e-04  (NA) &  -7e-04  (NA) &  98.3  \\ \cline{2-9}
                               & $\eta$   &  1  &  1.0137  ( 1.0044 ) &  0.0478  &  (0.9324, 1.1179)  &  0.0025  ( 0.0021 ) &  0.0137  ( 0.0044 ) &  95.4  \\ \cline{2-9}
                               & $\beta$  &  5  &  4.9287  ( 4.9877 ) &  0.3016  &  (4.2241, 5.432)  &  0.1068  ( 0.1062 ) &  -0.0713  ( -0.0123 ) &  91.9  \\ \cline{2-9}
                               & $u$      &  10.9861  &  11.646  ( 11.4842 ) &  1.4605  &  (9.1183, 14.727)  &  2.7888  ( 2.5617 ) &  0.6599  ( 0.4981 ) &  91.1  \\ \cline{2-9}
                               & $\xi$    &  0  &  -0.0018  ( -0.0259 ) &  0.1947  &  (-0.3378, 0.4128)  &  0.0377  ( 0.0393 ) &  -0.0018  ( -0.0259 ) &  91.1  \\ \cline{2-9}
                               & $\sigma$ &  5  &  5.1047  ( 5.2195 ) &  1.0369  &  (2.3039, 6.1576)  &  1.5795  ( 1.9047 ) &  0.1047  ( 0.2195 ) &  90.2  \\ \hline
    \end{tabular}
    \captionsetup{
  labelfont=bf, 
  textfont=normal 
}
     \caption{Simulation results based on \( N = 2000 \) replications for sample sizes \( n = 300, 400, 500, 750, \) and \( 1000 \), with an inlier proportion of 10\% and exponential-tailed generalized Pareto distributions (GPD). The table presents the sample mean of estimated parameters, bootstrap standard error (BSE), 95\% bootstrap confidence interval (BCI), mean square error (MSE), bias, and convergence probability (CP).}
   \label{Table-1500}
\end{table*}
\end{landscape}

\begin{landscape} 
\begin{table*}[htbp]
    \centering
    \scriptsize 
    \setlength{\tabcolsep}{3pt} 
    \renewcommand{\arraystretch}{1.3} 
    \begin{tabular}{|p{2.5cm}|p{1.8cm}|p{1.8cm}|p{2.3cm}|p{1.8cm}|p{2.3cm}|p{2.3cm}|p{2.3cm}|p{1.8cm}|} 
        \hline
        \textbf{Sample Size} & \textbf{Parameters} & \textbf{True Value} & \textbf{Sample Mean} & \textbf{BSE} & \textbf{BCI} & \textbf{MSE} & \textbf{Bias} & \textbf{CP} \\ \hline

   \multirow{6}{*}{ 300 } & $\alpha$ &  0.3  &  0.2998  (NA) &  0.0273  &  (0.2679, 0.3745)  &  7e-04  (NA) &  -2e-04  (NA) &  96.1  \\ \cline{2-9}
                               & $\eta$   &  1  &  1.0221  ( 1.0171 ) &  0.1184  &  (0.9971, 1.4622)  &  0.0097  ( 0.0097 ) &  0.0221  ( 0.0171 ) &  97.2  \\ \cline{2-9}
                               & $\beta$  &  5  &  4.9016  ( 4.9524 ) &  0.536  &  (3.0424, 5.1601)  &  0.4146  ( 0.4579 ) &  -0.0984  ( -0.0476 ) &  91.3  \\ \cline{2-9}
                               & $u$      &  9.7295  &  9.5766  ( 11.2167 ) &  0.8949  &  (7.2611, 10.8713)  &  0.9534  ( 6.4922 ) &  -0.153  ( 1.4872 ) &  94.8  \\ \cline{2-9}
                               & $\xi$    &  0  &  -0.0868  ( -0.1075 ) &  0.2737  &  (-0.7358, 0.3839)  &  0.0905  ( 0.2823 ) &  -0.0868  ( -0.1075 ) &  93.5  \\ \cline{2-9}
                               & $\sigma$ &  5  &  5.5018  ( 6.1499 ) &  2.0387  &  (2.4432, 10.4816)  &  4.4596  ( 13.8242 ) &  0.5018  ( 1.1499 ) &  94.1  \\ \hline
      \multirow{6}{*}{ 400 } & $\alpha$ &  0.3  &  0.2996  (NA) &  0.0238  &  (0.2722, 0.3646)  &  5e-04  (NA) &  -4e-04  (NA) &  97.3  \\ \cline{2-9}
                               & $\eta$   &  1  &  1.0169  ( 1.0117 ) &  0.0927  &  (0.9304, 1.2994)  &  0.0069  ( 0.007 ) &  0.0169  ( 0.0117 ) &  97.0  \\ \cline{2-9}
                               & $\beta$  &  5  &  4.9227  ( 4.9766 ) &  0.5117  &  (3.3609, 5.3739)  &  0.3138  ( 0.3561 ) &  -0.0773  ( -0.0234 ) &  93.6  \\ \cline{2-9}
                               & $u$      &  9.7295  &  9.5712  ( 11.3274 ) &  0.841  &  (7.307, 10.6563)  &  0.8071  ( 7.1992 ) &  -0.1583  ( 1.5978 ) &  93.3  \\ \cline{2-9}
                               & $\xi$    &  0  &  -0.0552  ( -0.0792 ) &  0.2151  &  (-0.6276, 0.2254)  &  0.0571  ( 0.2047 ) &  -0.0552  ( -0.0792 ) &  95.3  \\ \cline{2-9}
                               & $\sigma$ &  5  &  5.2892  ( 5.8434 ) &  1.5918  &  (2.9479, 9.2052)  &  2.7892  ( 9.8186 ) &  0.2892  ( 0.8434 ) &  94.1  \\ \hline
  \multirow{6}{*}{ 500 } & $\alpha$ &  0.3  &  0.2996  (NA) &  0.0212  &  (0.267, 0.3494)  &  4e-04  (NA) &  -4e-04  (NA) &  96.8  \\ \cline{2-9}
                               & $\eta$   &  1  &  1.0129  ( 1.0089 ) &  0.0761  &  (0.908, 1.2033)  &  0.0054  ( 0.0053 ) &  0.0129  ( 0.0089 ) &  96.2  \\ \cline{2-9}
                               & $\beta$  &  5  &  4.9415  ( 4.9773 ) &  0.471  &  (3.8489, 5.6909)  &  0.2583  ( 0.2703 ) &  -0.0585  ( -0.0227 ) &  92.7  \\ \cline{2-9}
                               & $u$      &  9.7295  &  9.57  ( 11.3408 ) &  0.7616  &  (7.8512, 10.8358)  &  0.6861  ( 6.6199 ) &  -0.1595  ( 1.6112 ) &  94.1  \\ \cline{2-9}
                               & $\xi$    &  0  &  -0.0398  ( -0.0706 ) &  0.1897  &  (-0.445, 0.3035)  &  0.0388  ( 0.1425 ) &  -0.0398  ( -0.0706 ) &  95.1  \\ \cline{2-9}
                               & $\sigma$ &  5  &  5.1865  ( 5.6851 ) &  1.1675  &  (2.5403, 6.9438)  &  1.926  ( 7.6904 ) &  0.1865  ( 0.6851 ) &  93.7  \\ \hline
     \multirow{6}{*}{ 750 } & $\alpha$ &  0.3  &  0.2994  (NA) &  0.0174  &  (0.2754, 0.3428)  &  3e-04  (NA) &  -6e-04  (NA) &  98.2  \\ \cline{2-9}
                               & $\eta$   &  1  &  1.0091  ( 1.005 ) &  0.0668  &  (0.9756, 1.2366)  &  0.0035  ( 0.0035 ) &  0.0091  ( 0.005 ) &  96.3  \\ \cline{2-9}
                               & $\beta$  &  5  &  4.9493  ( 4.9874 ) &  0.3638  &  (3.6508, 5.0603)  &  0.1733  ( 0.1818 ) &  -0.0507  ( -0.0126 ) &  93.3  \\ \cline{2-9}
                               & $u$      &  9.7295  &  9.5313  ( 11.4993 ) &  0.6311  &  (7.841, 10.3399)  &  0.5519  ( 6.5483 ) &  -0.1982  ( 1.7697 ) &  92.3  \\ \cline{2-9}
                               & $\xi$    &  0  &  -0.0216  ( -0.0497 ) &  0.1371  &  (-0.4071, 0.1519)  &  0.0233  ( 0.0796 ) &  -0.0216  ( -0.0497 ) &  93.4  \\ \cline{2-9}
                               & $\sigma$ &  5  &  5.0777  ( 5.4418 ) &  0.9646  &  (3.2103, 6.9568)  &  1.1818  ( 4.4643 ) &  0.0777  ( 0.4418 ) &  94.2  \\ \hline
       \multirow{6}{*}{ 1000 } & $\alpha$ &  0.3  &  0.2994  (NA) &  0.015  &  (0.2714, 0.3294)  &  2e-04  (NA) &  -6e-04  (NA) &  97.9  \\ \cline{2-9}
                               & $\eta$   &  1  &  1.0073  ( 1.0026 ) &  0.0536  &  (0.9336, 1.1429)  &  0.0026  ( 0.0025 ) &  0.0073  ( 0.0026 ) &  97.1  \\ \cline{2-9}
                               & $\beta$  &  5  &  4.9534  ( 4.9945 ) &  0.339  &  (3.9847, 5.3078)  &  0.1314  ( 0.1283 ) &  -0.0466  ( -0.0055 ) &  93.2  \\ \cline{2-9}
                               & $u$      &  9.7295  &  9.5273  ( 11.5121 ) &  0.5917  &  (8.0346, 10.3676)  &  0.4743  ( 5.8316 ) &  -0.2022  ( 1.7825 ) &  91.7  \\ \cline{2-9}
                               & $\xi$    &  0  &  -0.0134  ( -0.0366 ) &  0.1117  &  (-0.3471, 0.0923)  &  0.0167  ( 0.0506 ) &  -0.0134  ( -0.0366 ) &  93.0  \\ \cline{2-9}
                               & $\sigma$ &  5  &  5.0407  ( 5.3181 ) &  0.8359  &  (3.6651, 6.8606)  &  0.8225  ( 2.7629 ) &  0.0407  ( 0.3181 ) &  93.7  \\ \hline
    \end{tabular}
    \captionsetup{
  labelfont=bf, 
  textfont=normal 
}
     \caption{Simulation results based on \( N = 2000 \) replications for sample sizes \( n = 300, 400, 500, 750, \) and \( 1000 \), with an inlier proportion of 30\% and exponential-tailed generalized Pareto distributions (GPD). The table presents the sample mean of estimated parameters, bootstrap standard error (BSE), 95\% bootstrap confidence interval (BCI), mean square error (MSE), bias, and convergence probability (CP).}
   \label{Table-1600}
\end{table*}
\end{landscape}
\appendix

\addcontentsline{toc}{section}{Appendix A}

\end{document}